[ english version / português abaixo ]

# **Anti-woke agenda, gender issues, revisionism and hate speech communities on Brazilian Telegram:** from harmful reactionary speech to the crime of glorifying Nazism and Hitler

*Ergon Cugler de Moraes Silva*

Brazilian Institute of Information in
Science and Technology (IBICT)
Brasília, Federal District, Brazil

contato@ergoncugler.com
www.ergoncugler.com

## Abstract

Resistance to progressive policies and hate speech have been consolidating on Brazilian Telegram, with anti-woke communities rejecting diversity and promoting a worldview that sees these social changes as a threat. Therefore, this study aims to address the research question: **how are Brazilian conspiracy theory communities on anti-woke agenda, gender issues, revisionism and hate speech topics characterized and articulated on Telegram?** It is worth noting that this study is part of a series of seven studies whose main objective is to understand and characterize Brazilian conspiracy theory communities on Telegram. This series of seven studies is openly and originally available on arXiv at Cornell University, applying a mirrored method across the seven studies, changing only the thematic object of analysis and providing investigation replicability, including with proprietary and authored codes, adding to the culture of free and open-source software. Regarding the main findings of this study, the following were observed: Anti-woke communities emerge as central forces in the Brazilian conspiracy ecosystem; During crises, mentions of hate speech and revisionism have increased significantly, reflecting polarization; Nazi communities on Telegram propagate extremist ideologies, glorifying Hitler; The interconnectivity between anti-woke, anti-gender and revisionism strengthens the ecosystem of hate; Anti-gender speech facilitates the spread of anti-vaccine disinformation, creating an intersection between health and conspiracy.

### Key findings

➔ Anti-woke communities emerge as central forces in the Brazilian conspiratorial ecosystem, with 1,293,430 posts and 154,391 users, rejecting progressive policies and promoting a worldview where diversity is seen as a threat to the established order, consolidating organized resistance against social changes;

➔ During global crises, such as the 2018 elections in Brazil and the COVID-19 pandemic, mentions of hate speech and revisionism increased by 1,100% between 2019 and 2021, reflecting exacerbated polarization and the expansion of disinformation in communities;

➔ Brazilian Nazi communities on Telegram propagate extremist ideologies, glorifying Hitler and disseminating hate speech against minorities, representing a grave threat to social security by promoting radicalization and normalizing extremist ideologies;

- ➔ The interconnectivity between anti-woke, anti-gender, and revisionism strengthens an ecosystem of hate and exclusion, with 1,729 links between these communities, creating a cohesive network that amplifies extremist ideologies and normalizes hate and exclusion;
- ➔ Anti-woke communities function as gateways to other conspiracy theories, such as the New World Order and revisionism, with 965 links to NWO and 599 to anti-vaccine narratives, introducing members to a broader spectrum of disinformation and radicalization;
- ➔ The impact of anti-woke and anti-gender communities on Telegram is significant, with 1,293,430 posts and 154,391 users, acting as disinformation hubs that amplify narratives against diversity and inclusion beyond their digital borders;
- ➔ There is a strong correlation between historical revisionism and anti-gender discourse, with 416 invitation links revealing this interconnectivity, suggesting that the forces that deny historical facts, such as the Holocaust, also reject gender equality policies;
- ➔ Anti-woke discourse accelerates social radicalization, with a 1,000% increase in mentions between 2020 and 2021, reflecting how global crises are used to intensify polarization and promote hate narratives within these communities;
- ➔ Anti-woke communities position themselves as resistance to globalization, with 978 links related to Globalism, presenting themselves as a defense against the supposed imposition of globalist values that threaten national sovereignty and identity;
- ➔ Anti-gender discourse facilitates the spread of anti-vaccine disinformation, with 599 identified links, using the anti-vaccine narrative as an extension of their opposition to a supposed global threat to individual freedom, creating an intersection between health and conspiracy.

## 1. Introduction

After analyzing thousands of Brazilian conspiracy theory communities on Telegram and extracting tens of millions of content pieces from these communities, created and/or shared by millions of users, this study aims to compose a series of seven studies that address the phenomenon of conspiracy theories on Telegram, focusing on Brazil as a case study. Through the identification approaches implemented, it was possible to reach a total of 109 Brazilian conspiracy theory communities on Telegram on anti-woke agenda, gender issues, revisionism and hate speech topics, summing up 1,159,678 content pieces published between July 2016 (initial publications) and August 2024 (date of this study), with 188,771 users aggregated from within these communities. Thus, this study aims to understand and characterize the communities focused on anti-woke agenda, gender issues, revisionism and hate speech present in this Brazilian network of conspiracy theories identified on Telegram.

To this end, a mirrored method will be applied across all seven studies, changing only the thematic object of analysis and providing investigation replicability. In this way, we will adopt technical approaches to observe the connections, temporal series, content, and overlaps of themes within the conspiracy theory communities. In addition to this study, the other six are openly and originally available on arXiv at Cornell University. This series paid particular attention to ensuring data integrity and respecting user privacy, as provided by Brazilian legislation (Law No. 13,709/2018 / Brazilian law from 2018).

Therefore, the question arises: **how are Brazilian conspiracy theory communities on anti-woke agenda, gender issues, revisionism and hate speech topics characterized and articulated on Telegram?**

## 2. Materials and methods

The methodology of this study is organized into three subsections: **2.1. Data extraction**, which describes the process and tools used to collect information from Telegram communities; **2.2. Data processing**, which discusses the criteria and methods applied to classify and anonymize the collected data; and **2.3. Approaches to data analysis**, which details the techniques used to investigate the connections, temporal series, content, and thematic overlaps within conspiracy theory communities.

### 2.1. Data extraction

This project began in February 2023 with the publication of the first version of TelegramScrap (Silva, 2023), a proprietary, free, and open-source tool that utilizes Telegram's Application Programming Interface (API) by Telethon library and organizes data extraction cycles from groups and open channels on Telegram. Over the months, the database was expanded and refined using four approaches to identifying conspiracy theory communities:

**(i) Use of keywords:** at the project's outset, keywords were listed for direct identification in the search engine of Brazilian groups and channels on Telegram, such as "apocalypse", "survivalism", "climate change", "flat earth", "conspiracy theory", "globalism", "new world order", "occultism", "esotericism", "alternative cures", "qAnon" "reptilians", "revisionism", "aliens", among others. This initial approach provided some communities whose titles and/or descriptions of groups and channels explicitly contained terms related to conspiracy theories. However, over time, it was possible to identify many other communities that the listed keywords did not encompass, some of which deliberately used altered characters to make it difficult for those searching for them on the network.

**(ii) Telegram channel recommendation mechanism:** over time, it was identified that Telegram channels (except groups) have a recommendation tab called "similar channels", where Telegram itself suggests ten channels that have some similarity with the channel being observed. Through this recommendation mechanism, it was possible to find more Brazilian conspiracy theory communities, with these being recommended by the platform itself.

**(iii) Snowball approach for invitation identification:** after some initial communities were accumulated for extraction, a proprietary algorithm was developed to identify URLs containing "t.me/", the prefix for any invitation to Telegram groups and channels. Accumulating a frequency of hundreds of thousands of links that met this criterion, the unique addresses were listed, and their repetitions counted. In this way, it was possible to investigate new Brazilian groups and channels mentioned in the messages of those already investigated, expanding the network. This process was repeated periodically to identify new communities aligned with conspiracy theory themes on Telegram.

**(iv) Expansion to tweets published on X mentioning invitations:** to further diversify the sources of Brazilian conspiracy theory communities on Telegram, a proprietary search query was developed to identify conspiracy theory-themed keywords using tweets published on X (formerly Twitter) that, in addition to containing one of the keywords, also included "t.me/", the prefix for any invitation to Telegram groups and channels, "https://x.com/search?q=lang%3Apt%20%22t.me%2F%22%20SEARCH-TERM".

With the implementation of community identification approaches for conspiracy theories developed over months of investigation and method refinement, it was possible to build a project database encompassing a total of 855 Brazilian conspiracy theory communities on Telegram (including other themes not covered in this study). These communities have collectively published 27,227,525 pieces of content from May 2016 (the first publications) to August 2024 (the period of this study), with a combined total of 2,290,621 users across the Brazilian communities. It is important to note that this volume of users includes two elements: first, it is a variable figure, as users can join and leave communities daily, so this value represents what was recorded on the publication extraction date; second, it is possible that the same user is a member of more than one group and, therefore, is counted more than once. In this context, while the volume remains significant, it may be slightly lower when considering the deduplicated number of citizens within these Brazilian conspiracy theory communities.

### 2.2. Data processing

With all the Brazilian conspiracy theory groups and channels on Telegram extracted, a manual classification was conducted considering the title and description of the community. If there was an explicit mention in the title or description of the community related to a specific theme, it was classified into one of the following categories: (i) "Anti-Science"; (ii) "Anti-Woke and Gender"; (iii) "Antivax"; (iv) "Apocalypse and Survivalism"; (v) "Climate Changes"; (vi) "Flat Earth"; (vii) "Globalism"; (viii) "New World Order"; (ix) "Occultism and Esotericism"; (x) "Off Label and Quackery"; (xi) "QAnon"; (xii) "Reptilians and Creatures"; (xiii) "Revisionism and Hate Speech"; (xiv) "UFO and Universe". If there was no explicit mention related to the themes in the title or description of the community, it was classified as (xv) "General Conspiracy". In the following table, we can observe the metrics related to the classification of these conspiracy theory communities in Brazil.

**Table 01.** Conspiracy theory communities in Brazil (metrics up to August 2024)

| Categories | Groups | Users | Contents | Comments | Total |
|---|---|---|---|---|---|
| **Anti-Science** | 22 | 58,138 | 187,585 | 784,331 | 971,916 |
| **Anti-Woke and Gender** | 43 | 154,391 | 276,018 | 1,017,412 | 1,293,430 |
| **Antivax** | 111 | 239,309 | 1,778,587 | 1,965,381 | 3,743,968 |
| **Apocalypse and Survivalism** | 33 | 109,266 | 915,584 | 429,476 | 1,345,060 |
| **Climate Changes** | 14 | 20,114 | 269,203 | 46,819 | 316,022 |
| **Flat Earth** | 33 | 38,563 | 354,200 | 1,025,039 | 1,379,239 |
| **General Conspiracy** | 127 | 498,190 | 2,671,440 | 3,498,492 | 6,169,932 |
| **Globalism** | 41 | 326,596 | 768,176 | 537,087 | 1,305,263 |
| **NWO** | 148 | 329,304 | 2,411,003 | 1,077,683 | 3,488,686 |
| **Occultism and Esotericism** | 39 | 82,872 | 927,708 | 2,098,357 | 3,026,065 |
| **Off Label and Quackery** | 84 | 201,342 | 929,156 | 733,638 | 1,662,794 |
| **QAnon** | 28 | 62,346 | 531,678 | 219,742 | 751,420 |
| **Reptilians and Creatures** | 19 | 82,290 | 96,262 | 62,342 | 158,604 |
| **Revisionism and Hate Speech** | 66 | 34,380 | 204,453 | 142,266 | 346,719 |
| **UFO and Universe** | 47 | 58,912 | 862,358 | 406,049 | 1,268,407 |
| **Total** | **855** | **2,296,013** | **13,183,411** | **14,044,114** | **27,227,525** |

Source: Own elaboration (2024).

With this volume of extracted data, it was possible to segment and present in this study only communities and content related to anti-woke agenda, gender issues, revisionism and hate speech themes. In parallel, other themes of Brazilian conspiracy theory communities were also addressed with studies aimed at characterizing the extent and dynamics of the network, which are openly and originally available on arXiv at Cornell University.

Additionally, it should be noted that only open communities were extracted, meaning those that are not only publicly identifiable but also do not require any request to access the content, being available to any user with a Telegram account who needs to join the group or channel. Furthermore, in compliance with Brazilian legislation, particularly the General Data Protection Law (LGPD), or Law No. 13,709/2018 (Brazilian law from 2018), which deals with privacy control and the use/treatment of personal data, all extracted data were anonymized for the purposes of analysis and investigation. Therefore, not even the identification of the communities is possible through this study, thus extending the user's privacy by anonymizing their data beyond the community itself to which they submitted by being in a public and open group or channel on Telegram.

### 2.3. Approaches to data analysis

A total of 109 selected communities focused on anti-woke agenda, gender issues, revisionism and hate speech themes, containing 1,159,678 publications and 188,771 combined users, will be analyzed. Four approaches will be used to investigate the conspiracy theory communities selected for the scope of this study. These metrics are detailed in the following table:

**Table 02.** Selected communities for analysis (metrics up to August 2024)

| Categories | Groups | Users | Contents | Comments | Total |
|---|---|---|---|---|---|
| **Anti-Woke and Gender** | 43 | 154,391 | 276,018 | 1,017,412 | 1,293,430 |
| **Revisionism and Hate Speech** | 66 | 34,380 | 204,453 | 142,266 | 346,719 |
| **Total** | **109** | **188,771** | **480,471** | **1,159,678** | **1,640,149** |

Source: Own elaboration (2024).

**(i) Network:** by developing a proprietary algorithm to identify messages containing the term "t.me/" (inviting users to join other communities), we propose to present volumes and connections observed on how **(a)** communities within the anti-woke agenda, gender issues, revisionism and hate speech theme circulate invitations for their users to explore more groups and channels within the same theme, reinforcing shared belief systems; and how **(b)** these same communities circulate invitations for their users to explore groups and channels dealing with other conspiracy theories, distinct from their explicit purpose. This approach is valuable for observing whether these communities use themselves as a source of legitimation and reference and/or rely on other conspiracy theory themes, even opening doors for their users to explore other conspiracies. Furthermore, it is worth mentioning the study by Rocha *et al.* (2024), where a network identification approach was also applied in Telegram communities, but by observing similar content based on an ID generated for each unique message and its similar ones;

**(ii) Time series:** the "Pandas" library (McKinney, 2010) is used to organize the investigation data frames, observing **(a)** the volume of publications over the months; and **(b)** the volume of engagement over the months, considering metadata of views, reactions, and comments collected during extraction. In addition to volumetry, the "Plotly" library (Plotly Technologies Inc., 2015) enabled the graphical representation of this variation;

**(iii) Content analysis:** in addition to the general word frequency analysis, time series are applied to the variation of the most frequent words over the semesters—observing from July 2016 (initial publications) to August 2024 (when this study was conducted). With the "Pandas" (McKinney, 2010) and "WordCloud" (Mueller, 2020) libraries, the results are presented both volumetrically and graphically;

**(iv) Thematic agenda overlap:** following the approach proposed by Silva & Sátiro (2024) for identifying thematic agenda overlap in Telegram communities, we used the "BERTopic" model (Grootendorst, 2020). BERTopic is a topic modeling algorithm that facilitates the generation of thematic representations from large amounts of text. First, the algorithm extracts document embeddings using sentence transformer models, such as "all-MiniLM-L6-v2". These embeddings are then reduced in dimensionality using techniques like "UMAP", facilitating the clustering process. Clustering is performed using "HDBSCAN", a density-based technique that identifies clusters of different shapes and sizes, as well as outliers. Subsequently, the documents are tokenized and represented in a bag-of-words structure, which is normalized (L1) to account for size differences between clusters. The topic representation is refined using a modified version of "TF-IDF", called "Class-TF-IDF", which considers the importance of words within each cluster (Grootendorst, 2020). It is important to note that before applying BERTopic, we cleaned the dataset by removing Portuguese "stopwords" using "NLTK" (Loper & Bird, 2002). For topic modeling, we used the "loky" backend to optimize performance during data fitting and transformation.

In summary, the methodology applied ranged from data extraction using the own tool TelegramScrap (Silva, 2023) to the processing and analysis of the collected data, employing various approaches to identify and classify Brazilian conspiracy theory communities on Telegram. Each stage was carefully implemented to ensure data integrity and respect for user privacy, as mandated by Brazilian legislation. The results of this data will be presented below, aiming to reveal the dynamics and characteristics of the studied communities.

## 3. Results

The results are detailed below in the order outlined in the methodology, beginning with the characterization of the network and its sources of legitimation and reference, progressing to the time series, incorporating content analysis, and concluding with the identification of thematic agenda overlap among the conspiracy theory communities.

### 3.1. Network

The analysis of the networks involving the themes of anti-woke, anti-gender, revisionism, and hate speech reveals the interconnectivity and cohesion of these communities within the conspiratorial ecosystem. The first figure maps the internal network among these communities, highlighting how their narratives not only overlap but mutually reinforce each other, creating an environment conducive to the propagation of extremist ideologies. The large spheres representing the main nodes in the network demonstrate the centrality of certain communities that act as primary channels for disseminating these ideas. The close connection between anti-woke and revisionism, for example, suggests that these discourses often go hand in hand, amplifying the impact and resonance of these ideologies both within and outside the network. In the second figure, we observe how the anti-woke, anti-gender, and revisionist communities act as gateways to new themes within the conspiratorial universe. These communities not only introduce followers to specific ideas but also connect them to a broader network of extremist narratives, such as hate speech. The interconnection of these communities with other narratives suggests that, once exposed to these discourses, individuals are quickly led to explore other areas of disinformation, increasing their engagement and radicalization within the ecosystem.

The third figure illustrates how communities centered on anti-woke and anti-gender discourses serve as exit points to other conspiracy theories, such as the New World Order and Historical Revisionism. The strong interconnectivity of these communities with areas like General Conspiracy and Occultism reflects the capacity of these themes to both feed and be fed by a broader spectrum of disinformation. This indicates that the polarization surrounding the anti-woke theme not only attracts but also facilitates the transition to other conspiratorial narratives, creating a continuous cycle of radicalization.

The invitation link flow charts highlight the importance of these themes within the conspiratorial ecosystem. The analysis of invitations between anti-woke and anti-gender communities, for example, reveals a strong connection with NWO and anti-vaccine narratives, indicating that these narratives are embedded in a broader context of resistance to progressive social changes. More intriguing is the flow of invitations sent to General Conspiracy and Globalism, demonstrating how these themes not only support but are also supported by broader narratives of global control. This suggests that anti-woke and anti-gender discourse operates as a key piece in a broader resistance against any state intervention perceived as a threat to individual freedom.

Finally, the invitation flow chart in revisionism and hate speech communities highlights how historical revisionism is often introduced as part of a broader narrative of distrust and opposition to the established order. The strong interconnection between revisionism and other forms of extremism, such as anti-woke and NWO, suggests a conscious strategy of alignment between different forms of extremism, where revisionism serves to solidify a worldview opposed to social progress. This convergence of narratives demonstrates that revisionism is not merely a denial of historical facts but an active attempt to rewrite history to legitimize contemporary forms of oppression, being an integral part of a broader disinformation agenda aimed at undermining trust in democratic institutions and science.

[ english version / português abaixo ]

**Figure 01.** Internal network between anti-woke, revisionism and hate speech communities

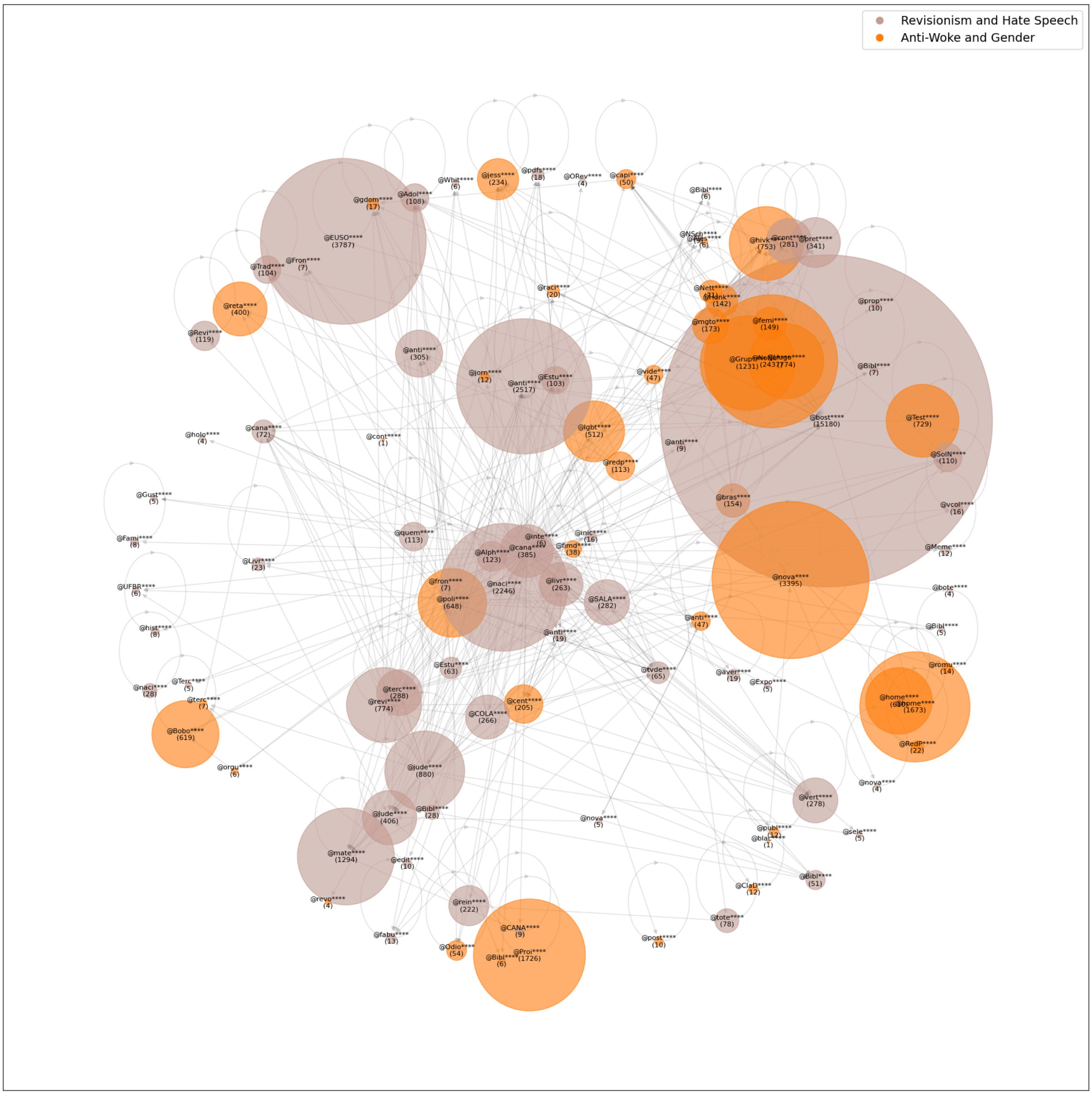

Source: Own elaboration (2024).

This figure maps the interconnection between communities that promote anti-woke, anti-gender, revisionist, and hate speech discourses. The network demonstrates a cohesive structure, where communities sharing these themes are closely connected. This suggests that narratives rejecting progressive and inclusive agendas are deeply intertwined with revisionist and hateful ideas, creating an environment where these ideologies mutually reinforce each other. The large nodes represent the main convergence points of these ideas, functioning as primary dissemination channels. The overlap between anti-woke and revisionism communities indicates that these ideologies are often presented together, increasing the strength of their propagation. The environment of this network suggests strong resistance to change and acceptance of diversity, contributing to the radicalization and maintenance of prejudiced and discriminatory beliefs.

[ english version / português abaixo ]

**Figure 02.** Network of communities that open doors to the theme (gateway)

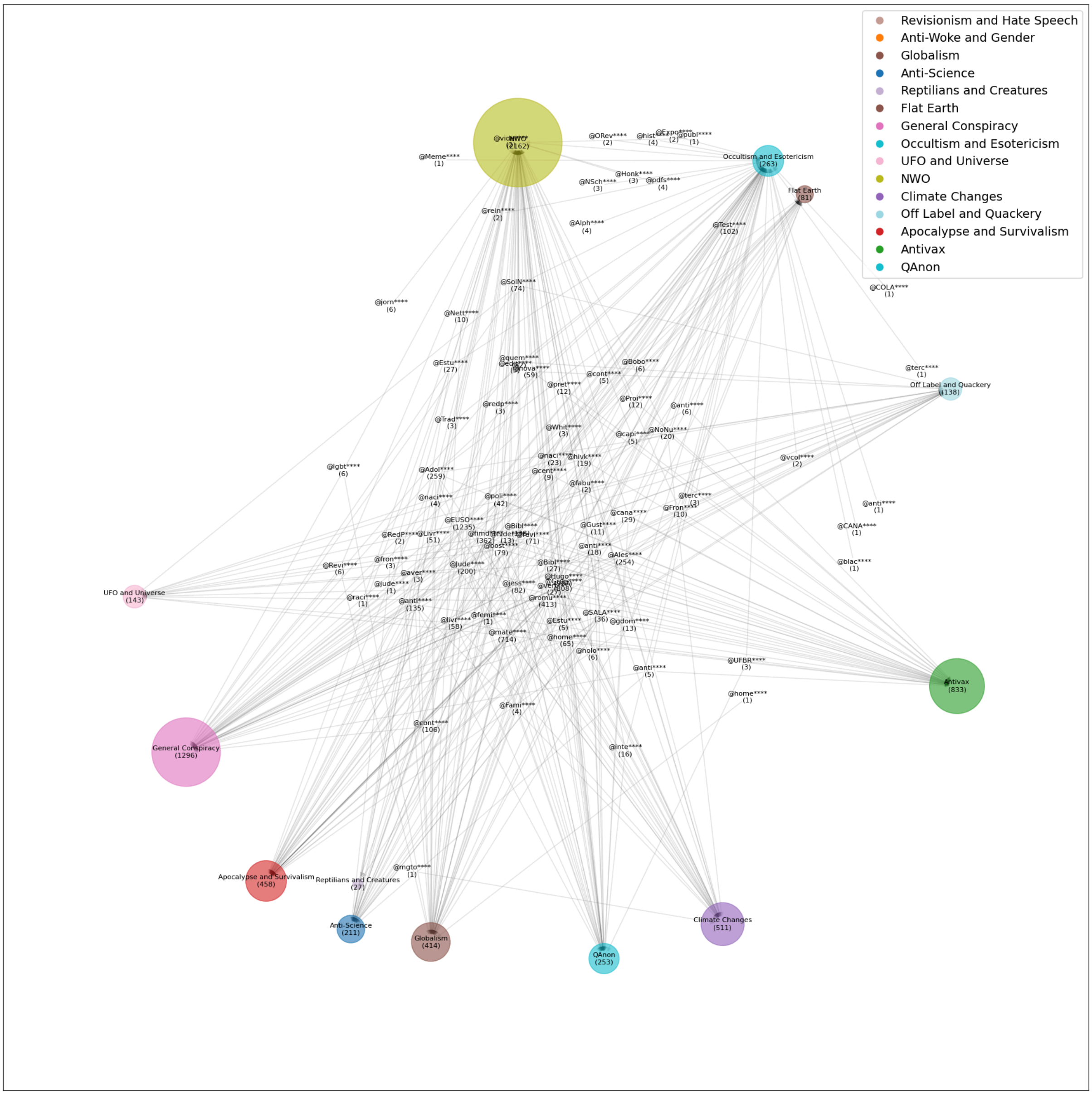

Source: Own elaboration (2024).

This figure reveals how communities promoting anti-woke, anti-gender, and revisionist discourses serve as entry points for new themes within the conspiratorial network. The interconnection between these communities and others, such as hate speech, reflects an environment where these ideas not only reinforce each other but also serve as gateways for deeper engagement in other extreme narratives. The chart suggests that individuals attracted by anti-woke discourse may quickly find themselves involved in broader discussions about revisionism and even hate narratives, making these communities central to the dissemination of a broader spectrum of conspiratorial ideologies.

**Figure 03.** Network of communities whose theme opens doors (exit point)

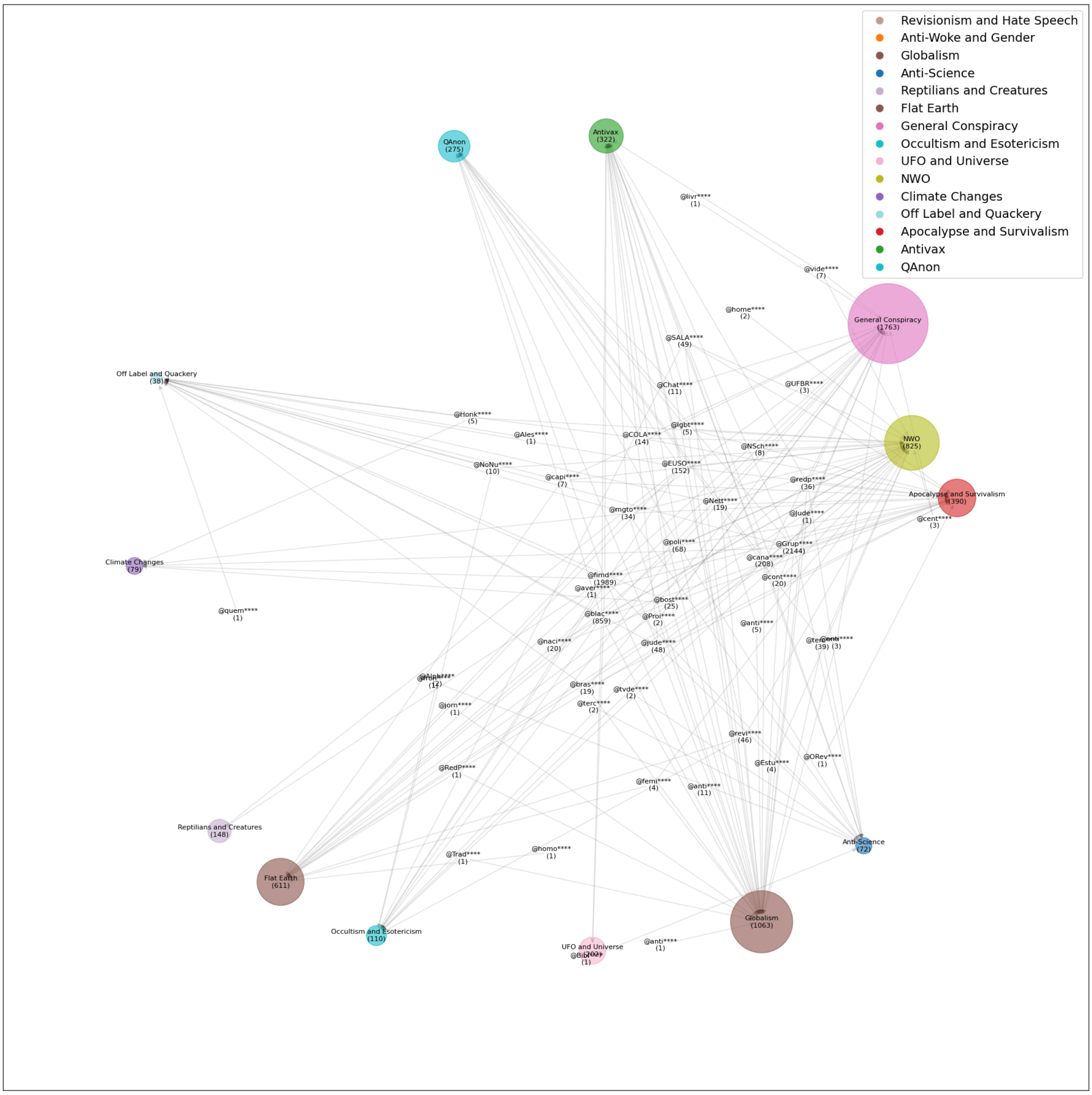

Source: Own elaboration (2024).

In the chart illustrating the connections around anti-woke themes, it is observed that these communities have significant ties to a range of other conspiratorial narratives. The figure suggests that anti-woke discussions often serve as a starting point for the introduction of other theories, such as the New World Order and Historical Revisionism. The interconnectivity of anti-woke communities with other areas like General Conspiracy and Occultism reflects how polarization around the anti-woke theme can feed and be fed by a broader spectrum of conspiratorial theories. These connections indicate that once inside the anti-woke universe, individuals have easy access to a wider network of disinformation, where radical beliefs can be reinforced and expanded, leading to increasing radicalization within the conspiratorial ecosystem.

[ english version / português abaixo ]

**Figure 04.** Flow of invitation links between anti-woke and gender issues communities

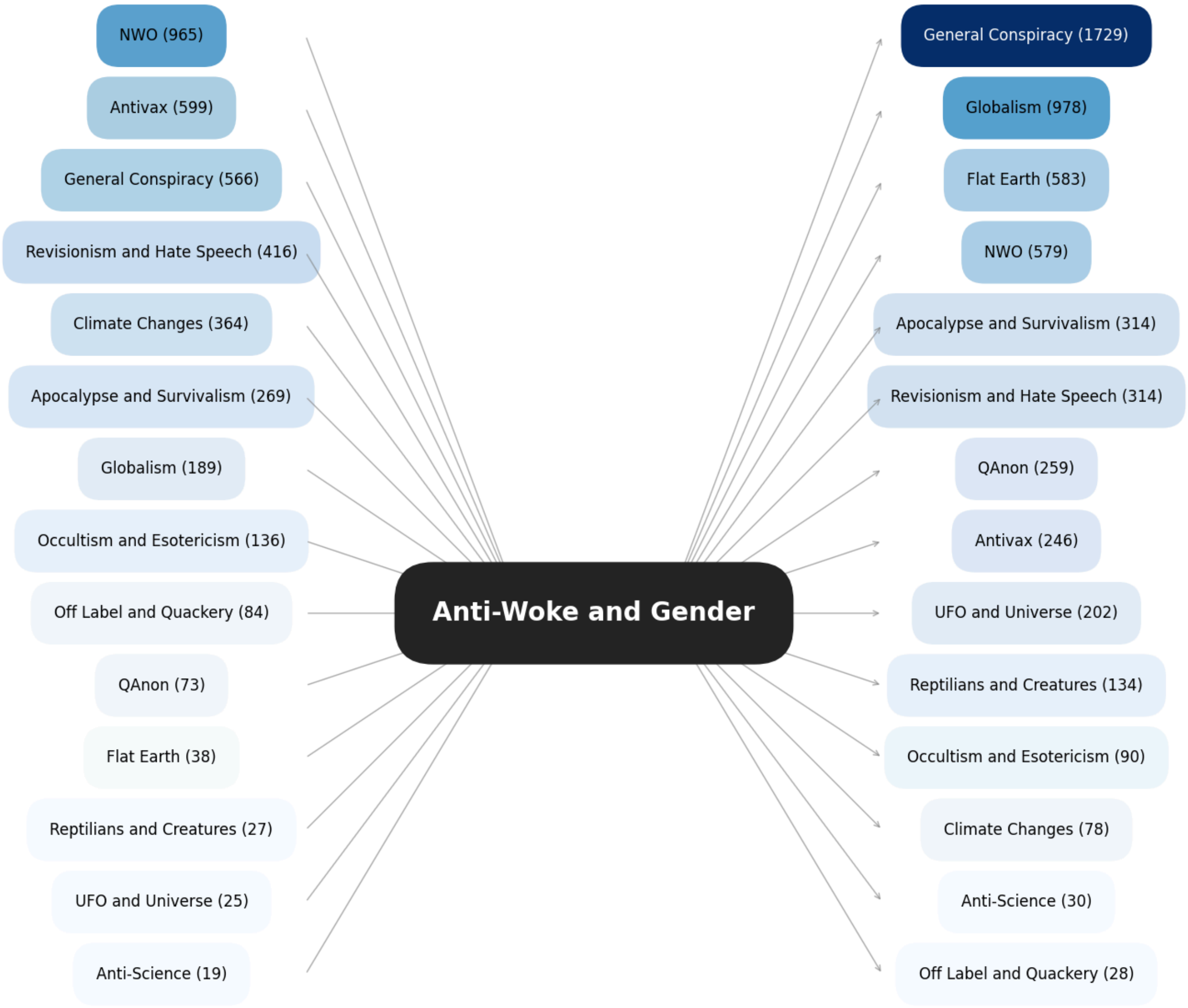

Source: Own elaboration (2024).

The chart related to the Anti-Woke and Gender theme highlights that this theme attracts a diverse set of other theories, with NWO (965 links) and Anti-Vaccines (599 links) among the main ones. This interconnection suggests that Anti-Woke and Gender discourse is embedded in a broader context of resistance to progressive social changes, where narratives of distrust and opposition to public health policies, such as anti-vaccines, find fertile ground. More interesting is the direction of invitations issued by these communities to General Conspiracies (1,729 links) and Globalism (978 links). This reveals a dynamic in which ideas opposing diversity and inclusion policies are not only supported by but also support broader narratives of global control and manipulation. Anti-Woke and Gender, therefore, cannot be seen as simply resistance to identity politics, but as a key piece in a broader mosaic of opposition to any form of state intervention perceived as a threat to individual freedom, as is often the case in anti-vaccine and globalism narratives.

**Figure 05.** Flow of invitation links between revisionism and hate speech communities

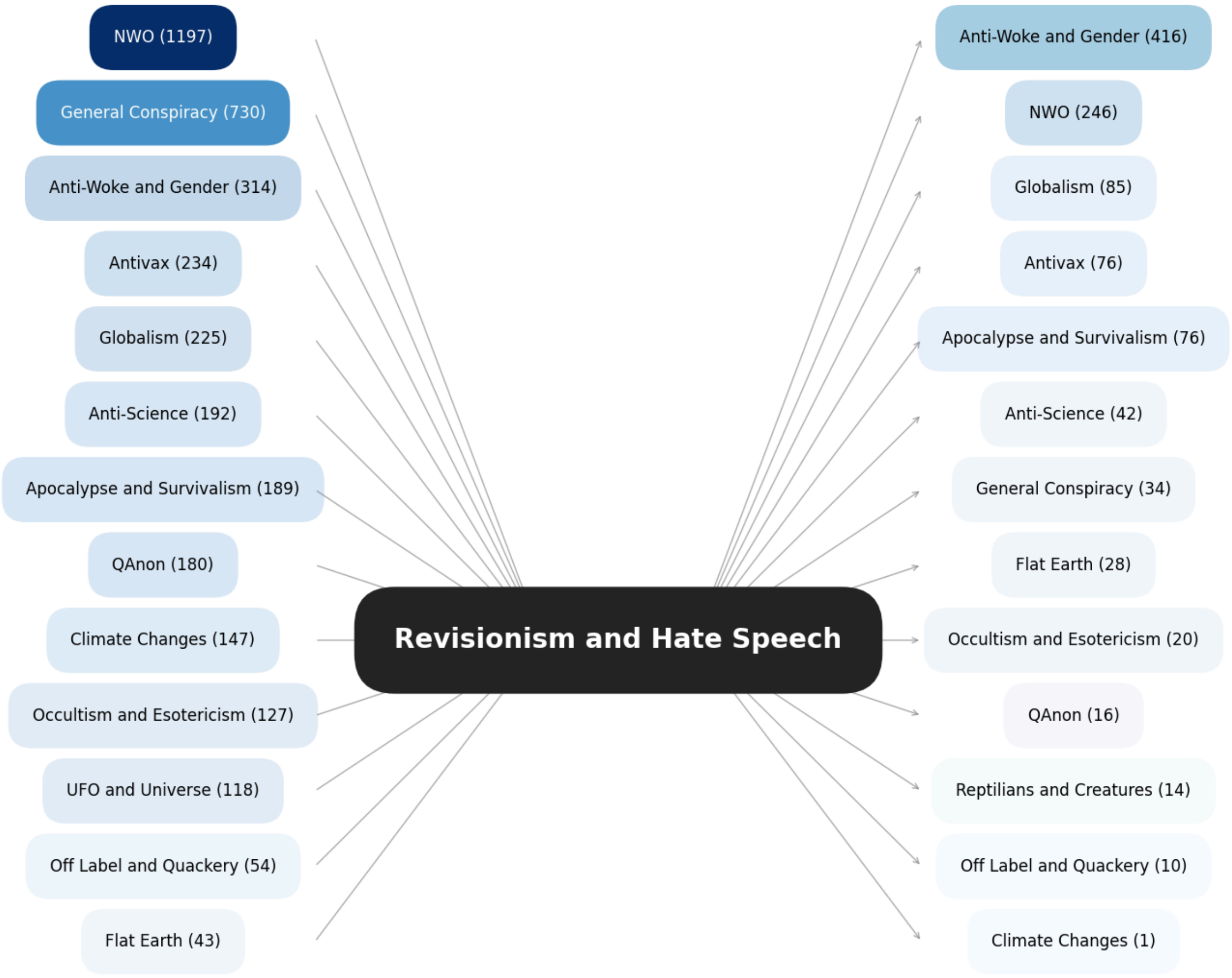

Source: Own elaboration (2024).

The Revisionism and Hate chart highlights the role of this community as a space for articulating different forms of extremism. NWO (1,197 links) and General Conspiracies (730 links) appear as the main sources, suggesting that historical revisionism, including the denial of events like the Holocaust, is often introduced to individuals as part of a broader narrative of distrust and opposition to the established order. The direction of invitations to Anti-Woke and Gender (416 links) and NWO (246 links) points to a conscious strategy of alignment between different forms of extremism, where revisionism serves to solidify a worldview that is, by definition, opposed to any form of social progress. This convergence of narratives suggests that revisionism is not simply a denial of facts but an active attempt to rewrite history to legitimize contemporary forms of oppression, being part of a broader disinformation agenda aimed at undermining trust in democratic institutions and science.

### 3.2. Time series

The analysis of time series for the categories Anti-Woke and Gender, as well as Revisionism and Hate, reveals a significant increase in mentions over the past few years. As we will see in the following graph, the rise in discussions around identity politics and social justice, especially following the emergence of movements like Black Lives Matter, catalyzed polarized debates, reflected in the sharp growth of mentions of Anti-Woke and Revisionism. From 2022 onward, the pace of these discussions stabilizes, indicating a lower intensity but a consistent presence. This phenomenon demonstrates how cultural battles have become central to social and political discussions, absorbing narratives of hate and historical revisionism.

**Figure 06.** Line graph over the period

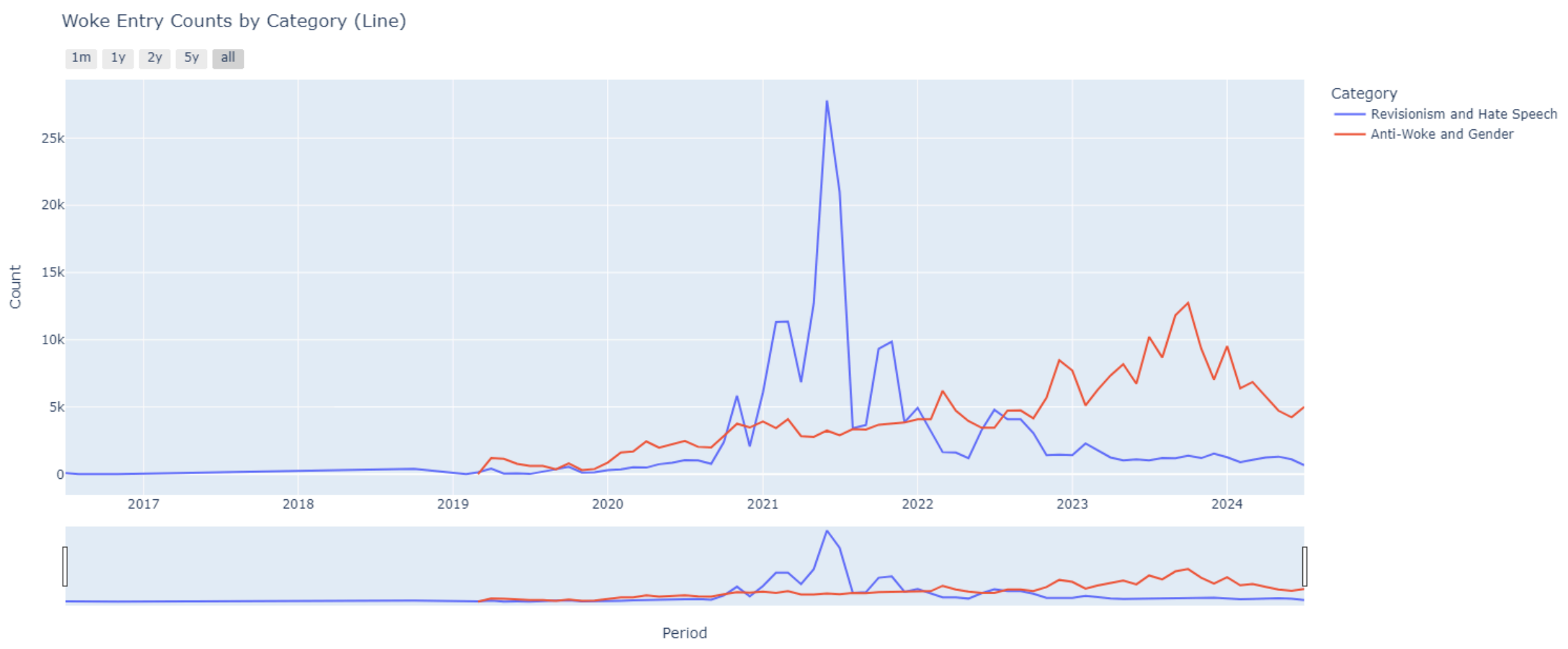

Source: Own elaboration (2024).

Posts related to Anti-Woke and Gender saw a 1,000% increase in mentions between 2019 and 2021, with peaks ranging from 10,000 to 20,000 mentions, driven by intense debates on identity and social justice issues. Compared to early 2020, when mentions were around 2,000, the growth was massive, reaching 20,000 mentions at its peak. Revisionism and Hate followed a similar pattern, growing from approximately 1,000 mentions to 12,000 in March 2021, an increase of 1,100%. This growth reveals how historical revisionism and hate speech feed off narratives opposing progressive policies. From 2022 onward, both categories show a stabilization with a decrease of about 20% compared to the maximum peaks, indicating that although the debate has lost some of its intensity, it remains present.

[ english version / português abaixo ]

**Figure 07.** Absolute area chart over the period

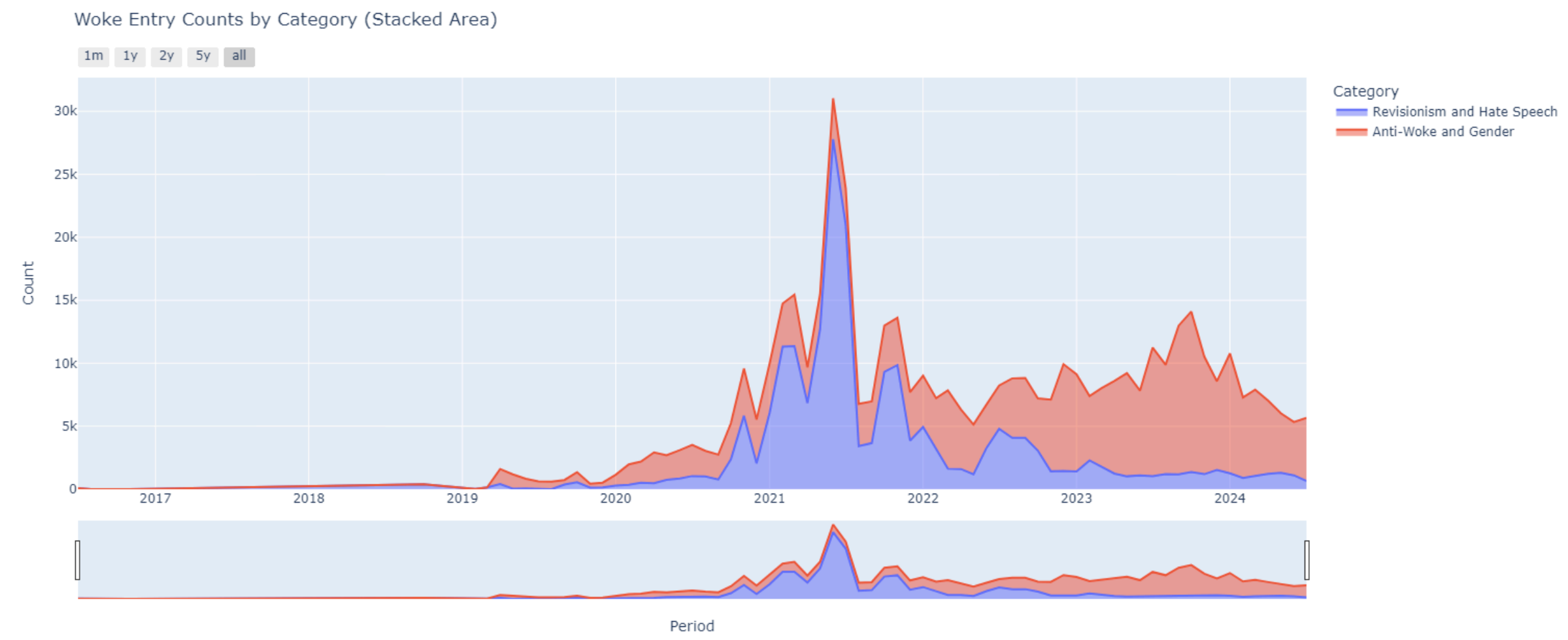

Source: Own elaboration (2024).

This graph shows a significant increase in discussions related to Revisionism and Hate and Anti-Woke starting in 2020, with a sharp peak in 2021. The Anti-Woke and Gender movement, which dominates in terms of absolute volume, reflects a growing backlash against inclusion and social justice policies that gained prominence globally during and after the pandemic. This peak coincides with events such as the Black Lives Matter protests and discussions on gender identity, which polarized public debate, especially in the United States and other Western countries. The increase in mentions of Revisionism and Hate is also significant, indicating a resurgence of revisionist narratives that seek to rewrite or deny aspects of history, often to justify ideologies of supremacy or exclusion. These peaks can be associated with the reaction against perceived social changes with the rise of neo-Nazi movements, illustrating how crises and social aspects can fuel radicalism and polarization.

**Figure 08.** Relative area chart over the period

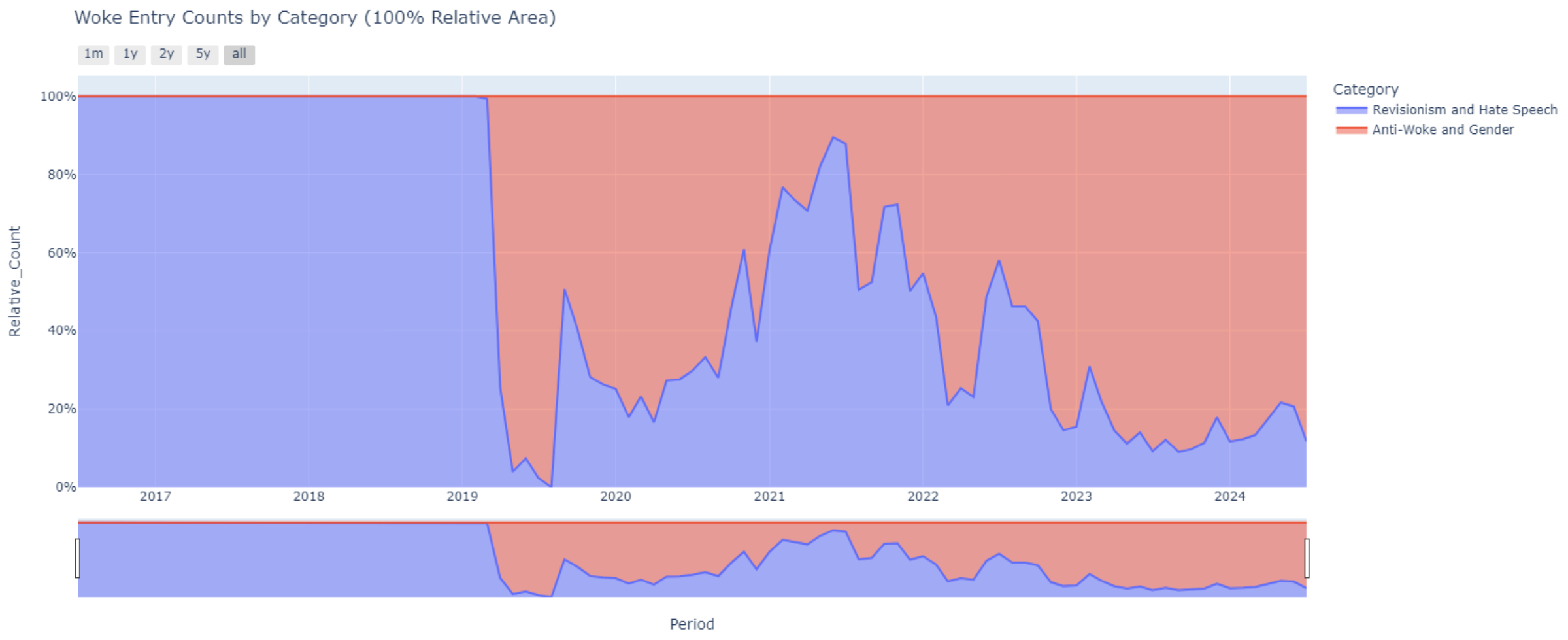

Source: Own elaboration (2024).

In the relative area graph, the Anti-Woke and Gender category progressively dominates from 2020 onward, reflecting the growing attention and controversy surrounding these issues. The relative decline in mentions of Revisionism and Hate, especially compared to the Anti-Woke movement, may suggest that while historical revisionism remains a concern, it is increasingly subsumed within a larger narrative of resistance to cultural and social changes promoted by the Woke movement. This shift may also indicate that as debates on gender and social justice become more prominent, they absorb some of the discursive space previously occupied by revisionist narratives. The graph suggests that cultural battles have become the most visible battleground, while historical revisionism continues to play a secondary but persistent role within the same spheres of radicalization.

### 3.3. Content analysis

The content analysis of communities related to anti-woke, revisionism, and hate speech, through word clouds, reveals the main themes and discursive dynamics that sustain these narratives within the digital environment. By observing the most frequent terms, such as “Brazil,” “world,” “God,” and “Jew,” it is possible to perceive an intersection of topics that combine nationalist sentiments with a religious and socially exclusionary rhetoric. The presence of these terms indicates how these communities seek to build a collective identity based on principles of opposition to what they consider cultural and social threats, especially those linked to progressive movements and external influences, often associated with global conspiracies. Through the highlighted words, it becomes evident that these communities are actively involved in constructing and disseminating narratives aimed at reinforcing divisions and promoting a polarized worldview, where otherness is constantly villainized and marginalized. The analysis of word clouds not only identifies the most recurrent terms but also offers an overview of how these discussions evolve and adapt to contextual changes, revealing the discursive strategies used to maintain and expand these ideologies.

**Figure 09.** Consolidated word cloud for anti-woke, revisionism and hate speech

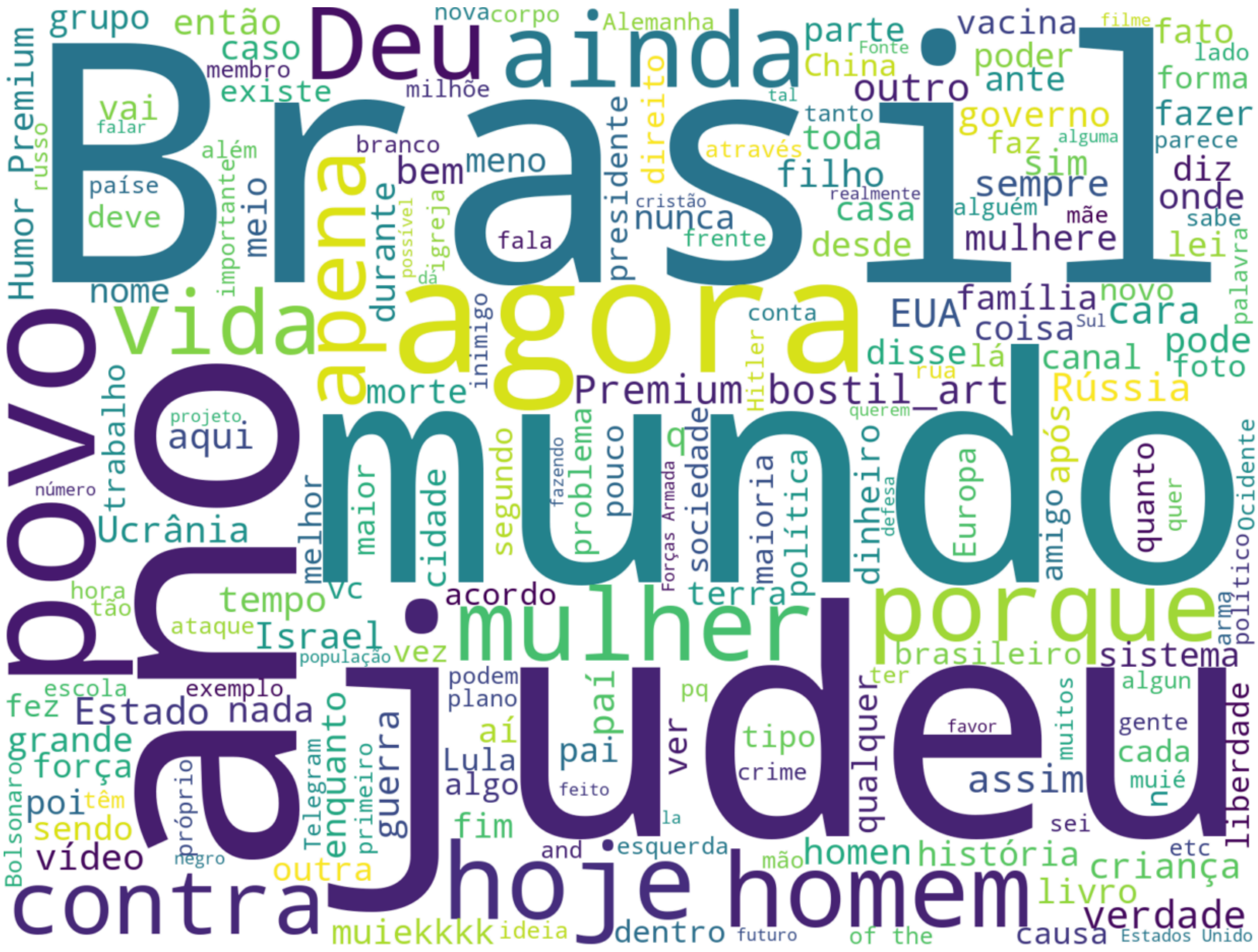

Source: Own elaboration (2024).

The consolidated word cloud of discussions involving anti-woke, revisionism, and hate speech highlights central terms such as “Brazil,” “world,” “God,” “now,” and “Jew.” The word “Brazil” holds a prominent position, indicating a strong presence of nationalist sentiments permeating these communities. This term is frequently associated with narratives defending the country against external or internal influences perceived as threats. “World” and “God” appear frequently, suggesting that discussions in these groups are not limited to the national context but also involve a spiritual and global dimension, where religious issues are mobilized to justify ideological positions. The presence of “Jew” in the word cloud reflects a strong antisemitic component within these communities, where hatred against Jews is often articulated as part of a broader conspiracy theory. Terms like “now” and “woman” indicate a concern with the present time and gender issues, respectively, highlighting that these discussions are deeply rooted in contemporary social themes. The combination of these terms reveals how anti-woke and revisionist narratives are constructed on a foundation that mixes nationalism, religion, and social exclusion, creating a discourse aimed at reinforcing divisions and polarizations in society.

[ english version / português abaixo ]

**Chart 01.** Temporal word cloud series for anti-woke agenda and gender issues

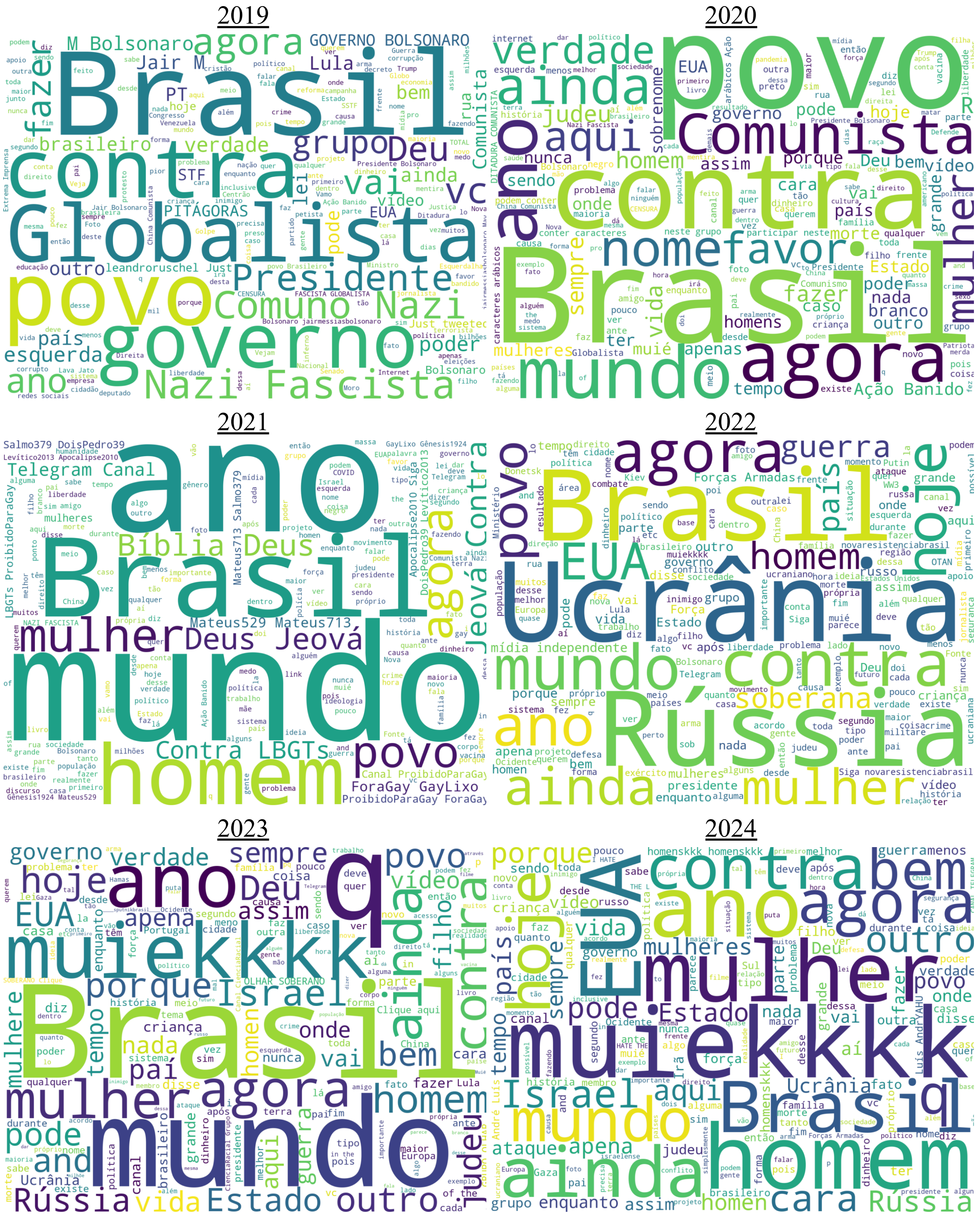

Source: Own elaboration (2024).

In Chart 01, which analyzes the temporal evolution of anti-woke and anti-gender discussions, it is observed that over the years, the term “Brazil” maintains a constant and prominent presence, reflecting a continuous concern with national issues and internal politics. In 2019, “against” and “people” emerge as significant terms, suggesting a rhetoric of opposition and resistance against progressive movements. From 2020 onwards, there is an increase in the presence of terms like "communist" and "globalist," indicating a growth in theories linking gender issues to global conspiracies. In 2021, the word "year" gains prominence, possibly reflecting a year of particular importance for discussions, perhaps related to significant political or social events. In the subsequent years, especially in 2023 and 2024, "world" and "woman" become more prominent, indicating an expansion of the focus of discussions to include global themes and gender issues, reinforcing the idea that these groups see gender issues as part of a larger struggle against a supposed globalist agenda.

**Chart 02.** Temporal word cloud series for revisionism and hate speech

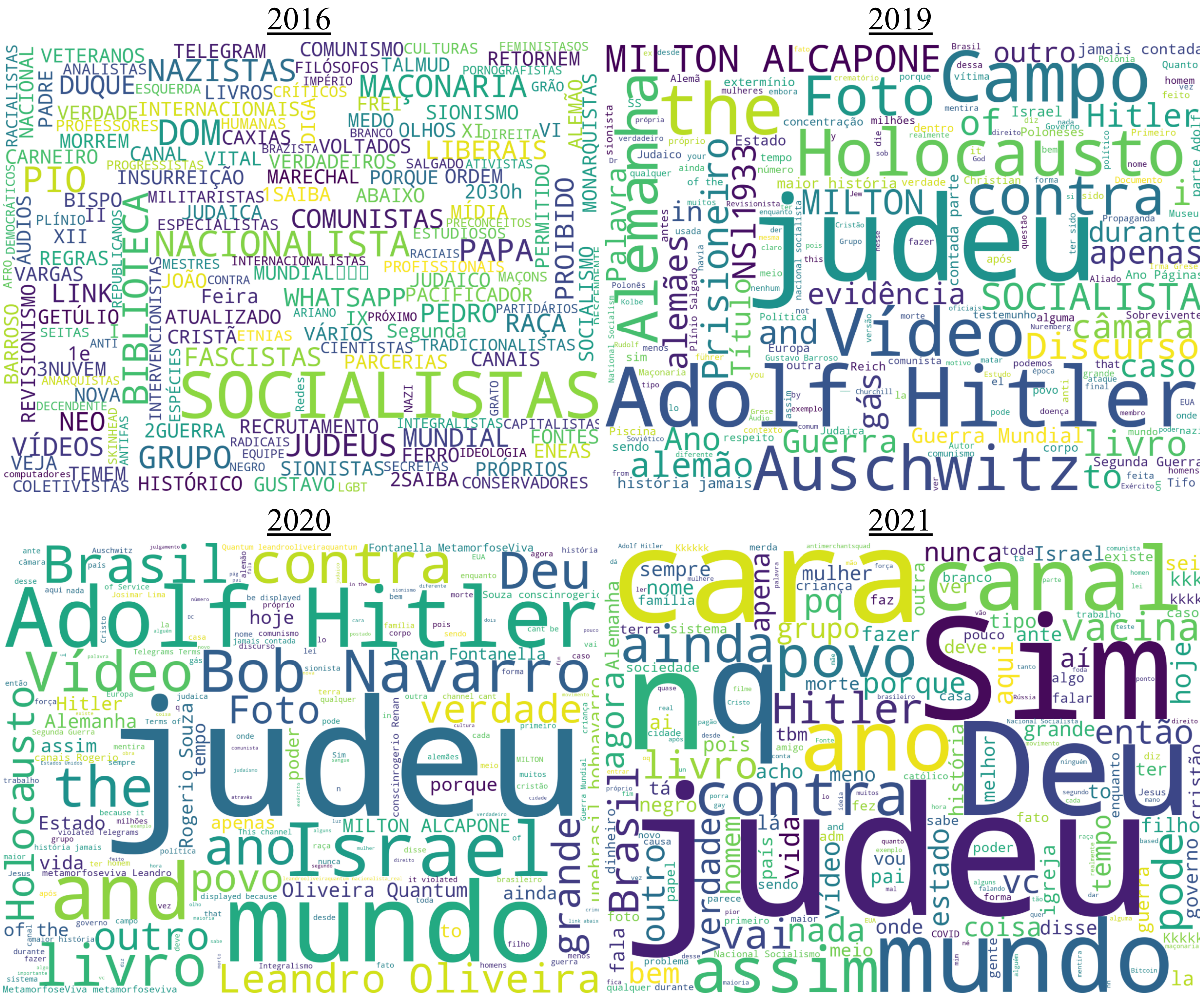

[ english version / português abaixo ]

2022

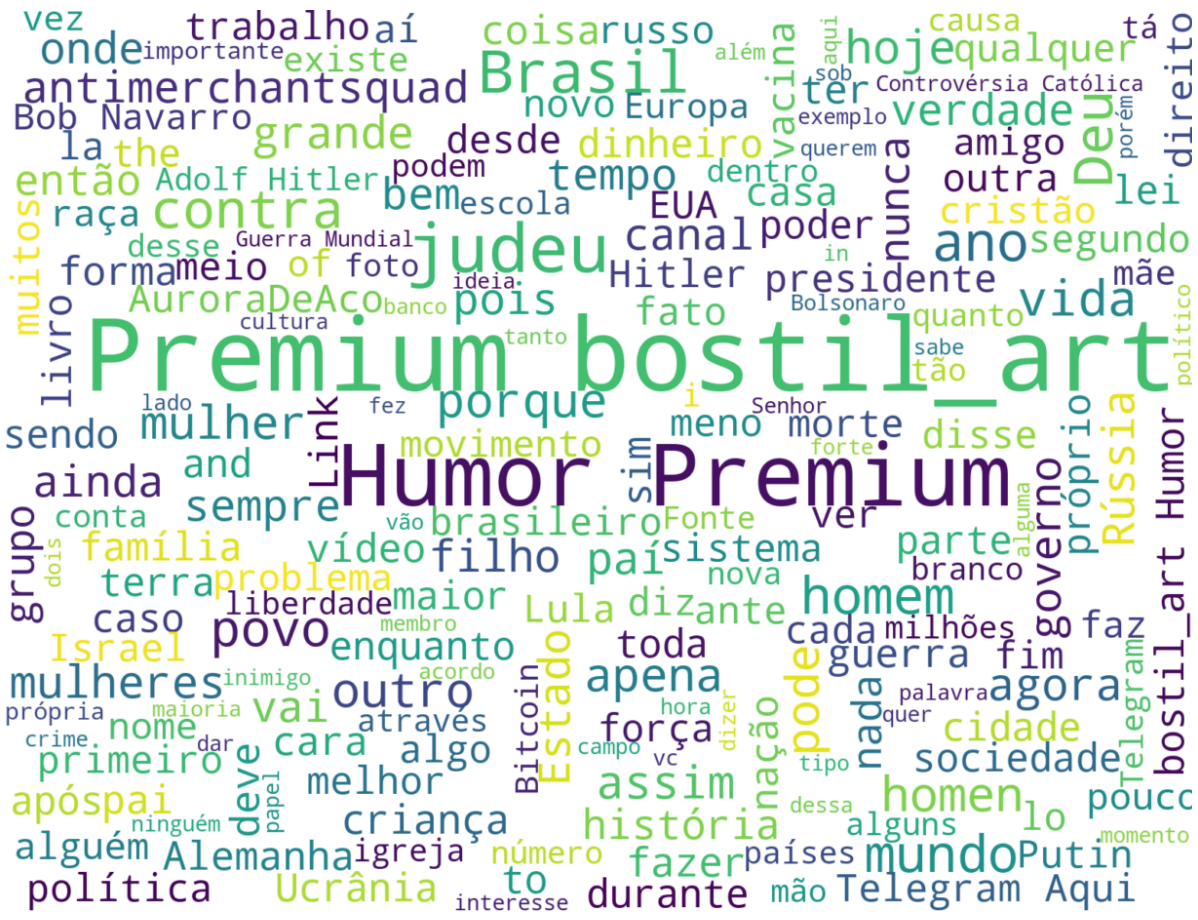

2023

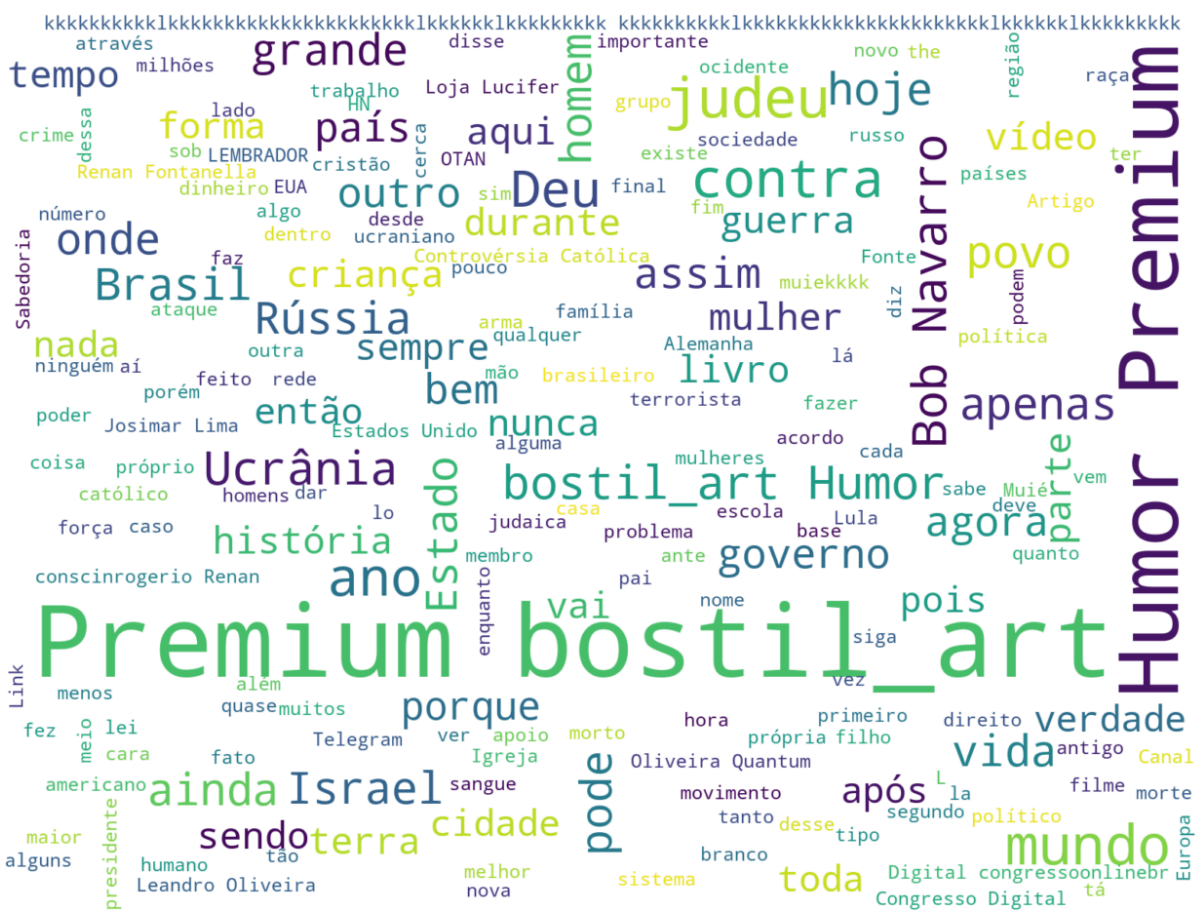

2024

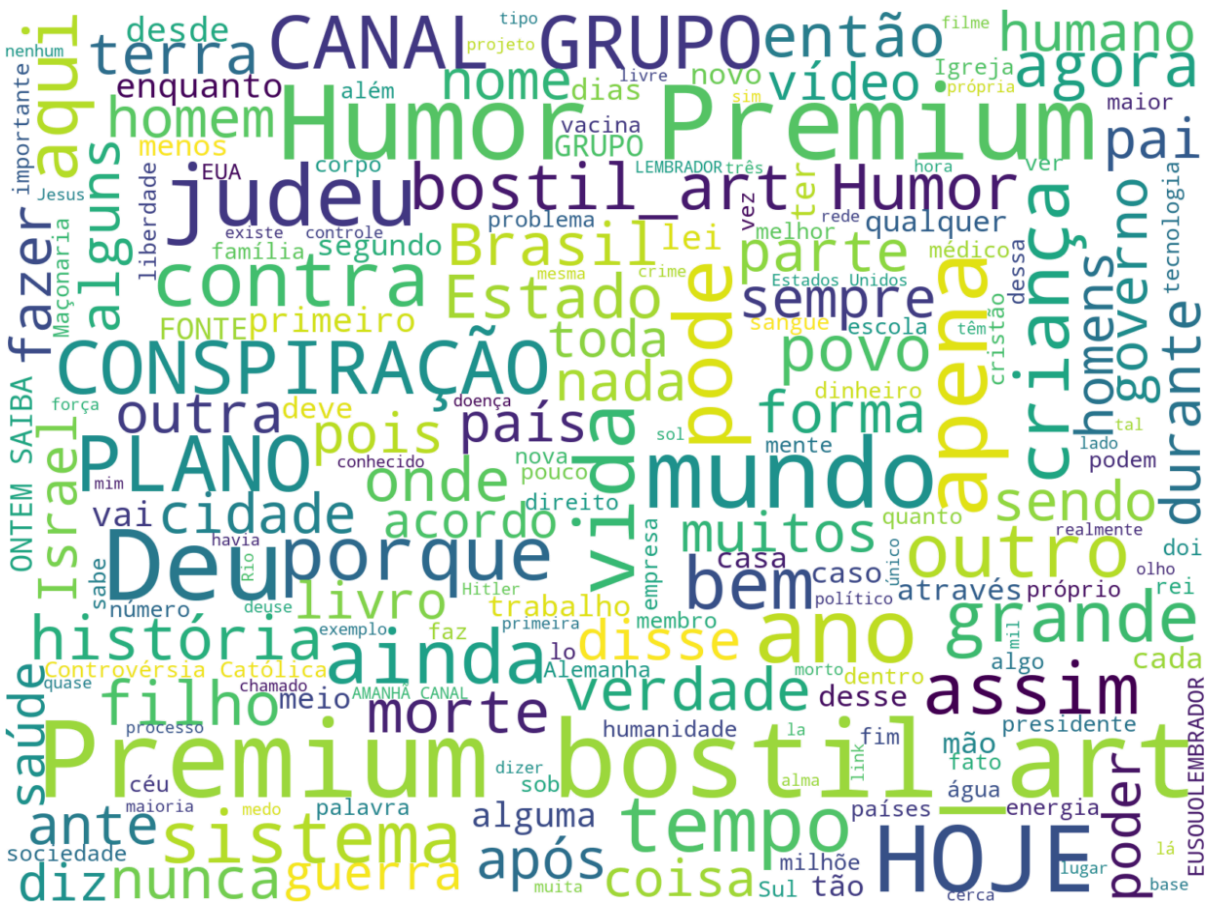

Source: Own elaboration (2024).

Chart 02, focused on revisionism and hate speech, reveals that since 2016, terms like “Jew” and “Hitler” are recurrent, highlighting the persistence of antisemitic and revisionist narratives within these communities. In 2019, the term “Holocaust” appears prominently, indicating a year of particular importance for revisionist discussions about World War II. Over the years, “world” and “Brazil” remain constant, suggesting a connection between global and national issues within these hate discussions. In 2022, the presence of terms like “Premiu” and “bostil” (a pejorative slang term referring to Brazil) suggests a discourse of devaluation and hate that goes beyond antisemitism, encompassing a critique of the country itself. In 2023 and 2024, “Jew” continues to be a central term, reflecting the persistence of hate speech against this community, while “world” maintains its relevance, indicating that these revisionist and hate discussions have a global dimension, where Brazil is just one of the contexts in which these narratives are applied.

### 3.4. Thematic agenda overlap

The following figures analyze how conspiracy theory communities surrounding Anti-Woke and Gender, as well as Revisionism and Hate Speech, interact and overlap with other conspiracy narratives. The analysis of these themes reveals the centrality of these discourses within the communities and how they are used to reinforce extremist beliefs and disinformation, connecting with broader topics such as faith, geopolitics, and traditionalist moralism. These connections expand the reach of these theories, making intervention and factual correction more complex.

**Figure 10.** Faith and religion themes

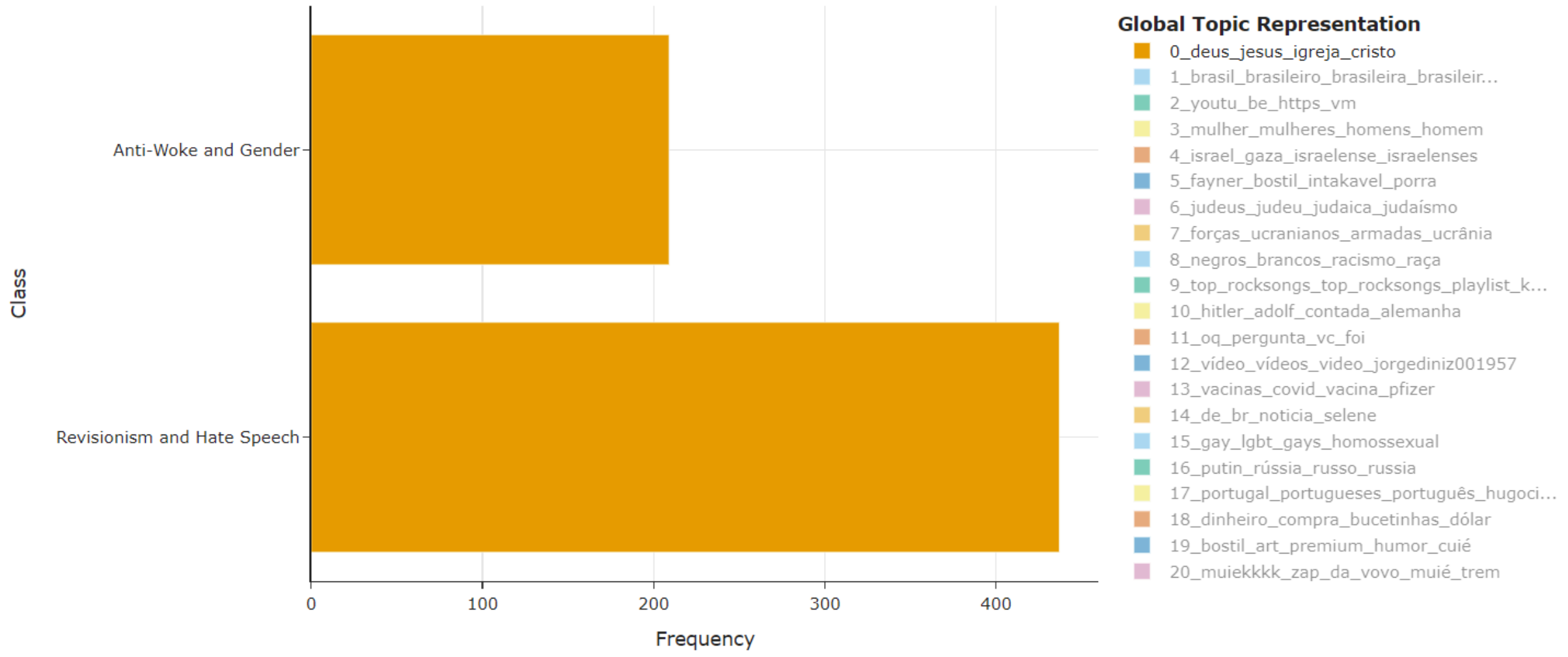

Source: Own elaboration (2024).

In Figure 10, we observe the intersection between themes of faith and religion with Anti-Woke and Revisionism discourses. Topics such as “Deus”, “Jesus”, and “igreja” are widely discussed, suggesting that these communities use religious elements to legitimize their discourses against social inclusion and gender equality movements. The association of these themes with Revisionism and Hate Speech reinforces the idea that faith is instrumentalized to justify resistance to social changes, positioning as a defense of traditional religious values.

**Figure 11.** Geopolitical disputes and globalism themes

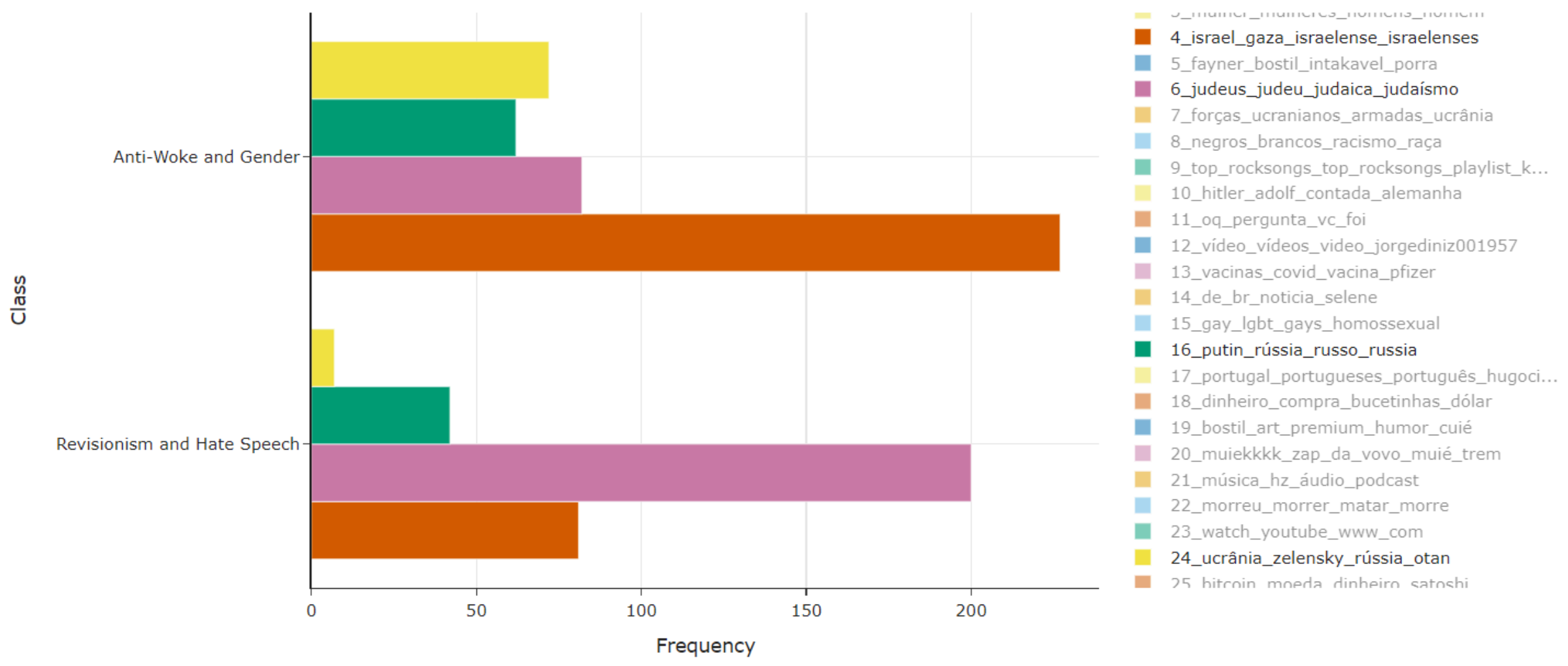

Source: Own elaboration (2024).

Figure 11 shows the interconnection between Anti-Woke and Revisionism discourses with themes of geopolitics and globalism. Topics such as “Putin”, “Rússia”, and “Israel” stand out, showing that these communities frequently link their narratives to global events, using these issues to strengthen their conspiratorial agendas. The overlap with globalism suggests that these communities view geopolitical conflicts as part of a larger conspiracy to implement a globalist agenda, which they perceive as threatening their values and ideologies.

**Figure 12.** Misogyny, racism, and anti-LGBT themes

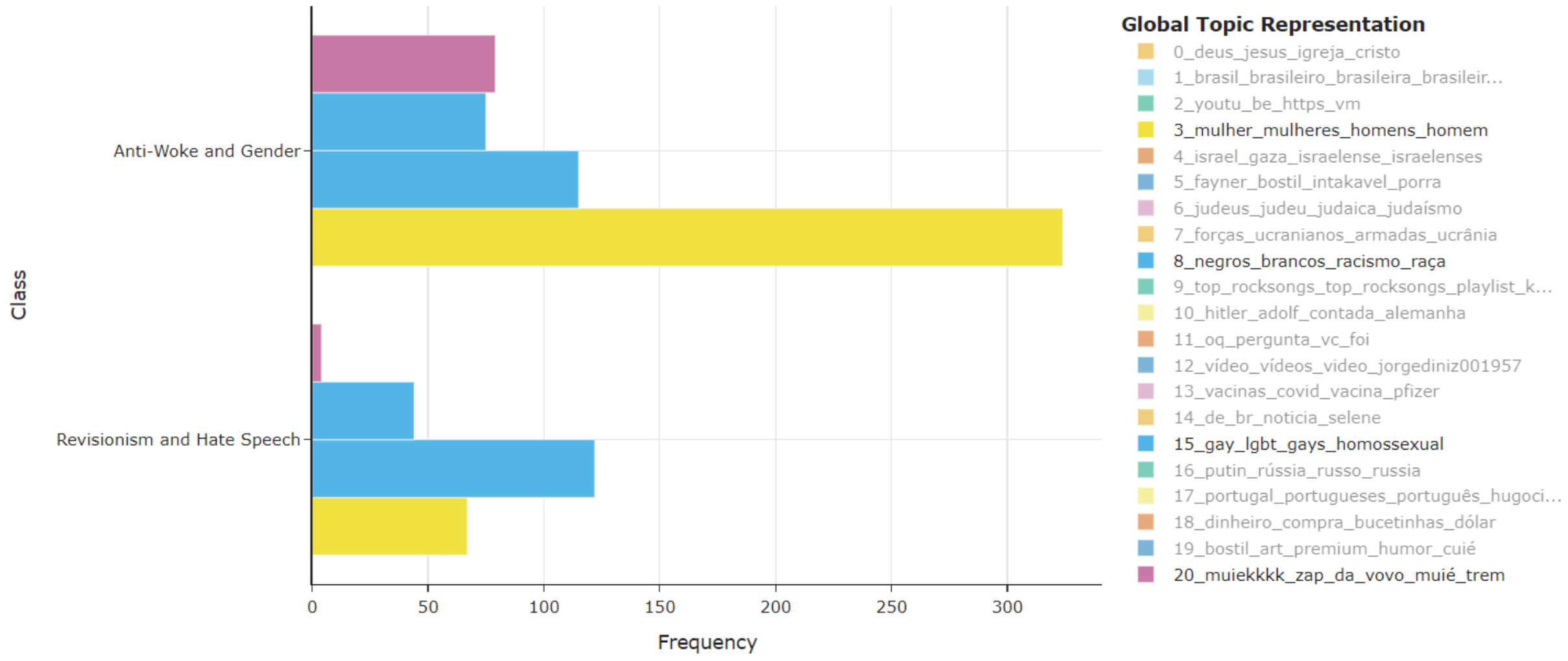

Source: Own elaboration (2024).

In Figure 12, themes of misogyny, racism, and anti-LGBT discourse within Anti-Woke and Revisionism communities are analyzed. Topics such as “mulher”, “homens”, and “raça” indicate a strong presence of discourses that attack minorities and promote gender and racial inequality. This overlap demonstrates how hatred and discrimination are central to these communities' narratives, using issues of identity and sexual orientation as primary targets to reinforce their discourses of hatred and exclusion.

**Figure 13.** Vaccine denialism themes

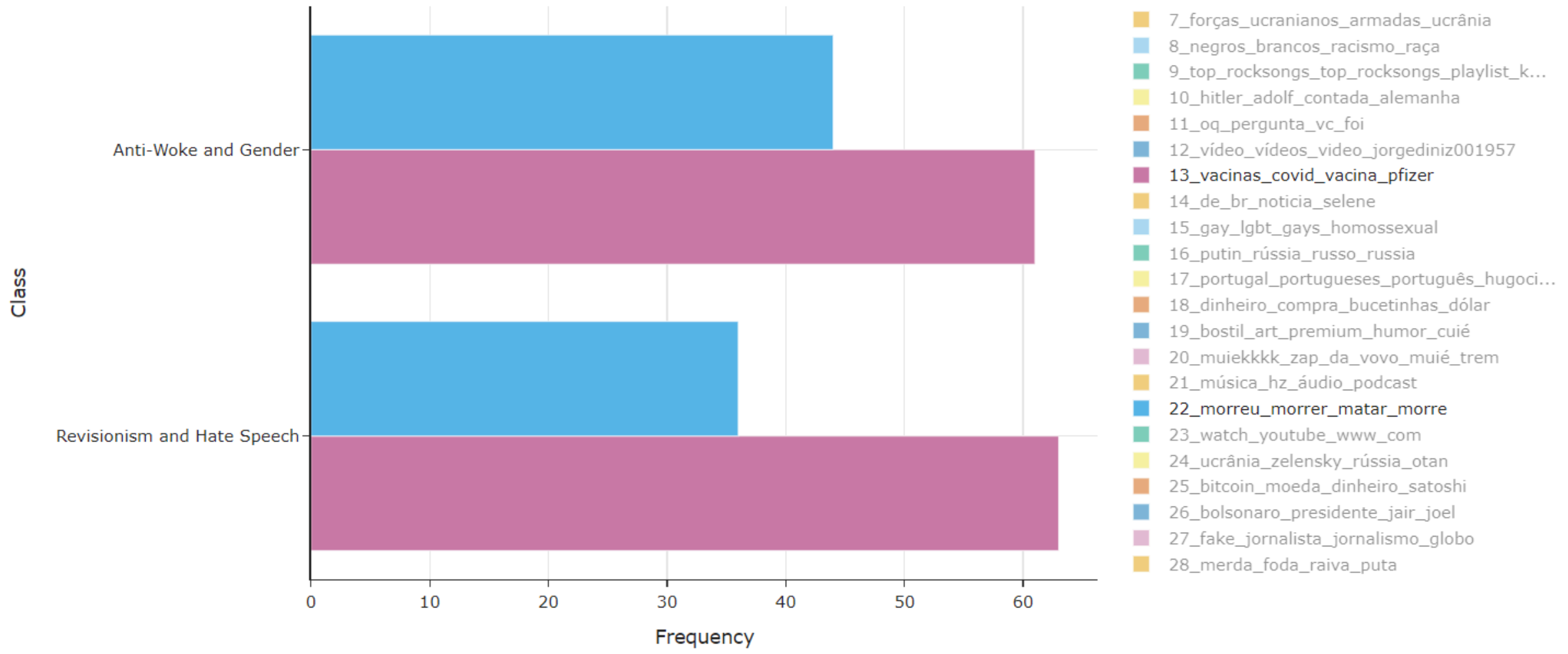

Source: Own elaboration (2024).

Figure 13 highlights how vaccine denialism is related to Anti-Woke and Revisionism discourses within these communities. Topics such as "vacinas", "Pfizer", and "covid" are prevalent, suggesting that these communities use vaccine denialism as an extension of their agendas against science and public health policies. The connection with Revisionism indicates an attempt to rewrite the scientific narrative, promoting disinformation that fuels distrust of vaccination and science.

**Figure 14.** Traditionalist moralism themes

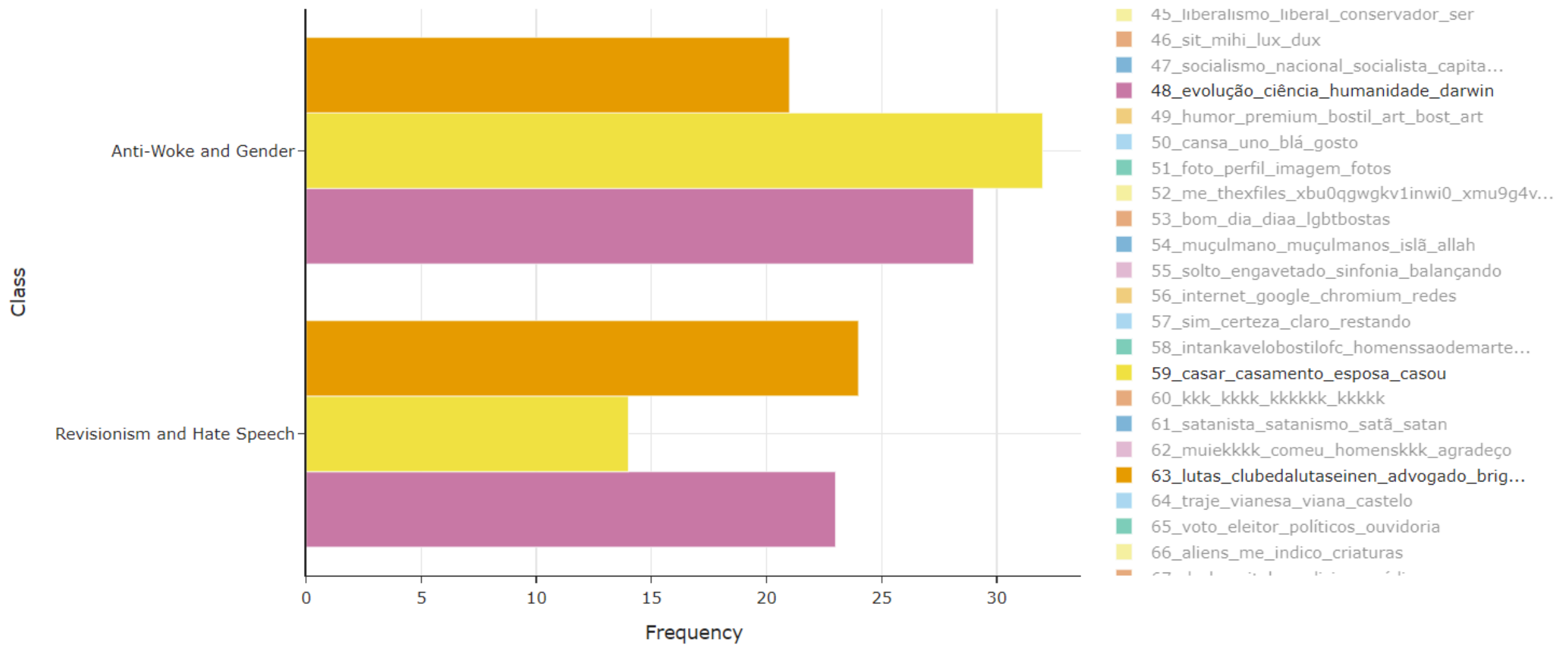

Source: Own elaboration (2024).

In Figure 14, we observe how themes of traditionalist moralism overlap with Anti-Woke and Revisionism discourses. Topics such as "evolução", "darwin", and "liberalismo" indicate that these communities use a moralistic narrative to justify their opposition to progressive and scientific ideas. The presence of themes related to morality and social order suggests that these discourses are used to promote a conservative view of society, where any deviation from traditional norms is seen as a threat to the established order.

**Figure 15.** Satanism and faith themes

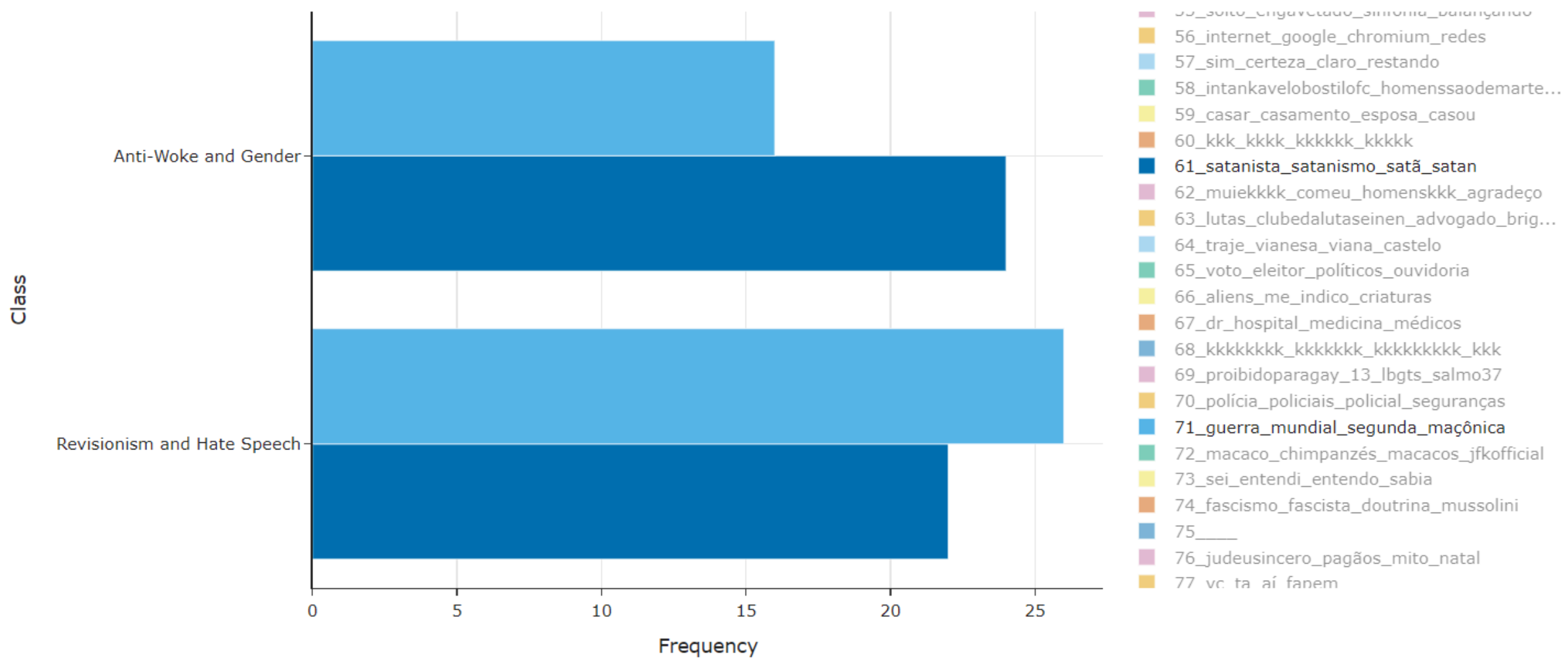

Source: Own elaboration (2024).

Figure 15 explores the association between themes of satanism and faith with Anti-Woke and Revisionism discourses. Topics such as “satanismo”, “fé”, and “igreja” are frequently discussed, showing that these communities use the fear of satanism and religious manipulation to reinforce their narratives. The overlap with Revisionism and Hate Speech suggests that these communities not only reject modern social movements but also seek to demonize their opponents by associating them with evil religious practices.

## 4. Reflections and future works

To answer the research question, **“how are Brazilian conspiracy theory communities on anti-woke agenda, gender issues, revisionism and hate speech topics characterized and articulated on Telegram?”**, this study adopted mirrored techniques in a series of seven publications aimed at characterizing and describing the phenomenon of conspiracy theories on Telegram, focusing on Brazil as a case study. After months of investigation, it was possible to extract a total of 109 Brazilian conspiracy theory communities on Telegram focused on anti-woke agenda, gender issues, revisionism and hate speech topics, amounting to 1,640,149 pieces of content published between July 2016 (initial publications) and August 2024 (when this study was conducted), with 188,771 users combined across the communities.

Four main approaches were adopted: **(i)** Network, which involved the creation of an algorithm to map connections between communities through invitations circulated among groups and channels; **(ii)** Time series, which used libraries like “Pandas” (McKinney, 2010) and “Plotly” (Plotly Technologies Inc., 2015) to analyze the evolution of publications and engagements over time; **(iii)** Content analysis, where textual analysis techniques were applied to identify patterns and word frequencies in the communities over the semesters; and **(iv)** Thematic agenda overlap, which utilized the BERTopic model (Grootendorst, 2020) to group and interpret large volumes of text, generating coherent topics from the analyzed publications. The main reflections are detailed below, followed by suggestions for future works.

### 4.1. Main reflections

**The anti-woke agenda consolidates as one of the main forces of resistance in the Brazilian conspiracy ecosystem:** The study identified that anti-woke communities are central within the conspiracy universe on Brazilian Telegram. With 1,293,430 posts and 154,391 active users, these communities not only reject progressive social inclusion policies but also promote a worldview where diversity is seen as a threat to the established order. The consolidation of this agenda reflects an organized resistance against social change, becoming a key element in disinformation narratives;

**Exponential growth in mentions of hate speech and revisionism during global crises:** During critical events such as the 2018 elections in Brazil and the peak of the COVID-19 pandemic, there was a sharp increase in mentions of revisionism and hate speech, with a 1,100% rise between 2019 and 2021. This increase reflects the exacerbated polarization during these periods, where disinformation found ground to expand within communities;

**Brazilian Nazi communities on Telegram propagate extremist ideologies and glorify Hitler:** One of the most alarming findings of the study was the identification of dozens of Brazilian communities on Telegram that explicitly advocate Nazi ideologies, including the glorification of Hitler and the dissemination of hate against minority groups. These communities represent a severe threat as they promote radicalization and normalize extremist ideologies, compromising social security and fostering an environment of hate;

**Interconnectivity between anti-woke, anti-gender, and revisionism strengthens an ecosystem of hate and exclusion:** Anti-woke and anti-gender communities frequently overlap with hate speech and historical revisionism, creating a cohesive network that amplifies extremist ideologies. Network analysis shows that 1,729 links between these communities indicate a constant flow of ideas that reinforce a narrative of resistance against social changes, where hate and exclusion are normalized;

**Gateways: anti-woke as a pathway to other conspiracy narratives:** Anti-woke communities function as gateways to other conspiracy theories, such as the New World Order and revisionism. With 965 links directed to NWO and 599 to anti-vaccine content, these communities not only disseminate their own ideologies but also introduce their members to a broader spectrum of disinformation, creating an interconnected radicalization network;

**Total volume and influence: the impact of anti-woke and anti-gender communities on Telegram:** The total volume of posts related to anti-woke and anti-gender on Telegram reached 1,293,430, demonstrating the significant influence of these communities. With 43 groups and 154,391 users, these communities act as hubs of disinformation, where narratives against diversity and inclusion are not only promoted but also amplified beyond their digital borders;

**The strong correlation between historical revisionism and anti-gender discourse:** The study reveals that historical revisionism is closely linked to anti-gender discussions, with 416 invitation links showing this interconnectivity. This alignment suggests that the same forces driving the denial of historical facts, such as the Holocaust, are those that reject gender equality policies, forming a coherent agenda of opposition to social progress;

**Accelerated radicalization: the role of anti-woke discourse in social polarization:** The temporal analysis of anti-woke and revisionism discussions shows an acceleration in the radicalization of discourse, especially during the period from 2020 to 2021, where there was a 1,000% increase in mentions of these themes. This growth reflects how global crises are used to intensify polarization and promote hate narratives within these communities;

**Anti-woke as a narrative of resistance against globalization:** Anti-woke communities are strongly interconnected with narratives of resistance to globalization, showing significant overlap with Globalism and NWO themes. With 978 links related to globalism, these communities position the anti-woke agenda as a defense against a supposed imposition of globalist values that threaten national sovereignty and identity;

**Anti-gender discourse as facilitators of anti-vaccine disinformation dissemination:** The study highlights that anti-gender communities not only reject equality policies but also facilitate the dissemination of anti-vaccine disinformation. With 599 identified links, these communities use the anti-vaccine narrative as an extension of their opposition to what they perceive as a global threat to individual freedom, creating an intersection between public health and extremist ideologies.

### 4.2. Future works

Based on the findings of this study, future research should focus on exploring the internal dynamics of anti-woke and revisionist communities, particularly regarding the instrumentalization of faith and morality to promote narratives of hate and exclusion. Understanding how these communities use religious and moralistic rhetoric to justify their positions could reveal the mechanisms that strengthen resistance to social change and diversity. Additionally, it would be interesting to investigate how these narratives are adapted and reformulated in response to different cultural contexts and global crises, such as the COVID-19 pandemic or elections, to understand the plasticity and resilience of ideologies.

Another crucial area for future research is mapping the interconnections between anti-woke, revisionism, and other conspiracy theories, such as vaccine denialism and globalism. A detailed analysis of these connections could reveal patterns of radicalization that cross different narratives and identify which groups or individuals function as "super spreaders" of disinformation. This investigation could also explore how these disinformation networks are structured, operate, and adapt to external interventions, offering insights into how specific interventions might be more effective in dismantling these networks.

Additionally, identifying and monitoring explicitly Nazi communities on Brazilian Telegram should be a priority in future research. Given the seriousness of this discovery, studies focused on these communities could explore the strategies used to recruit and radicalize new members, as well as examine the impact of these ideologies on society at large. Research investigating the intersections between these communities and other hate networks, such as anti-woke and anti-gender groups, could provide a comprehensive view of the threats posed by these extremist ideologies.

Finally, future studies should consider the impact of global events on the intensification of these narratives, as observed during the COVID-19 pandemic and significant political crises. Research exploring the relationship between global crises and the growth of extremist discourse could help develop more effective strategies to mitigate the spread of disinformation during times of social vulnerability. Moreover, it would be relevant to investigate the effectiveness of digital interventions, such as content moderation and factual correction campaigns, in facing the spread of disinformation and hate speech on Telegram.

Other directions for research include longitudinal analysis of these communities to identify trends and changes in narratives over time. Studying how communities react to interventions, whether technological or political, could offer a deeper understanding of resistance and adaptation dynamics. Additionally, research could explore the role of international communities and how external influences shape discussions and practices within Brazilian communities. Understanding these global interactions could be crucial for developing culturally sensitive and more effective disinformation countermeasures.

Lastly, it is essential to investigate the offline consequences of these communities, including how narratives propagated online can influence behaviors and attitudes in the real world. This could encompass everything from the organization of extremist events to acts of violence motivated by ideologies promoted in these spaces. Research connecting online activities with tangible social impacts could help shape public policies and interventions aimed at mitigating the negative effects of these communities on society.

## 6. Author biography

**Ergon Cugler de Moraes Silva** has a Master’s degree in Public Administration and Government (FGV), Postgraduate MBA in Data Science & Analytics (USP) and Bachelor’s degree in Public Policy Management (USP). He is associated with the Bureaucracy Studies Center (NEB FGV), collaborates with the Interdisciplinary Observatory of Public Policies (OIPP USP), with the Study Group on Technology and Innovations in Public Management (GETIP USP) with the Monitor of Political Debate in the Digital Environment (Monitor USP) and with the Working Group on Strategy, Data and Sovereignty of the Study and Research Group on International Security of the Institute of International Relations of the University of Brasília (GEPSI UnB). He is also a researcher at the Brazilian Institute of Information in Science and Technology (IBICT), where he works for the Federal Government on strategies against disinformation. Brasília, Federal District, Brazil. Web site: https://ergoncugler.com/.

[ versão português / english above ]

# **Comunidades de anti-*woke*, anti-gênero, de revisionismo e discurso de ódio no Telegram brasileiro:** do grave discurso reacionário, ao crime de glorificação ao nazismo e à Hitler

*Ergon Cugler de Moraes Silva*

Instituto Brasileiro de Informação
em Ciência e Tecnologia (IBICT)
Brasília, Distrito Federal, Brasil

contato@ergoncugler.com
www.ergoncugler.com

## Resumo

A resistência às políticas progressistas e o discurso de ódio tem se consolidado no Telegram brasileiro, com comunidades anti-*woke* rejeitando a diversidade e promovendo uma visão de mundo que vê essas mudanças sociais como uma ameaça. Dessa forma, esse estudo busca responder à pergunta de pesquisa: **como são caracterizadas e articuladas as comunidades de teorias da conspiração brasileiras sobre temáticas de anti-*woke*, anti-gênero, revisionismo e discurso de ódio no Telegram?** Vale ressaltar que este estudo faz parte de uma série de um total de sete estudos que possuem como objetivo principal compreender e caracterizar as comunidades brasileiras de teorias da conspiração no Telegram. Esta série de sete estudos está disponibilizada abertamente e originalmente no arXiv da Cornell University, aplicando um método espelhado nos sete estudos, mudando apenas o objeto temático de análise e provendo uma replicabilidade de investigação, inclusive com códigos próprios e autorais elaborados, somando-se à cultura de software livre e de código aberto. No que diz respeito aos principais achados deste estudo, observa-se: Comunidades anti-*woke* emergem como forças centrais no ecossistema conspiratório brasileiro; Durante crises, menções a discursos de ódio e revisionismo cresceram significativamente, refletindo a polarização; Comunidades nazistas no Telegram propagam ideologias extremistas, glorificando Hitler; A interconectividade entre anti-*woke*, anti-gênero e revisionismo fortalece o ecossistema de ódio; Discursos anti-gênero facilitam a disseminação de desinformação antivacinas, criando uma interseção entre saúde e conspiração.

### Principais descobertas

➔ As comunidades anti-*woke* emergem como forças centrais no ecossistema conspiratório brasileiro, com 1.293.430 publicações e 154.391 usuários, rejeitando políticas progressistas e promovendo uma visão de mundo onde a diversidade é vista como ameaça à ordem estabelecida, consolidando a resistência organizada contra mudanças sociais;

➔ Durante crises globais, como as eleições de 2018 no Brasil e a Pandemia da COVID-19, menções a discursos de ódio e revisionismo cresceram 1.100% entre 2019 e 2021, refletindo a polarização exacerbada e a expansão da desinformação em comunidades digitais;

➔ Comunidades nazistas brasileiras no Telegram propagam ideologias extremistas, glorificando Hitler e disseminando discursos de ódio contra minorias, representando uma grave ameaça à segurança social ao promover a radicalização e normalizar ideologias extremistas;

- ➔ A interconectividade entre anti-*woke*, anti-gênero e revisionismo fortalece um ecossistema de ódio e exclusão, com 1.729 links entre essas comunidades, criando uma rede coesa que amplifica ideologias extremistas e normaliza o ódio e a exclusão;

- ➔ Comunidades anti-*woke* funcionam como portas de entrada para outras teorias da conspiração, como Nova Ordem Mundial e revisionismo, com 965 links para NOM e 599 para antivacinas, introduzindo membros a um espectro mais amplo de desinformação e radicalização;

- ➔ O impacto das comunidades anti-*woke* e anti-gênero no Telegram é significativo, com 1.293.430 publicações e 154.391 usuários, atuando como *hubs* de desinformação que amplificam narrativas contrárias à diversidade e inclusão além de suas fronteiras digitais;

- ➔ Há uma forte correlação entre revisionismo histórico e discursos anti-gênero, com 416 links de convites revelando essa interconectividade, sugerindo que as forças que negam fatos históricos, como o Holocausto, também rejeitam políticas de igualdade de gênero;

- ➔ Discursos anti-*woke* aceleram a radicalização social, com um aumento de 1.000% nas menções entre 2020 e 2021, refletindo como crises globais são utilizadas para intensificar a polarização e promover narrativas de ódio dentro dessas comunidades;

- ➔ Comunidades anti-*woke* se posicionam como resistência à globalização, com 978 links relacionados ao Globalismo, apresentando-se como uma defesa contra a suposta imposição de valores globalistas que ameaçam a soberania e identidade nacional;

- ➔ Discursos anti-gênero facilitam a disseminação de desinformação antivacinas, com 599 links identificados, utilizando a narrativa antivacinas como extensão de sua oposição a uma suposta ameaça global à liberdade individual, criando uma interseção entre saúde e conspiração.

## 1. Introdução

Após percorrer milhares de comunidades brasileiras de teorias da conspiração no Telegram, extrair dezenas de milhões de conteúdos dessas comunidades, elaborados e/ou compartilhados por milhões de usuários que as compõem, este estudo tem o objetivo de compor uma série de um total de sete estudos que tratam sobre o fenômeno das teorias da conspiração no Telegram, adotando o Brasil como estudo de caso. Com as abordagens de identificação implementadas, foi possível alcançar um total de 109 comunidades de teorias da conspiração brasileiras no Telegram sobre temáticas de anti-*woke*, anti-gênero, revisionismo e discurso de ódio , estas somando 1.640.149 de conteúdos publicados entre julho de 2016 (primeiras publicações) até agosto de 2024 (realização deste estudo), com 188.771 usuários somados dentre as comunidades. Dessa forma, este estudo tem como objetivo compreender e caracterizar as comunidades sobre temáticas de anti-*woke*, anti-gênero, revisionismo e discurso de ódio presentes nessa rede brasileira de teorias da conspiração identificada no Telegram.

Para tal, será aplicado um método espelhado em todos os sete estudos, mudando apenas o objeto temático de análise e provendo uma replicabilidade de investigação. Assim, abordaremos técnicas para observar as conexões, séries temporais, conteúdos e sobreposições temáticas das comunidades de teorias da conspiração. Além desse estudo, é possível encontrar os seis demais disponibilizados abertamente e originalmente no arXiv da Cornell University.

Essa série contou com a atenção redobrada para garantir a integridade dos dados e o respeito à privacidade dos usuários, conforme a legislação brasileira prevê (Lei nº 13.709/2018).

Portanto questiona-se: **como são caracterizadas e articuladas as comunidades de teorias da conspiração brasileiras sobre temáticas de anti-*woke*, anti-gênero, revisionismo e discurso de ódio no Telegram?**

## 2. Materiais e métodos

A metodologia deste estudo está organizada em três subseções, sendo: **2.1. Extração de dados**, que descreve o processo e as ferramentas utilizadas para coletar as informações das comunidades no Telegram; **2.2. Tratamento de dados**, onde são abordados os critérios e métodos aplicados para classificar e anonimizar os dados coletados; e **2.3. Abordagens para análise de dados**, que detalha as técnicas utilizadas para investigar as conexões, séries temporais, conteúdos e sobreposições temáticas das comunidades de teorias da conspiração.

### 2.1. Extração de dados

Este projeto teve início em fevereiro de 2023, com a publicação da primeira versão do TelegramScrap (Silva, 2023), uma ferramenta própria e autoral, de software livre e de código aberto, que faz uso da Application Programming Interface (API) da plataforma Telegram por meio da biblioteca Telethon e organiza ciclos de extração de dados de grupos e canais abertos no Telegram. Ao longo dos meses, a base de dados pôde ser ampliada e qualificada fazendo uso de quatro abordagens de identificação de comunidades de teorias da conspiração:

**(i) Uso de palavras chave:** no início do projeto, foram elencadas palavras-chave para identificação diretamente no buscador de grupos e canais brasileiros no Telegram, tais como “apocalipse”, “sobrevivencialismo”, “mudanças climáticas”, “terra plana”, “teoria da conspiração”, “globalismo”, “nova ordem mundial”, “ocultismo”, “esoterismo”, “curas alternativas”, “qAnon”, “reptilianos”, “revisionismo”, “alienígenas”, dentre outras. Essa primeira abordagem forneceu algumas comunidades cujos títulos e/ou descrições dos grupos e canais contassem com os termos explícitos relacionados a teorias da conspiração. Contudo, com o tempo foi possível identificar outras diversas comunidades cujas palavras-chave elencadas não davam conta de abarcar, algumas inclusive propositalmente com caracteres trocados para dificultar a busca de quem a quisesse encontrar na rede;

**(ii) Mecanismo de recomendação de canais do Telegram:** com o tempo, foi identificado que canais do Telegram (exceto grupos) contam com uma aba de recomendação chamada de “canais similares”, onde o próprio Telegram sugere dez canais que tenham alguma similaridade com o canal que se está observando. A partir desse mecanismo de recomendação do próprio Telegram, foi possível encontrar mais comunidades de teorias da conspiração brasileiras, com estas sendo recomendadas pela própria plataforma;

**(iii) Abordagem de bola de neve para identificação de convites:** após algumas comunidades iniciais serem acumuladas para a extração, foi elaborado um algoritmo próprio autoral de identificação de urls que contivessem "t.me/", sendo o prefixo de qualquer convite para grupos e canais do Telegram. Acumulando uma frequência de centenas de milhares de links que atendessem a esse critério, foram elencados os endereços únicos e contabilizadas as suas repetições. Dessa forma, foi possível fazer uma investigação de novos grupos e canais brasileiros mencionados nas próprias mensagens dos já investigados, ampliando a rede. Esse processo foi sendo repetido periodicamente, buscando identificar novas comunidades que tivessem identificação com as temáticas de teorias da conspiração no Telegram;

**(iv) Ampliação para tweets publicados no X que mencionassem convites:** com o objetivo de diversificar ainda mais a fonte de comunidades de teorias da conspiração brasileiras no Telegram, foi elaborada uma query de busca própria que pudesse identificar as palavras-chave de temáticas de teorias da conspiração, porém usando como fonte tweets publicados no X (antigo Twitter) e que, além de conter alguma das palavras-chave, contivesse também o "t.me/", sendo o prefixo de qualquer convite para grupos e canais do Telegram, "https://x.com/search?q=lang%3Apt%20%22t.me%2F%22%20TERMO-DE-BUSCA".

Com as abordagens de identificação de comunidades de teorias da conspiração implementadas ao longo de meses de investigação e aprimoramento de método, foi possível construir uma base de dados do projeto com um total de 855 comunidades de teorias da conspiração brasileiras no Telegram (considerando as demais temáticas também não incluídas nesse estudo), estas somando 27.227.525 de conteúdos publicados entre maio de 2016 (primeiras publicações) até agosto de 2024 (realização deste estudo), com 2.290.621 usuários somados dentre as comunidades brasileiras. Há de se considerar que este volume de usuários conta com dois elementos, o primeiro é que trata-se de uma variável, pois usuários podem entrar e sair diariamente, portanto este valor representa o registrado na data de extração de publicações da comunidade; além disso, é possível que um mesmo usuário esteja em mais de um grupo e, portanto, é contabilizado mais de uma vez. Nesse sentido, o volume ainda sinaliza ser expressivo, mas pode ser levemente menor quando considerado o volume de cidadãos deduplicados dentro dessas comunidades brasileiras de teorias da conspiração.

### 2.2. Tratamento de dados

Com todos os grupos e canais brasileiros de teorias da conspiração no Telegram extraídos, foi realizada uma classificação manual considerando o título e a descrição da comunidade. Caso houvesse menção explícita no título ou na descrição da comunidade a alguma temática, esta foi classificada entre: (i) "Anticiência"; (ii) "Anti-Woke e Gênero"; (iii) "Antivax"; (iv) "Apocalipse e Sobrevivencialismo"; (v) "Mudanças Climáticas"; (vi) Terra Plana; (vii) "Globalismo"; (viii) "Nova Ordem Mundial"; (ix) "Ocultismo e Esoterismo"; (x) "Off Label e Charlatanismo"; (xi) "QAnon"; (xii) "Reptilianos e Criaturas"; (xiii) "Revisionismo e Discurso de Ódio"; (xiv) "OVNI e Universo". Caso não houvesse nenhuma menção explícita relacionada às temáticas no título ou na descrição da comunidade, esta foi classificada como (xv) "Conspiração Geral". No Quadro a seguir, podemos observar as métricas relacionadas à classificação dessas comunidades de teorias da conspiração no Brasil.

**Tabela 01.** Comunidades de teorias da conspiração no Brasil (métricas até agosto de 2024)

| Categorias | Grupos | Usuários | Publicações | Comentários | Total |
|---|---|---|---|---|---|
| **Anticiência** | 22 | 58.138 | 187.585 | 784.331 | 971.916 |
| **Anti-*Woke* e Gênero** | 43 | 154.391 | 276.018 | 1.017.412 | 1.293.430 |
| **Antivacinas (*Antivax*)** | 111 | 239.309 | 1.778.587 | 1.965.381 | 3.743.968 |
| **Apocalipse e Sobrevivência** | 33 | 109.266 | 915.584 | 429.476 | 1.345.060 |
| **Mudanças Climáticas** | 14 | 20.114 | 269.203 | 46.819 | 316.022 |
| **Terraplanismo** | 33 | 38.563 | 354.200 | 1.025.039 | 1.379.239 |
| **Conspirações Gerais** | 127 | 498.190 | 2.671.440 | 3.498.492 | 6.169.932 |
| **Globalismo** | 41 | 326.596 | 768.176 | 537.087 | 1.305.263 |
| **Nova Ordem Mundial (NOM)** | 148 | 329.304 | 2.411.003 | 1.077.683 | 3.488.686 |
| **Ocultismo e Esoterismo** | 39 | 82.872 | 927.708 | 2.098.357 | 3.026.065 |
| **Medicamentos *off label*** | 84 | 201.342 | 929.156 | 733.638 | 1.662.794 |
| **QAnon** | 28 | 62.346 | 531.678 | 219.742 | 751.420 |
| **Reptilianos e Criaturas** | 19 | 82.290 | 96.262 | 62.342 | 158.604 |
| **Revisionismo e Ódio** | 66 | 34.380 | 204.453 | 142.266 | 346.719 |
| **OVNI e Universo** | 47 | 58.912 | 862.358 | 406.049 | 1.268.407 |
| **Total** | **855** | **2.296.013** | **13.183.411** | **14.044.114** | **27.227.525** |

Fonte: Elaboração própria (2024).

Com esse volume de dados extraídos, foi possível segmentar para apresentar neste estudo apenas comunidades e conteúdos referentes às temáticas de anti-*woke*, anti-gênero, revisionismo e discurso de ódio. Em paralelo, as demais temáticas de comunidades brasileiras de teorias da conspiração também contaram com estudos elaborados para a caracterização da extensão e da dinâmica da rede, estes sendo disponibilizados abertamente e originalmente no arXiv da Cornell University.

Além disso, cabe citar que apenas foram extraídas comunidades abertas, isto é, não apenas identificáveis publicamente, mas também sem necessidade de solicitação para acessar ao conteúdo, estando aberto para todo e qualquer usuário com alguma conta do Telegram sem que este necessite ingressar no grupo ou canal. Além disso, em respeito à legislação brasileira e especialmente da Lei Geral de Proteção de Dados Pessoais (LGPD), ou Lei nº 13.709/2018, que trata do controle da privacidade e do uso/tratamento de dados pessoais, todos os dados extraídos foram anonimizados para a realização de análises e investigações. Dessa forma, nem mesmo a identificação das comunidades é possível por meio deste estudo, estendendo aqui a privacidade do usuário ao anonimizar os seus dados para além da própria comunidade à qual ele se submeteu ao estar em um grupo ou canal público e aberto no Telegram.

### 2.3. Abordagens para análise de dados

Totalizando 109 comunidades selecionadas nas temáticas de anti-*woke*, anti-gênero, revisionismo e discurso de ódio, contendo 1.640.149 publicações e 188.771 usuários somados, quatro abordagens serão utilizadas para investigar as comunidades de teorias da conspiração selecionadas para o escopo do estudo. Tais métricas são detalhadas no Quadro a seguir:

**Tabela 02.** Comunidades selecionadas para análise (métricas até agosto de 2024)

| Categorias | Grupos | Usuários | Publicações | Comentários | Total |
|---|---|---|---|---|---|
| **Anti-*Woke* e Gênero** | 43 | 154.391 | 276.018 | 1.017.412 | 1.293.430 |
| **Revisionismo e Ódio** | 66 | 34.380 | 204.453 | 142.266 | 346.719 |
| **Total** | **109** | **188.771** | **480.471** | **1.159.678** | **1.640.149** |

Fonte: Elaboração própria (2024).

**(i) Rede:** com a elaboração de um algoritmo próprio para a identificação de mensagens que contenham o termo de "t.me/" (de convite para entrarem em outras comunidades), propomos apresentar volumes e conexões observadas sobre como **(a)** as comunidades da temática de anti-*woke*, anti-gênero, revisionismo e discurso de ódio circulam convites para que os seus usuários conheçam mais grupos e canais da mesma temática, reforçando os sistemas de crença que comungam; e como **(b)** essas mesmas comunidades circulam convites para que os seus usuários conheçam grupos e canais que tratem de outras teorias da conspiração, distintas de seu propósito explícito. Esta abordagem é interessante para observar se essas comunidades utilizam a si próprias como fonte de legitimação e referência e/ou se embasam-se em demais temáticas de teorias da conspiração, inclusive abrindo portas para que seus usuários conheçam outras conspirações. Além disso, cabe citar o estudo de Rocha *et al.* (2024) em que uma abordagem de identificação de rede também foi aplicada em comunidades do Telegram, porém observando conteúdos similares a partir de um ID gerado para cada mensagem única e suas similares;

**(ii) Séries temporais:** utiliza-se da biblioteca "Pandas" (McKinney, 2010) para organizar os data frames de investigação, observando **(a)** o volume de publicações ao longo dos meses; e **(b)** o volume de engajamento ao longo dos meses, considerando metadados de visualizações, reações e comentários coletados na extração; Além da volumetria, a biblioteca "Plotly" (Plotly Technologies Inc., 2015) viabilizou a representação gráfica dessa variação;

**(iii) Análise de conteúdo:** além da análise geral de palavras com identificação das frequências, são aplicadas séries temporais na variação das palavras mais frequentes ao longo dos semestres — observando entre julho de 2016 (primeiras publicações) até agosto de 2024 (realização deste estudo). E com as bibliotecas "Pandas" (McKinney, 2010) e "WordCloud" (Mueller, 2020), os resultados são apresentados tanto volumetricamente quanto graficamente;

**(iv) Sobreposição de agenda temática:** seguindo a abordagem proposta por Silva & Sátiro (2024) para identificação de sobreposição de agenda temática em comunidades do Telegram, utilizamos o modelo "BERTopic" (Grootendorst, 2020). O BERTopic é um algoritmo de modelagem de tópicos que facilita a geração de representações temáticas a partir de grandes quantidades de textos. Primeiramente, o algoritmo extrai embeddings dos documentos usando modelos transformadores de sentenças, como o "all-MiniLM-L6-v2". Em seguida, essas embeddings têm sua dimensionalidade reduzida por técnicas como "UMAP", facilitando o processo de agrupamento. A clusterização é realizada usando "HDBSCAN", uma técnica baseada em densidade que identifica clusters de diferentes formas e tamanhos, além de detectar outliers. Posteriormente, os documentos são tokenizados e representados em uma estrutura de bag-of-words, que é normalizada (L1) para considerar as diferenças de tamanho entre os clusters. A representação dos tópicos é refinada usando uma versão modificada do "TF-IDF", chamada "Class-TF-IDF", que considera a importância das palavras dentro de cada cluster (Grootendorst, 2020). Cabe destacar que, antes de aplicar o BERTopic, realizamos a limpeza da base removendo "stopwords" em português, por meio da biblioteca "NLTK" (Loper & Bird, 2002). Para a modelagem de tópicos, utilizamos o backend "loky" para otimizar o desempenho durante o ajuste e a transformação dos dados.

Em síntese, a metodologia aplicada compreendeu desde a extração de dados com a ferramenta própria autoral TelegramScrap (Silva, 2023), até o tratamento e a análise de dados coletados, utilizando diversas abordagens para identificar e classificar comunidades de teorias da conspiração brasileiras no Telegram. Cada uma das etapas foi cuidadosamente implementada para garantir a integridade dos dados e o respeito à privacidade dos usuários, conforme a legislação brasileira prevê. A seguir, serão apresentados os resultados desses dados, com o intuito de revelar as dinâmicas e as características das comunidades estudadas.

## 3. Resultados

A seguir, os resultados são detalhados na ordem prevista na metodologia, iniciando com a caracterização da rede e suas fontes de legitimação e referência, avançando para as séries temporais, incorporando a análise de conteúdo e concluindo com a identificação de sobreposição de agenda temática dentre as comunidades de teorias da conspiração.

### 3.1. Rede

A análise das redes que envolvem as temáticas de anti-*woke*, anti-gênero, revisionismo e discurso de ódio revela a interconectividade e a coesão dessas comunidades dentro do ecossistema conspiratório. A primeira figura mapeia a rede interna entre essas comunidades, destacando como suas narrativas não apenas se sobrepõem, mas se reforçam mutuamente, criando um ambiente propício para a propagação de ideologias extremistas. As grandes esferas que representam os principais nós da rede evidenciam a centralidade de algumas comunidades, que atuam como canais principais de disseminação dessas ideias. A conexão estreita entre anti-*woke* e revisionismo, por exemplo, sugere que esses discursos frequentemente caminham juntos, amplificando o impacto e a ressonância dessas ideologias dentro e fora da rede. Na segunda figura, observamos como as comunidades de anti-*woke*, anti-gênero e revisionismo atuam como portas de entrada para novas temáticas dentro do universo conspiratório. Essas comunidades não apenas introduzem os seguidores a ideias específicas, mas também os conectam a uma rede mais ampla de narrativas extremistas, como o discurso de ódio. A interconexão dessas comunidades com outras narrativas sugere que, uma vez expostos a esses discursos, os indivíduos são rapidamente levados a explorar outras áreas de desinformação, aumentando seu engajamento e radicalização dentro do ecossistema.

A terceira figura ilustra como as comunidades centradas em discursos anti-*woke* e anti-gênero servem como portas de saída para outras teorias da conspiração, como Nova Ordem Mundial e Revisionismo Histórico. A forte interconectividade dessas comunidades com áreas como Conspiração Geral e Ocultismo reflete a capacidade dessas temáticas de alimentar e serem alimentadas por um espectro mais amplo de desinformação. Isso indica que a polarização em torno do tema anti-*woke* não apenas atrai, mas também facilita a transição para outras narrativas conspiratórias, criando um ciclo contínuo de radicalização.

Os gráficos de fluxo de links de convites destacam a importância dessas temáticas dentro do ecossistema conspiratório. A análise dos convites entre comunidades de anti-*woke* e anti-gênero, por exemplo, revela uma forte conexão com NOM e Antivacinas, indicando que essas narrativas estão inseridas em um contexto mais amplo de resistência a mudanças sociais progressistas. Mais intrigante é o fluxo de convites emitidos para Conspirações Gerais e Globalismo, que demonstra como essas temáticas não apenas sustentam, mas também são sustentadas por narrativas mais amplas de controle global. Isso sugere que o discurso anti-*woke* e anti-gênero opera como uma peça-chave em uma resistência mais ampla contra qualquer intervenção estatal percebida como uma ameaça à liberdade individual.

Por fim, o gráfico de fluxos de convites em comunidades de revisionismo e discurso de ódio destaca como o revisionismo histórico é frequentemente introduzido como parte de uma narrativa mais ampla de desconfiança e oposição à ordem estabelecida. A forte interconexão entre revisionismo e outras formas de extremismo, como anti-*woke* e NOM, sugere uma estratégia consciente de alinhamento entre diferentes formas de extremismo, onde o revisionismo serve para solidificar uma visão de mundo adversa ao progresso social. Essa convergência de narrativas demonstra que o revisionismo não é apenas uma negação de fatos históricos, mas uma tentativa de reescrever a história para legitimar formas contemporâneas de opressão, sendo uma parte integral de uma agenda mais ampla de desinformação que visa minar a confiança nas instituições democráticas e na ciência.

[ versão português / english above ]

**Figura 01.** Rede interna entre anti-*woke*, anti-gênero, revisionismo e discurso de ódio

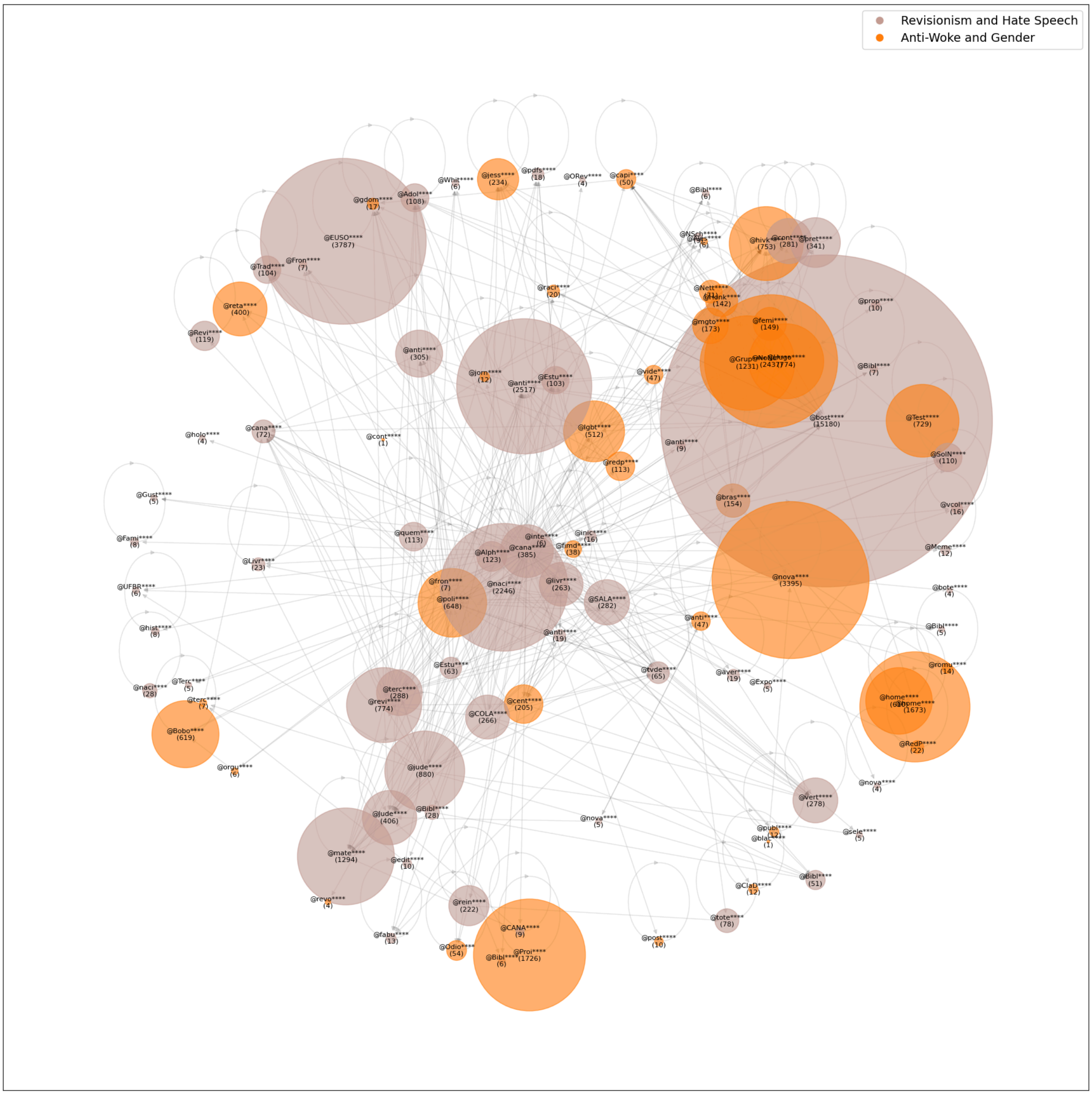

Fonte: Elaboração própria (2024).

Esta figura mapeia a interconexão entre comunidades que promovem discursos anti-*woke*, anti-gênero, revisionistas e ódio. A rede demonstra uma estrutura coesa, onde comunidades que compartilham essas temáticas estão intimamente conectadas. Isso sugere que as narrativas que rejeitam as pautas progressistas e de inclusão estão profundamente interligadas com ideias revisionistas e de ódio, criando um ambiente onde essas ideologias se reforçam mutuamente. Os grandes nós representam os principais pontos de convergência dessas ideias, funcionando como canais principais de disseminação. A sobreposição entre as comunidades de anti-*woke* e revisionismo indica que essas ideologias são muitas vezes apresentadas em conjunto, aumentando a força de sua propagação. O ambiente dessa rede sugere uma forte resistência à mudança e à aceitação da diversidade, o que contribui para a radicalização e a manutenção de crenças preconceituosas e discriminatórias.

**Figura 02.** Rede de comunidades que abrem portas para a temática (porta de entrada)

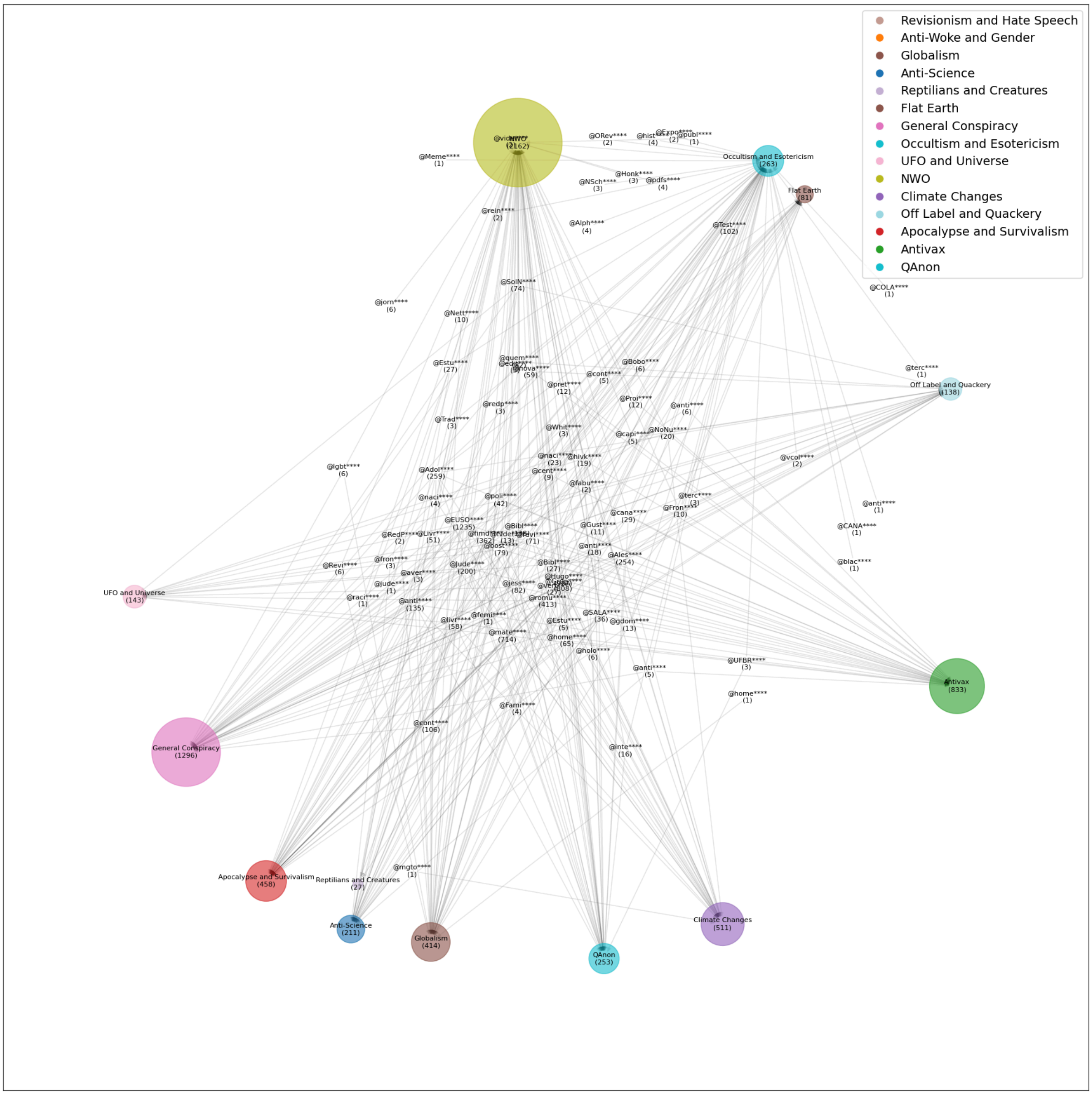

Fonte: Elaboração própria (2024).

Esta figura revela como comunidades que promovem discursos anti-*woke*, anti-gênero, e revisionistas servem como pontos de entrada para novas temáticas dentro da rede conspiratória. A interconexão entre essas comunidades e outras relacionadas, como o discurso de ódio, reflete um ambiente em que essas ideias não apenas se reforçam mutuamente, mas também servem como *gateways* para um engajamento mais profundo em outras narrativas extremas. O gráfico sugere que indivíduos atraídos por discursos anti-*woke* podem rapidamente se ver envolvidos em discussões mais amplas sobre revisionismo e até mesmo narrativas de ódio, tornando essas comunidades centrais para a disseminação de um espectro mais amplo de ideologias conspiratórias.

**Figura 03.** Rede de comunidades cuja temática abre portas (porta de saída)

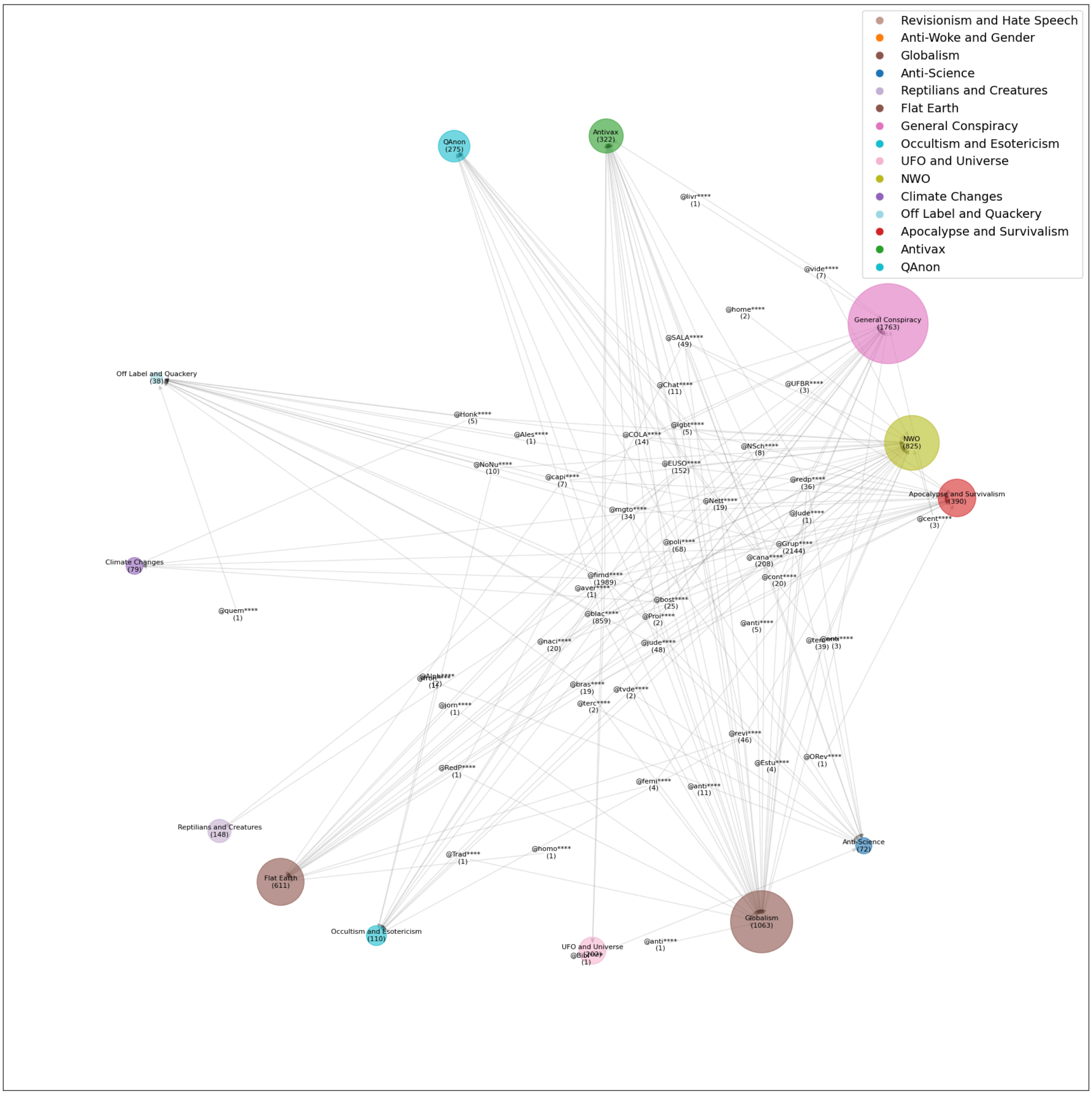

Fonte: Elaboração própria (2024).

No gráfico que ilustra as conexões em torno das temáticas anti-*woke*, observa-se que essas comunidades têm ligações significativas com uma série de outras narrativas conspiratórias. A figura sugere que as discussões anti-*woke* frequentemente servem como ponto de partida para a introdução de outras teorias, como Nova Ordem Mundial e Revisionismo Histórico. A interconectividade das comunidades anti-woke com outras áreas como Conspiração Geral e Ocultismo reflete como a polarização em torno do tema anti-*woke* pode alimentar e ser alimentada por um espectro mais amplo de teorias conspiratórias. Essas conexões indicam que uma vez dentro do universo anti-*woke*, os indivíduos têm fácil acesso a uma rede mais ampla de desinformação, onde as crenças radicais podem ser reforçadas e ampliadas, levando a uma radicalização crescente dentro do ecossistema conspiratório.

**Figura 04.** Fluxo de links de convites entre comunidades de anti-*woke* e anti-gênero

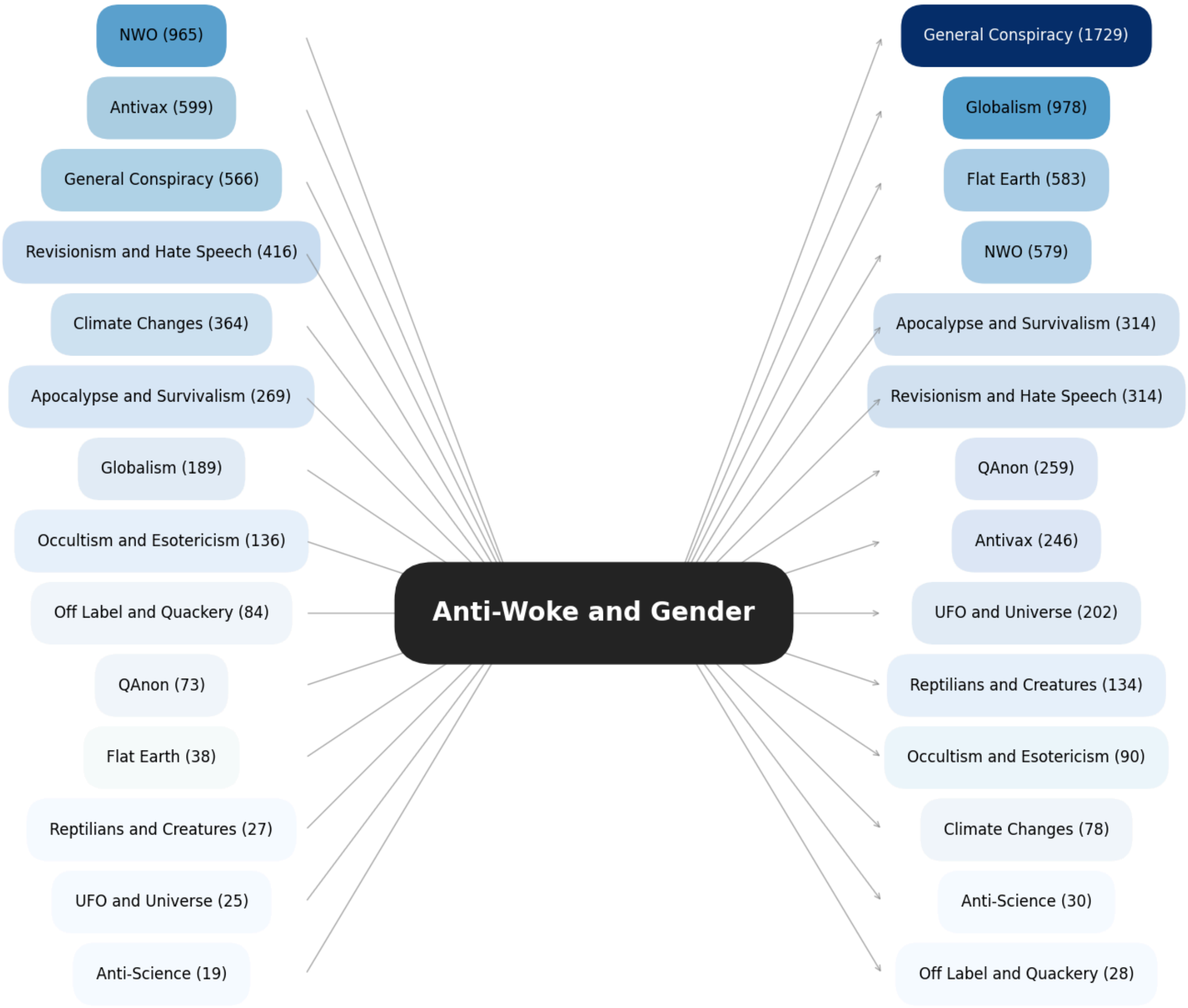

Fonte: Elaboração própria (2024).

O gráfico relacionado ao tema Anti-*Woke* e Gênero evidencia que essa temática atrai um conjunto diversificado de outras teorias, com NOM (965 links) e Antivacinas (599 links) entre as principais. Essa interconexão sugere que o discurso Anti-Woke e Gênero está inserido em um contexto mais amplo de resistência contra mudanças sociais progressistas, onde as narrativas de desconfiança e oposição a políticas de saúde pública, como as antivacinas, encontram terreno fértil. Mais interessante é a direção dos convites emitidos por essas comunidades para Conspirações Gerais (1.729 links) e Globalismo (978 links). Isso revela uma dinâmica em que as ideias contrárias às políticas de diversidade e inclusão são não apenas sustentadas por, mas também sustentam, narrativas mais amplas de controle e manipulação global. O Anti-*Woke* e Gênero, então, não pode ser visto como uma simples resistência a políticas identitárias, mas como uma peça-chave em um mosaico mais amplo de oposição a qualquer forma de intervenção estatal percebida como uma ameaça à liberdade individual, como é frequentemente o caso em narrativas antivacinas e de globalismo.

**Figura 05.** Fluxo de links de convites entre comunidades de revisionismo e discurso de ódio

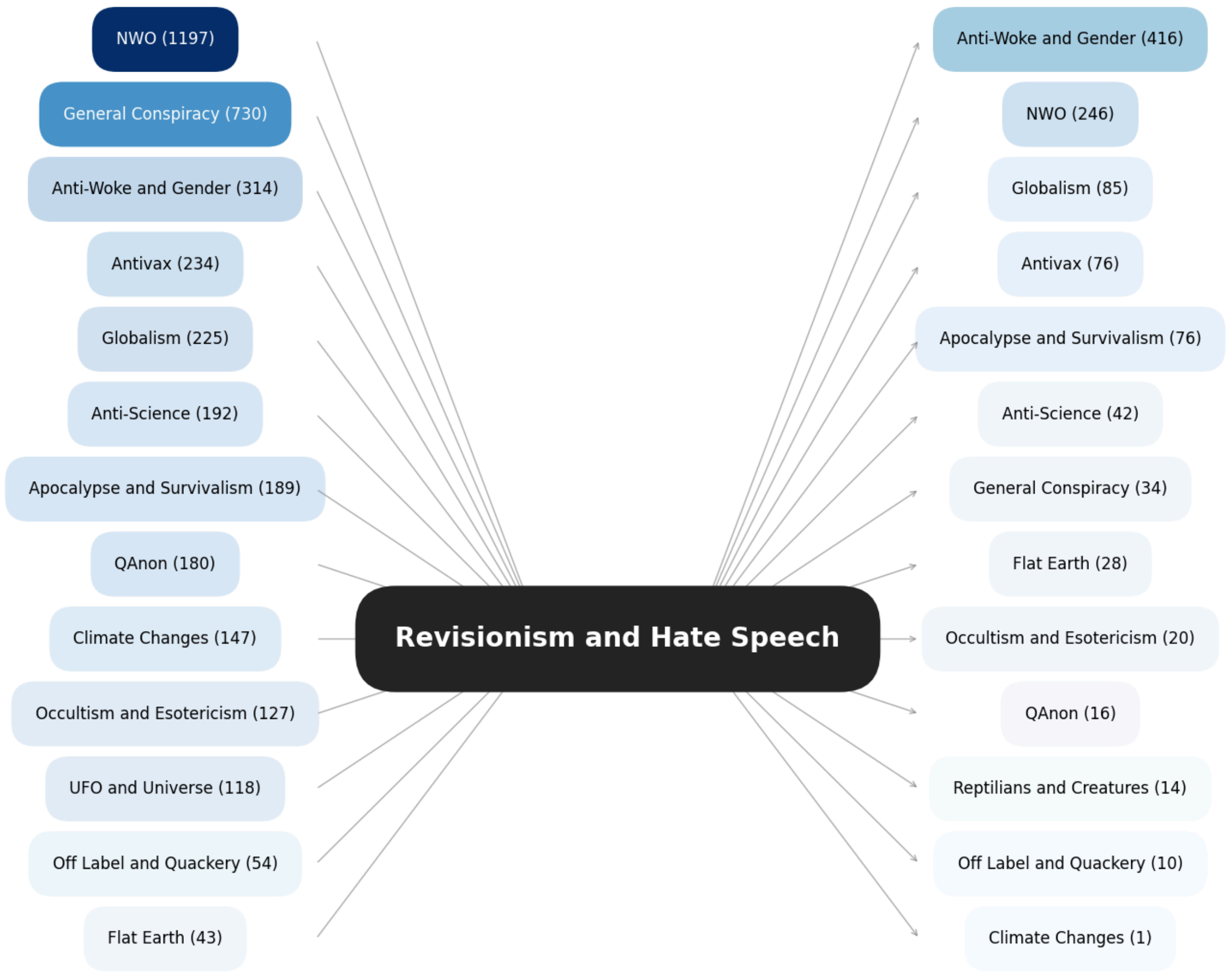

Fonte: Elaboração própria (2024).

O gráfico de Revisionismo e Ódio destaca o papel dessa comunidade como um espaço de articulação entre diferentes formas de extremismo. A NOM (1.197 links) e Conspirações Gerais (730 links) aparecem como as principais fontes, sugerindo que o revisionismo histórico, incluindo a negação de eventos como o Holocausto, é frequentemente introduzido aos indivíduos como parte de uma narrativa mais ampla de desconfiança e oposição à ordem estabelecida. O direcionamento dos convites para Anti-*Woke* e Gênero (416 links) e NOM (246 links) aponta para uma estratégia consciente de alinhamento entre diferentes formas de extremismo, onde o revisionismo serve para solidificar uma visão de mundo que é, por definição, adversa a qualquer forma de progresso social. Essa convergência de narrativas sugere que o revisionismo não é uma simples negação de fatos, mas uma tentativa ativa de reescrever a história para legitimar formas de opressão contemporâneas, sendo parte de uma agenda mais ampla de desinformação que busca minar a confiança nas instituições democráticas e na ciência.

### 3.2. Séries temporais

A análise das séries temporais das categorias Anti-*Woke* e Gênero, bem como Revisionismo e Ódio, revela um aumento significativo nas menções durante os últimos anos. Como veremos no gráfico a seguir, a ascensão das discussões em torno de políticas identitárias e justiça social, especialmente após a eclosão de movimentos como o Black Lives Matter, catalisou debates polarizados, refletidos no crescimento acentuado de menções a Anti-*Woke* e Revisionismo. A partir de 2022, o ritmo dessas discussões se estabiliza, indicando uma menor intensidade, mas ainda assim, uma presença constante. Este fenômeno demonstra como as batalhas culturais se tornaram centrais nas discussões sociais e políticas, absorvendo narrativas de ódio e revisionismo histórico em sua dinâmica.

**Figura 06.** Gráfico de linhas do período

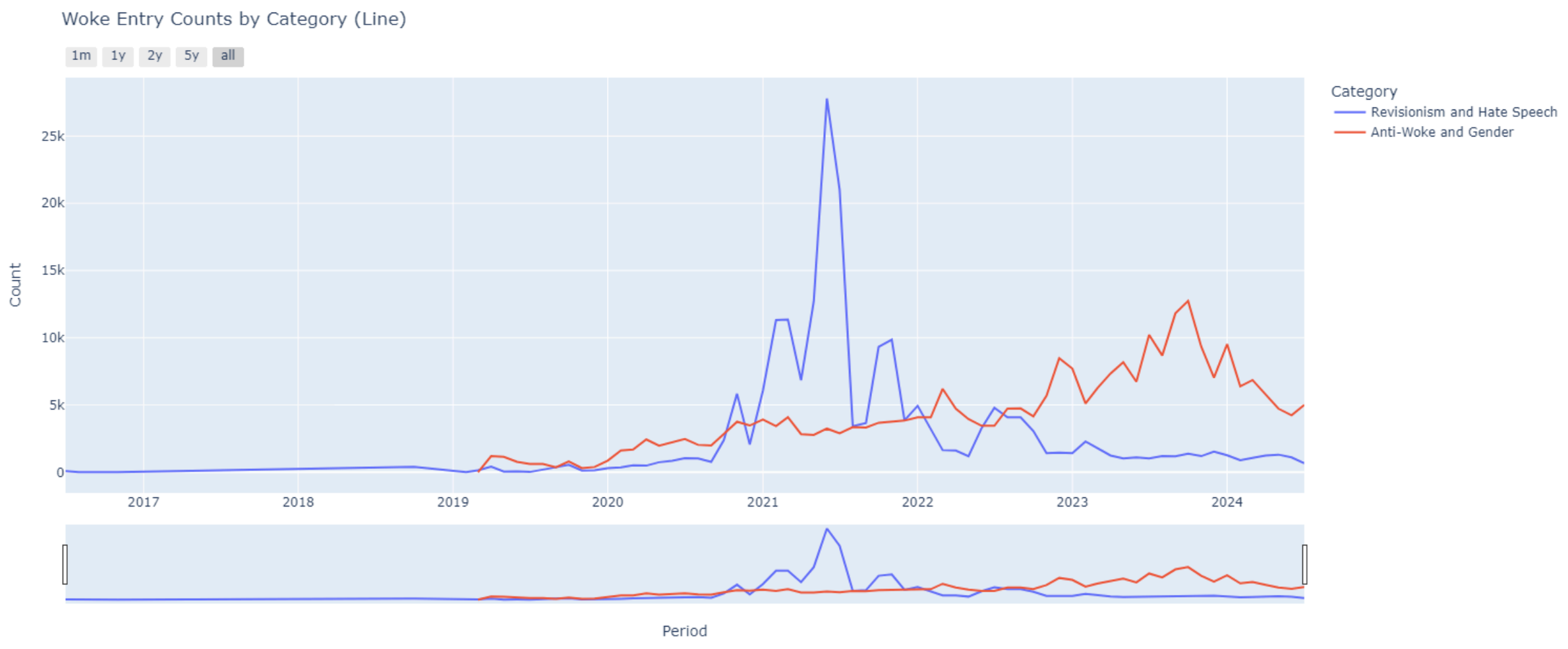

Fonte: Elaboração própria (2024).

As publicações de Anti-*Woke* e Gênero tiveram um aumento de 1.000% em menções entre 2019 e 2021, com picos de 10.000 a 20.000 menções, impulsionados por debates intensos sobre questões de identidade e justiça social. Comparado ao início de 2020, onde as menções estavam em torno de 2.000, o crescimento foi massivo, chegando a 20.000 menções em seu pico. Revisionismo e Ódio segue com um padrão semelhante, crescendo de aproximadamente 1.000 menções para 12.000 em março de 2021, um aumento de 1.100%. Esse crescimento revela como o revisionismo histórico e o discurso de ódio se alimentam das narrativas contra políticas progressistas. A partir de 2022, ambas as categorias mostram uma estabilização com uma queda de cerca de 20% em relação aos picos máximos, indicando que, embora o debate tenha perdido parte de sua intensidade, ele continua presente.

**Figura 07.** Gráfico de área absoluta do período

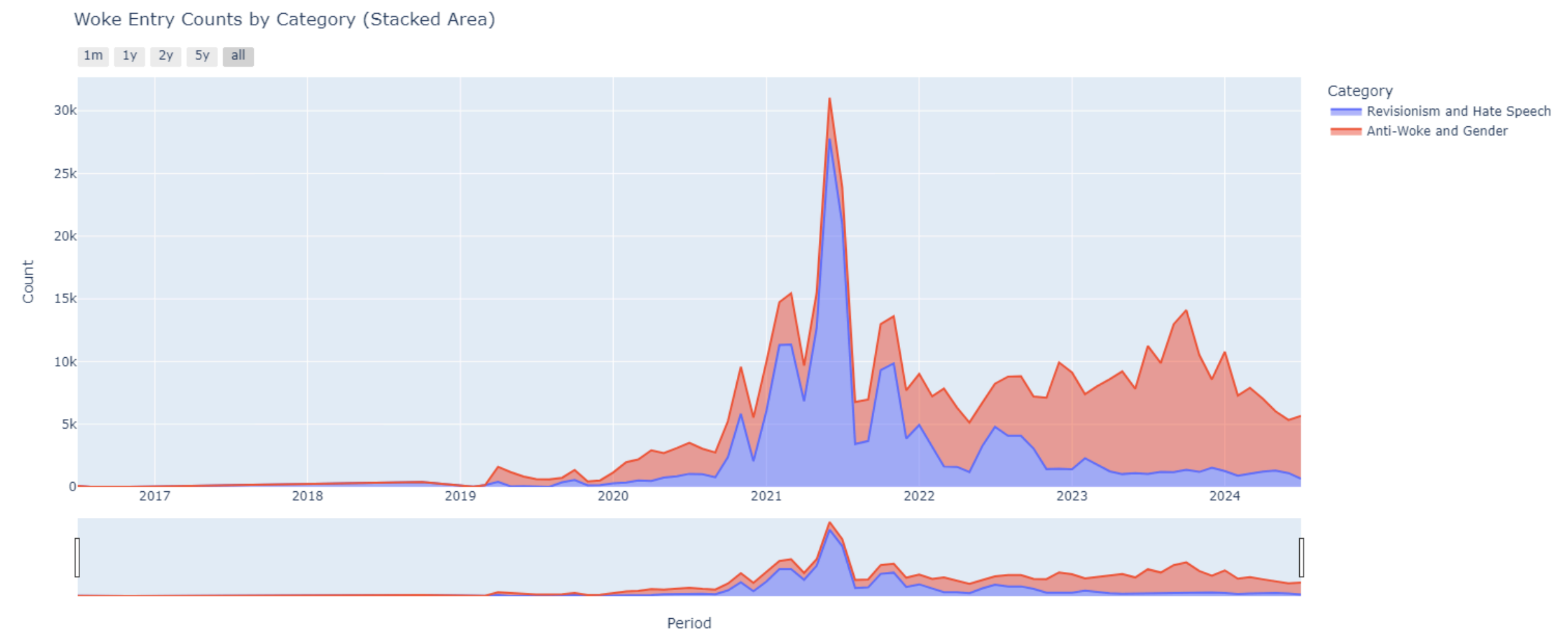

Fonte: Elaboração própria (2024).

Este gráfico mostra um aumento expressivo nas discussões relacionadas à Revisionismo e Ódio e Anti-*Woke* a partir de 2020, com um pico acentuado em 2021. O movimento Anti-*Woke* e Gênero, que domina em termos de volume absoluto, reflete uma crescente reação às políticas de inclusão e justiça social que ganharam destaque globalmente durante e após a Pandemia. Esse pico coincide com eventos como os protestos do *Black Lives Matter* e as discussões sobre identidade de gênero, que polarizaram o debate público, especialmente nos Estados Unidos e em outros países ocidentais. O aumento das menções a Revisionismo e Ódio também é significativo, indicando um ressurgimento de narrativas revisionistas que buscam reescrever ou negar aspectos da história, muitas vezes para justificar ideologias de supremacia ou exclusão. Esses picos podem ser associados à reação contra mudanças sociais percebidas com a ascensão de movimentos neonazistas, ilustrando como crises e transformações sociais podem alimentar tanto o radicalismo quanto a polarização.

**Figura 08.** Gráfico de área relativa do período

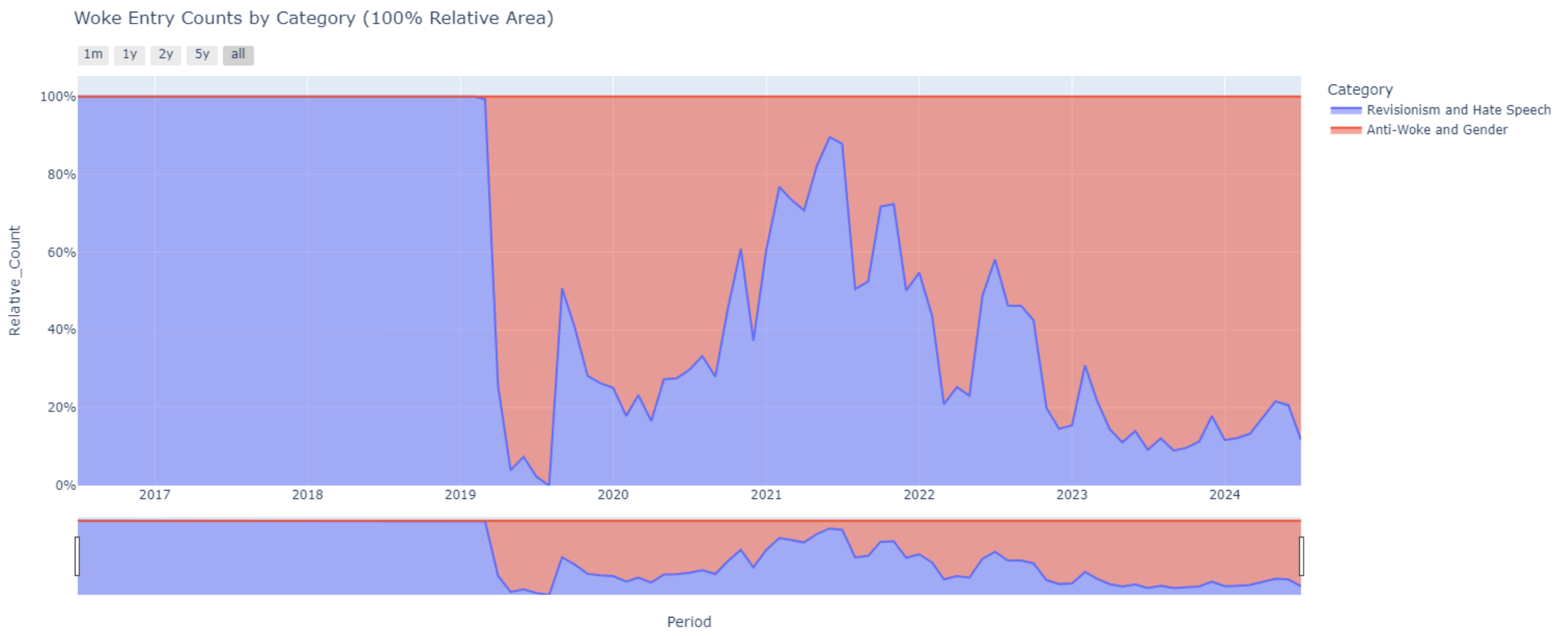

Fonte: Elaboração própria (2024).

No gráfico de área relativa, a categoria Anti-*Woke* e Gênero domina progressivamente a partir de 2020, refletindo a crescente atenção e controvérsia em torno dessas questões. O declínio relativo das menções a Revisionismo e Ódio, especialmente em comparação ao movimento Anti-*Woke*, pode sugerir que, embora o revisionismo histórico permaneça uma preocupação, ele é cada vez mais subsumido dentro de uma narrativa maior de resistência às mudanças culturais e sociais promovidas pelo movimento *Woke*. Essa mudança também pode indicar que, à medida que os debates sobre gênero e justiça social se tornam mais proeminentes, eles absorvem parte do espaço discursivo anteriormente ocupado por narrativas revisionistas. O gráfico sugere que as batalhas culturais se tornaram o campo de disputa mais visível, enquanto o revisionismo histórico continua a desempenhar um papel secundário, mas persistente, dentro das mesmas esferas de radicalização.

### 3.3. Análise de conteúdo

A análise de conteúdo das comunidades relacionadas ao anti-*woke*, revisionismo e discurso de ódio, por meio de nuvens de palavras, revela os principais temas e dinâmicas discursivas que sustentam essas narrativas dentro do ambiente digital. Observando os termos mais frequentes, como ”Brasil”, ”mundo”, ”Deus”, e ”judeu”, é possível perceber uma interseção de tópicos que combinam sentimentos nacionalistas com uma retórica religiosa e de exclusão social. A presença desses termos indica a maneira como essas comunidades buscam construir uma identidade coletiva baseada em princípios de oposição ao que consideram ameaças culturais e sociais, especialmente ligadas a movimentos progressistas e a influências externas, frequentemente associadas a conspirações globais. Através das palavras destacadas, torna-se evidente que essas comunidades estão ativamente envolvidas na construção e disseminação de narrativas que visam reforçar divisões e promover uma visão de mundo polarizada, onde a alteridade é constantemente vilanizada e marginalizada. A análise das nuvens de palavras não apenas identifica os termos mais recorrentes, mas também oferece um panorama sobre como essas discussões evoluem e se adaptam às mudanças contextuais, revelando as estratégias discursivas utilizadas para manter e expandir essas ideologias.

**Figura 09.** Nuvem de palavras consolidadas de anti-*woke*, revisionismo e discurso de ódio

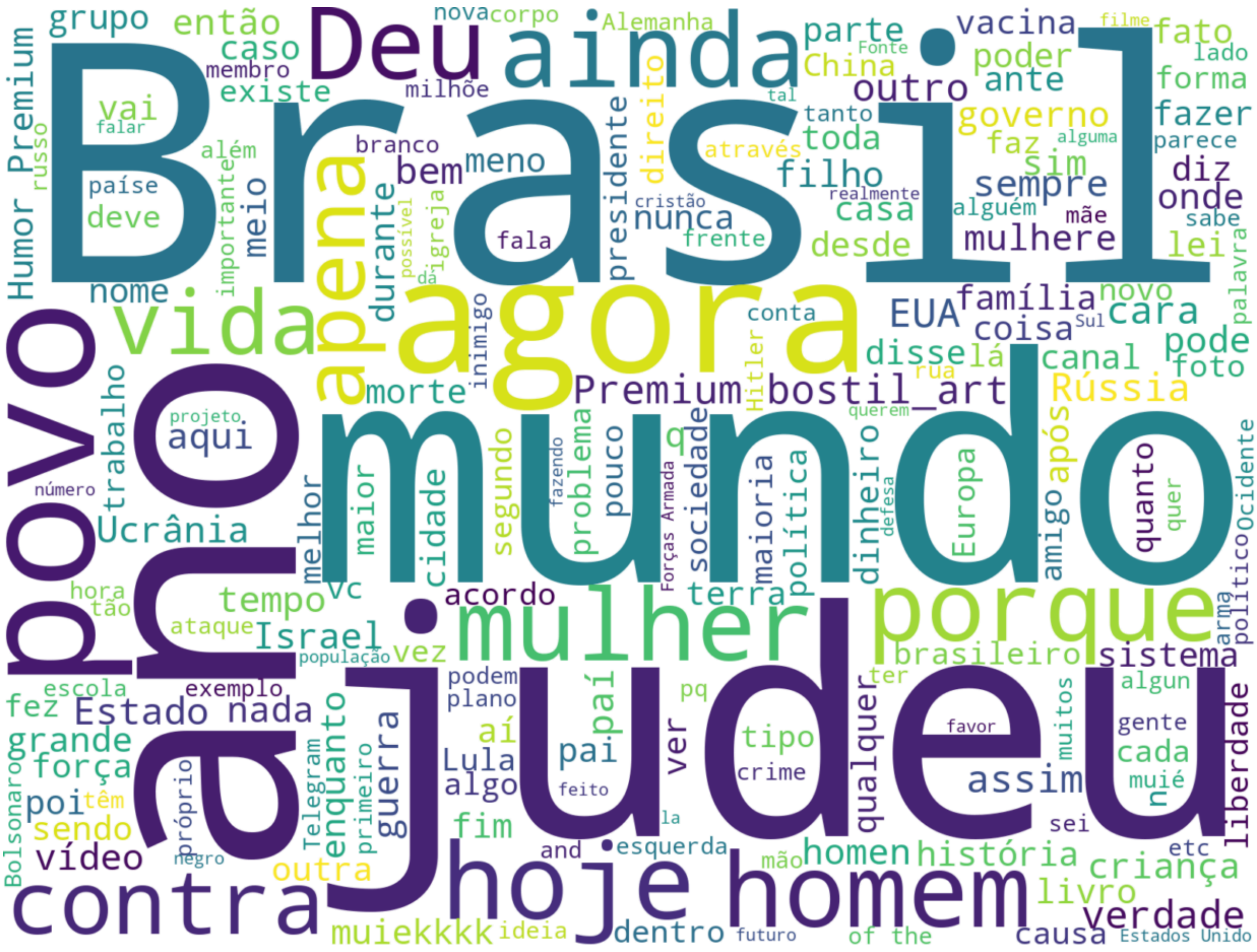

Fonte: Elaboração própria (2024).

A nuvem de palavras consolidada das discussões envolvendo anti-*woke*, revisionismo e discurso de ódio destaca termos centrais como ”Brasil”, ”mundo”, ”Deus”, ”agora”, e ”judeu”. A palavra ”Brasil” ocupa um lugar de destaque, indicando uma forte presença de sentimentos nacionalistas que permeiam essas comunidades. Esse termo é frequentemente associado a narrativas de defesa do país contra influências externas ou internas que são percebidas como ameaças. ”Mundo” e ”Deus” aparecem com frequência, sugerindo que as discussões nesses grupos não se limitam ao contexto nacional, mas também envolvem uma dimensão espiritual e global, onde questões religiosas são mobilizadas para justificar posições ideológicas. A presença de ”judeu” na nuvem de palavras reflete uma forte componente antissemita dentro dessas comunidades, onde o ódio contra judeus é frequentemente articulado como parte de uma teoria conspiratória mais ampla. Termos como ”agora” e ”mulher” indicam uma preocupação com o tempo presente e questões de gênero, respectivamente, evidenciando que essas discussões estão profundamente enraizadas em temas sociais contemporâneos. A combinação desses termos revela como as narrativas anti-*woke* e revisionistas são construídas sobre uma base que mistura nacionalismo, religião, e exclusão social, criando um discurso que busca reforçar divisões e polarizações na sociedade.

**Quadro 01.** Nuvem de palavras em série temporal de anti-*woke* e anti-gênero

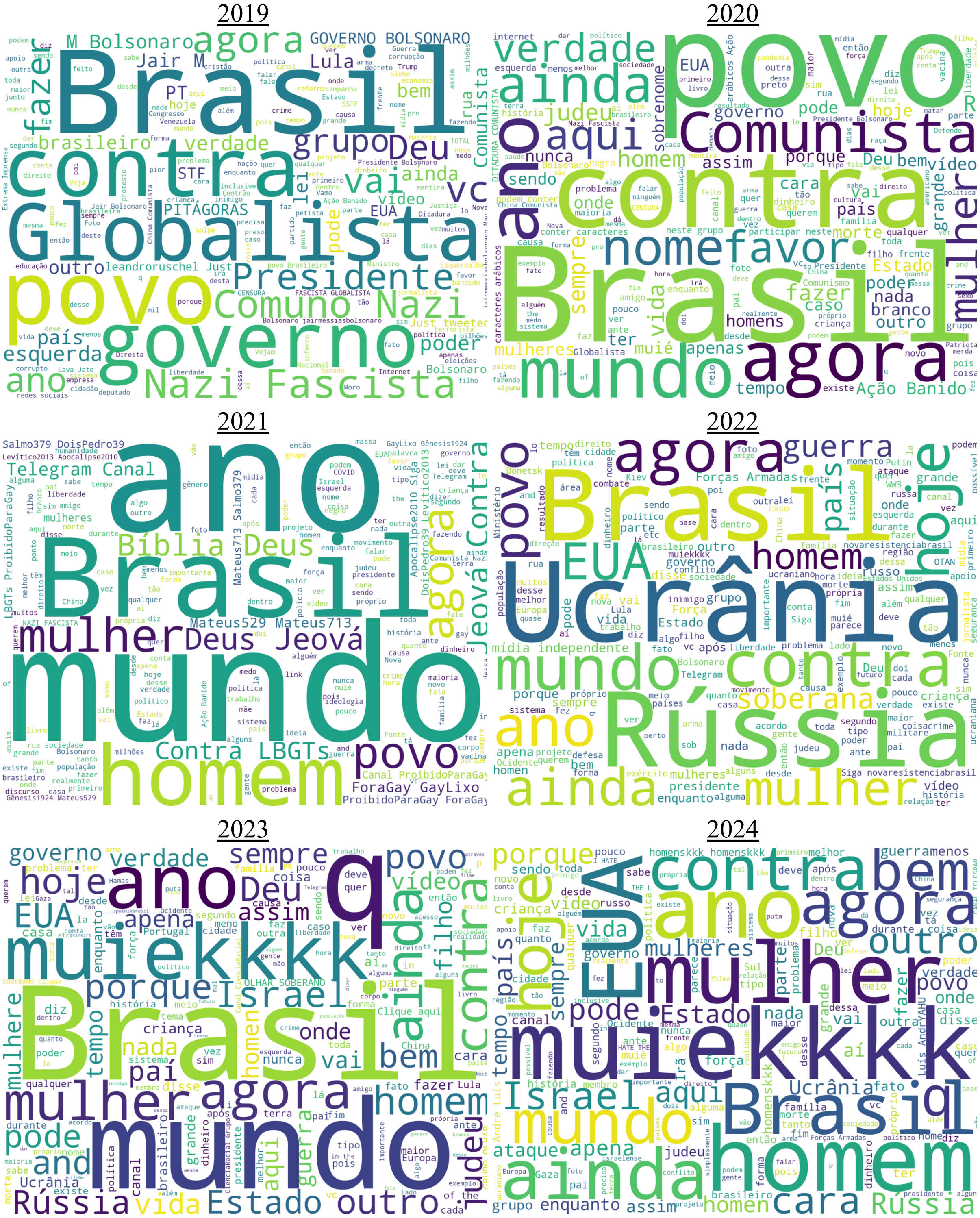

Fonte: Elaboração própria (2024).

No Quadro 01, que analisa a evolução temporal das discussões de anti-woke e anti-gênero, observa-se que, ao longo dos anos, o termo ”Brasil” mantém uma presença constante e destacada, refletindo uma contínua preocupação com questões nacionais e políticas internas. Em 2019, ”contra” e ”povo” emergem como termos significativos, sugerindo uma retórica de oposição e resistência contra movimentos progressistas. A partir de 2020, há um aumento na presença de termos como ”comunista” e ”globalista”, indicando um crescimento das teorias que ligam questões de gênero a conspirações globais. Em 2021, a palavra ”ano” ganha destaque, possivelmente refletindo um ano de particular importância para as discussões nesses grupos, talvez relacionado a eventos políticos ou sociais significativos. Nos anos subsequentes, especialmente em 2023 e 2024, ”mundo” e ”mulher” tornam-se mais proeminentes, indicando uma expansão do foco das discussões para incluir temas globais e questões de gênero, reforçando a ideia de que esses grupos veem as questões de gênero como parte de uma luta maior contra uma suposta agenda globalista.

**Quadro 02.** Nuvem de palavras em série temporal de revisionismo e discurso de ódio

2016

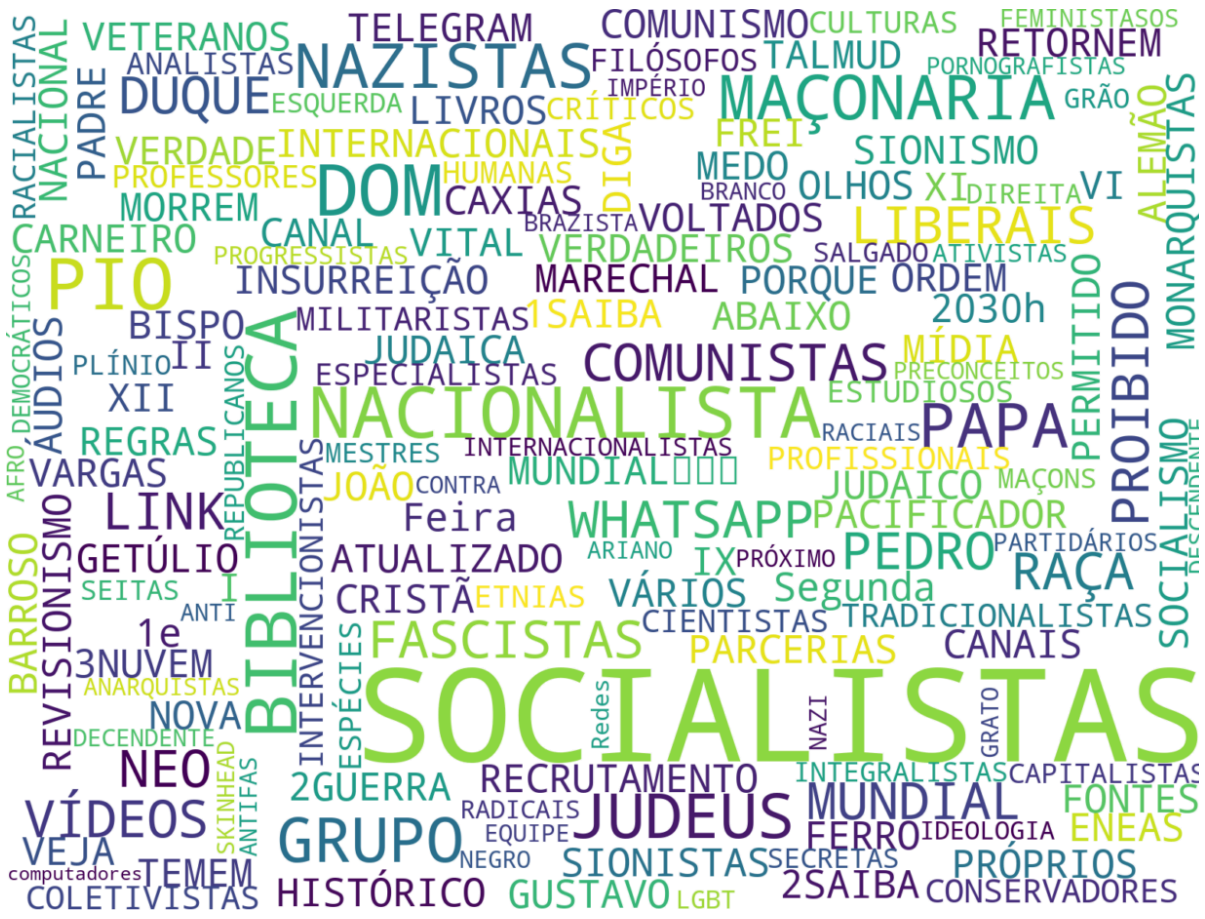

2019

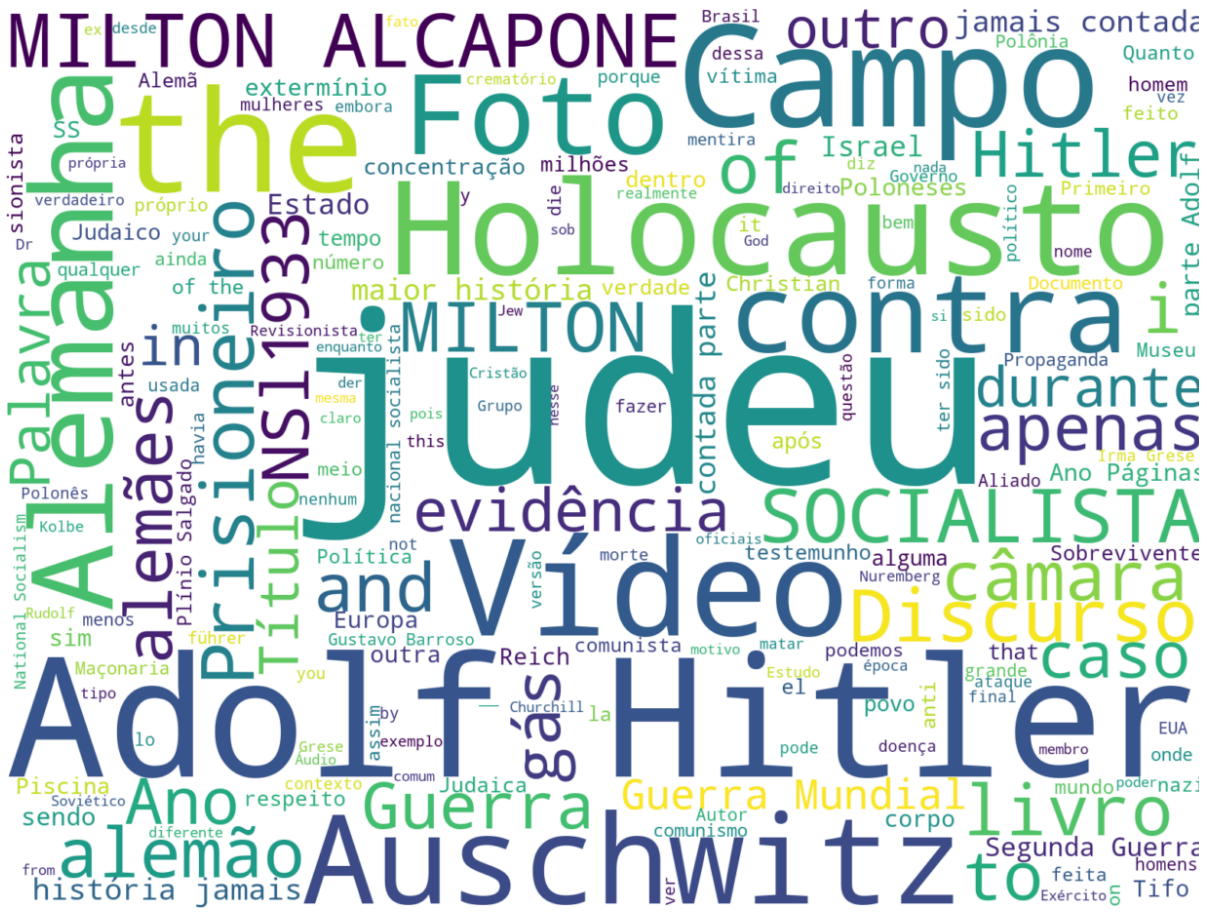

2020

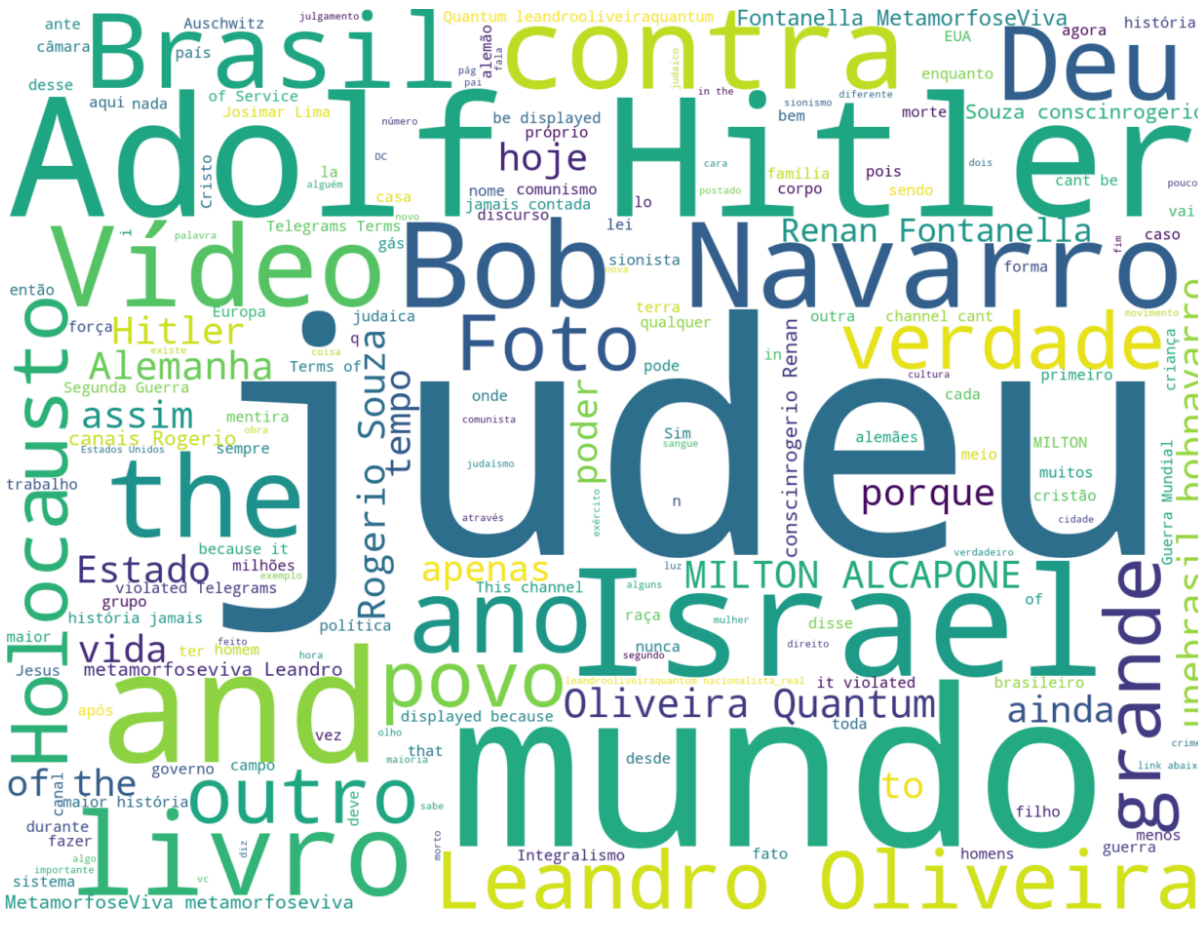

2021

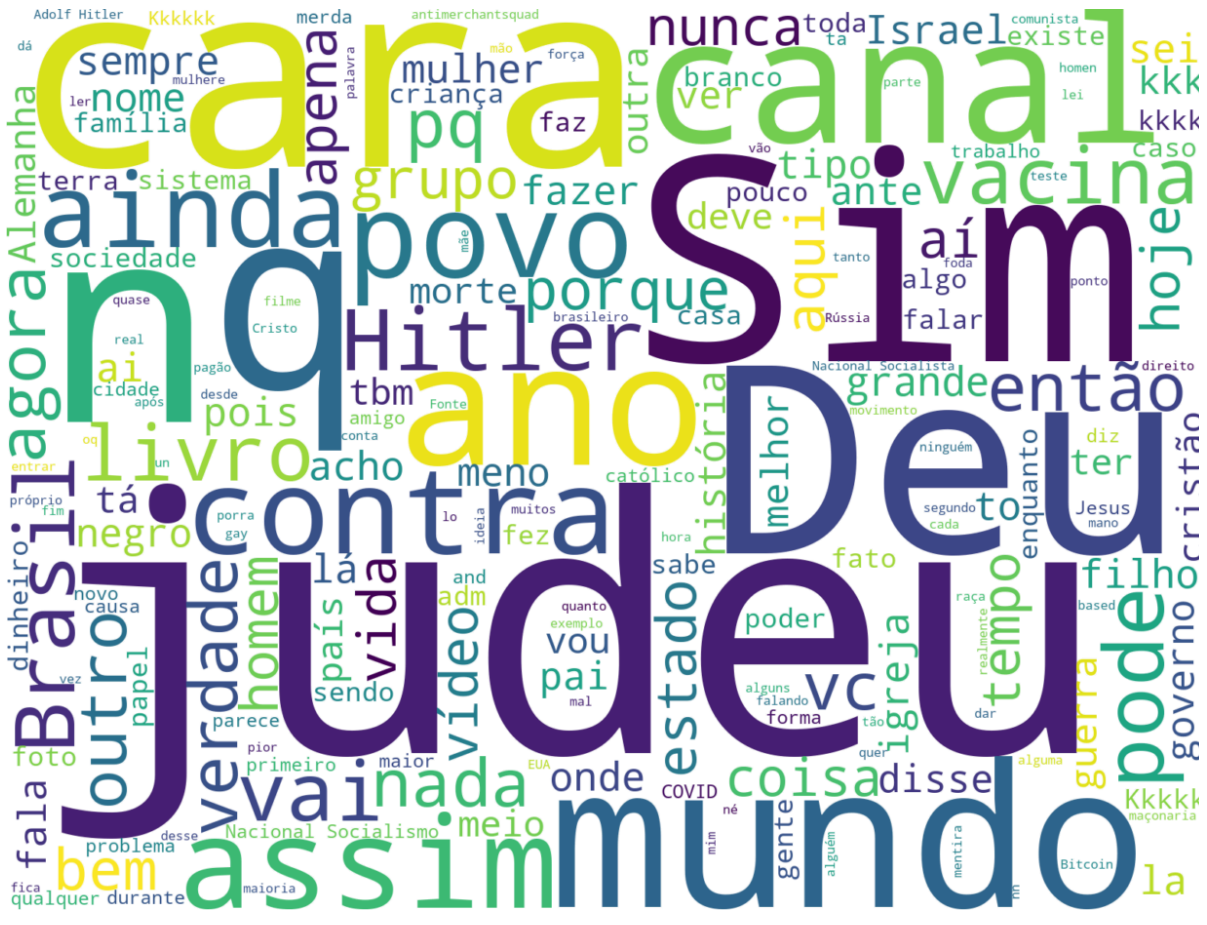

[ versão português / english above ]

2022

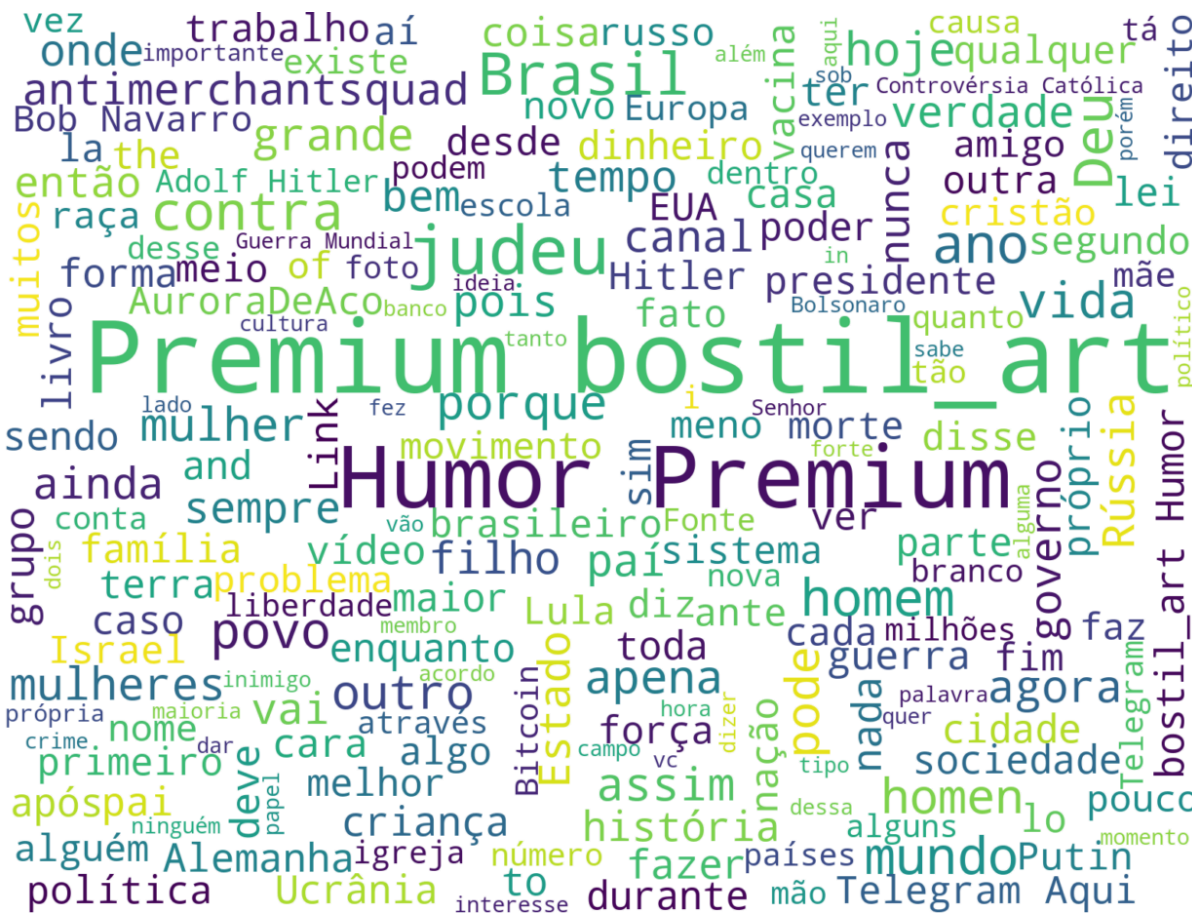

2023

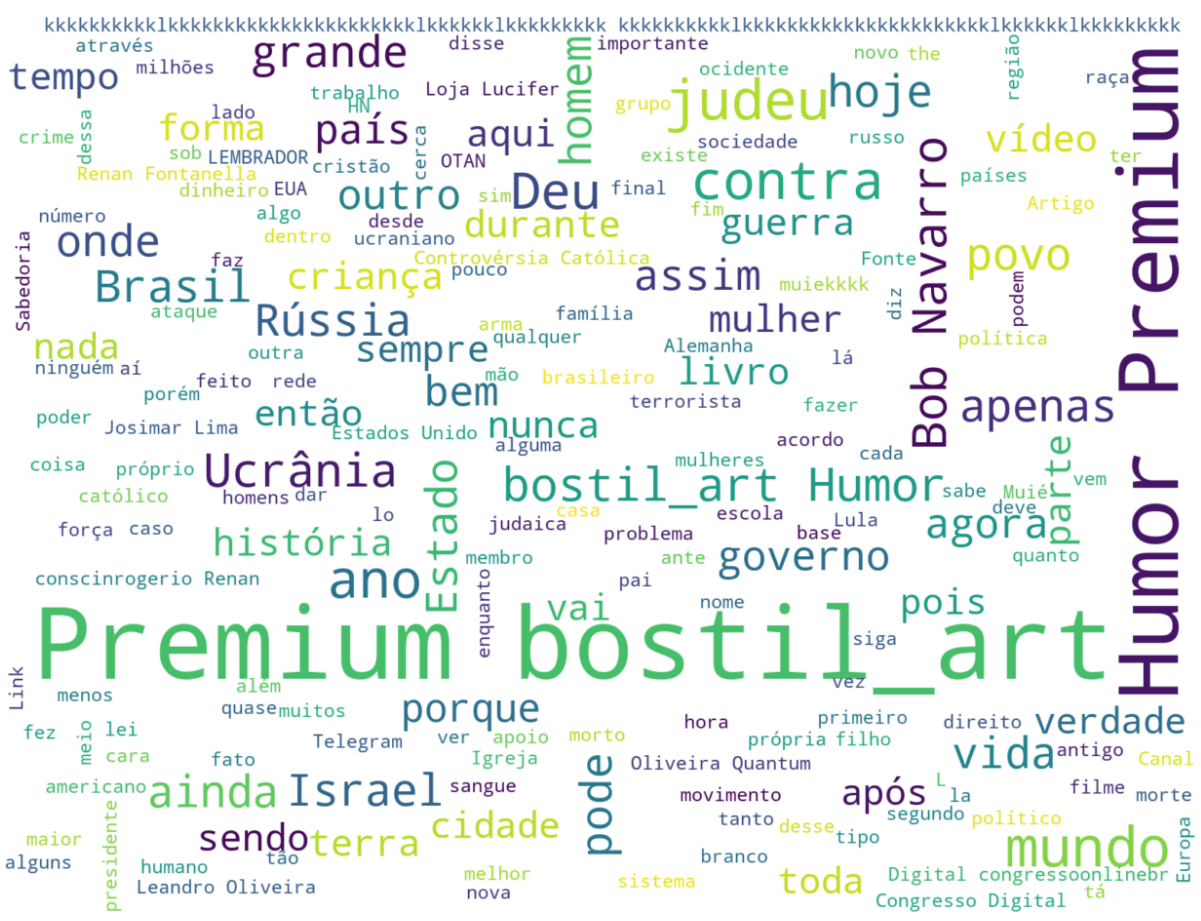

2024

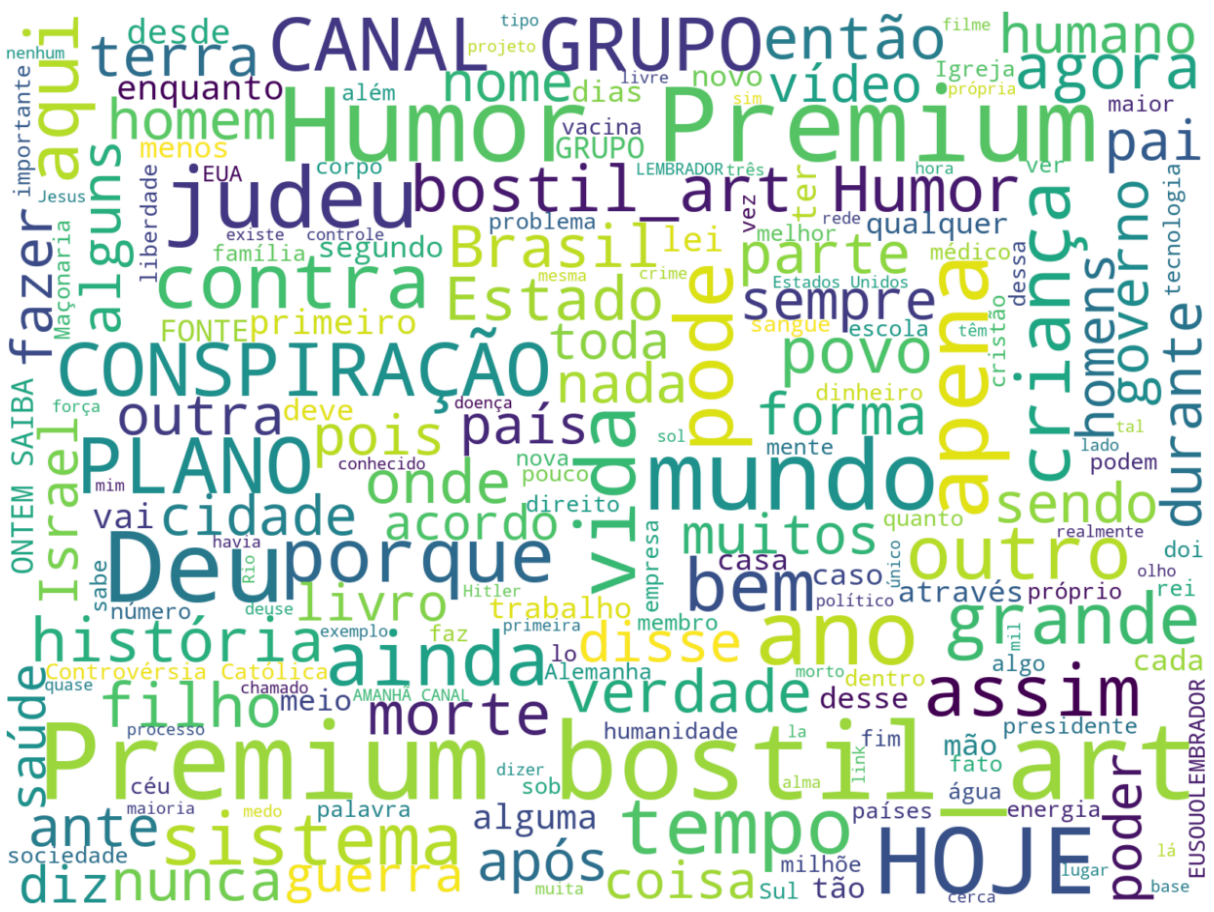

Fonte: Elaboração própria (2024).

O Quadro 02, focado no revisionismo e discurso de ódio, revela que, desde 2016, termos como ”judeu” e ”Hitler” são recorrentes, destacando a persistência de narrativas antissemitas e revisionistas dentro dessas comunidades. Em 2019, o termo ”Holocausto” aparece com destaque, indicando um ano de particular importância para as discussões revisionistas sobre a Segunda Guerra Mundial. Ao longo dos anos, ”mundo” e ”Brasil” permanecem constantes, sugerindo uma conexão entre questões globais e nacionais dentro dessas discussões de ódio. Em 2022, a presença de termos como ”Premiu” e ”bostil” (gíria pejorativa para se referir ao Brasil) sugere um discurso de desvalorização e ódio que vai além do antisemitismo, abrangendo também uma crítica ao próprio país. Em 2023 e 2024, ”judeu” continua a ser um termo central, refletindo a persistência de discursos de ódio contra essa comunidade, enquanto ”mundo” mantém sua relevância, indicando que essas discussões revisionistas e de ódio têm uma dimensão global, onde o Brasil é apenas um dos contextos em que essas narrativas são aplicadas.

### 3.4. Sobreposição de agenda temática

As figuras apresentadas a seguir analisam como as comunidades de teorias da conspiração sobre Anti-*Woke* e Gênero, além de Revisionismo e Discurso de Ódio, interagem e se sobrepõem a outras narrativas conspiratórias. A análise dessas temáticas revela a centralidade desses discursos dentro das comunidades e como eles são utilizados para reforçar crenças extremistas e desinformação, conectando-se com temas mais amplos como fé, geopolítica, e moralismo tradicionalista. Essas conexões ampliam o alcance dessas teorias, tornando mais complexa a intervenção e a correção factual.

**Figura 10.** Temáticas de fé e religião

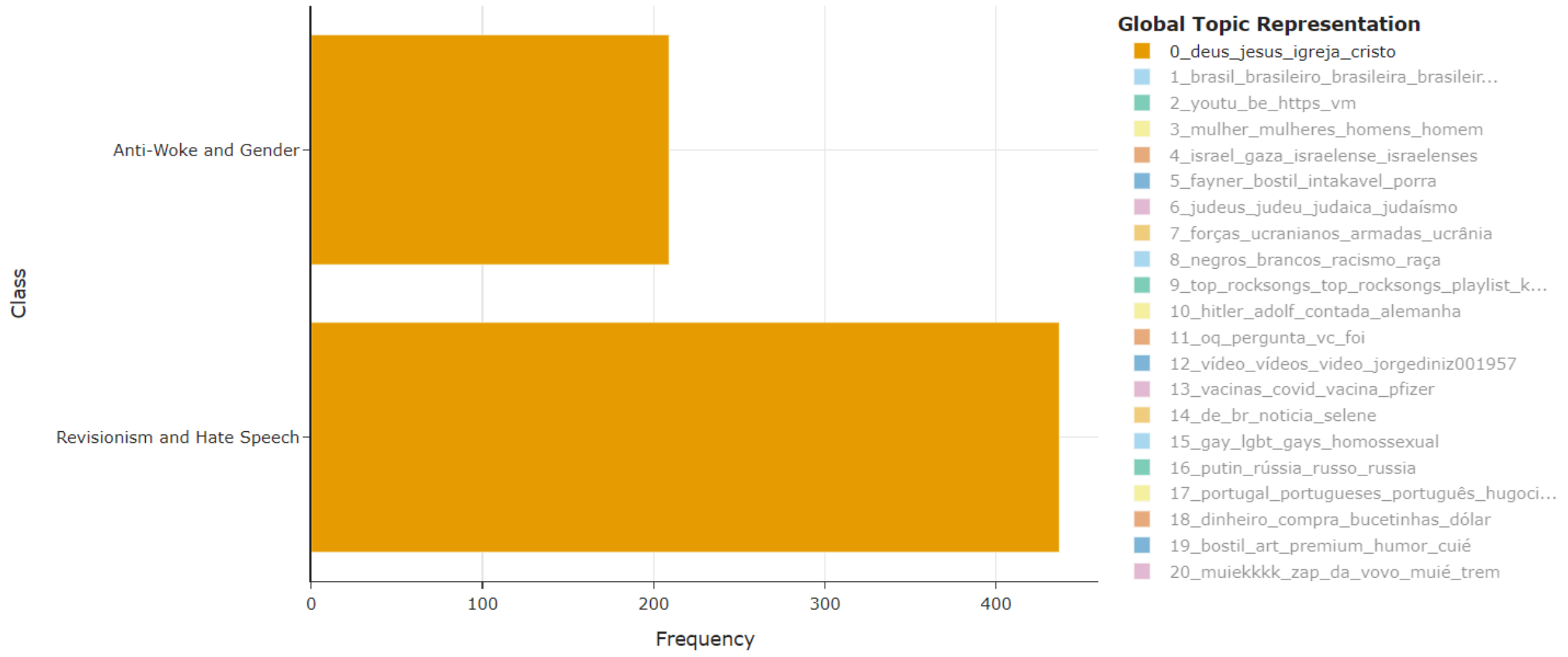

Fonte: Elaboração própria (2024).

Na Figura 10, observamos a interseção entre temáticas de fé e religião com discursos Anti-*Woke* e Revisionismo. Tópicos como “Deus”, “Jesus”, e “igreja” são amplamente discutidos, sugerindo que essas comunidades utilizam elementos religiosos para legitimar seus discursos contra movimentos de inclusão social e igualdade de gênero. A associação desses temas com Revisionismo e Discurso de Ódio reforça a ideia de que a fé é instrumentalizada para justificar a resistência a mudanças sociais, posicionando tais crenças como defesa de valores religiosos tradicionais.

**Figura 11.** Temáticas de disputas geopolíticas e globalismo

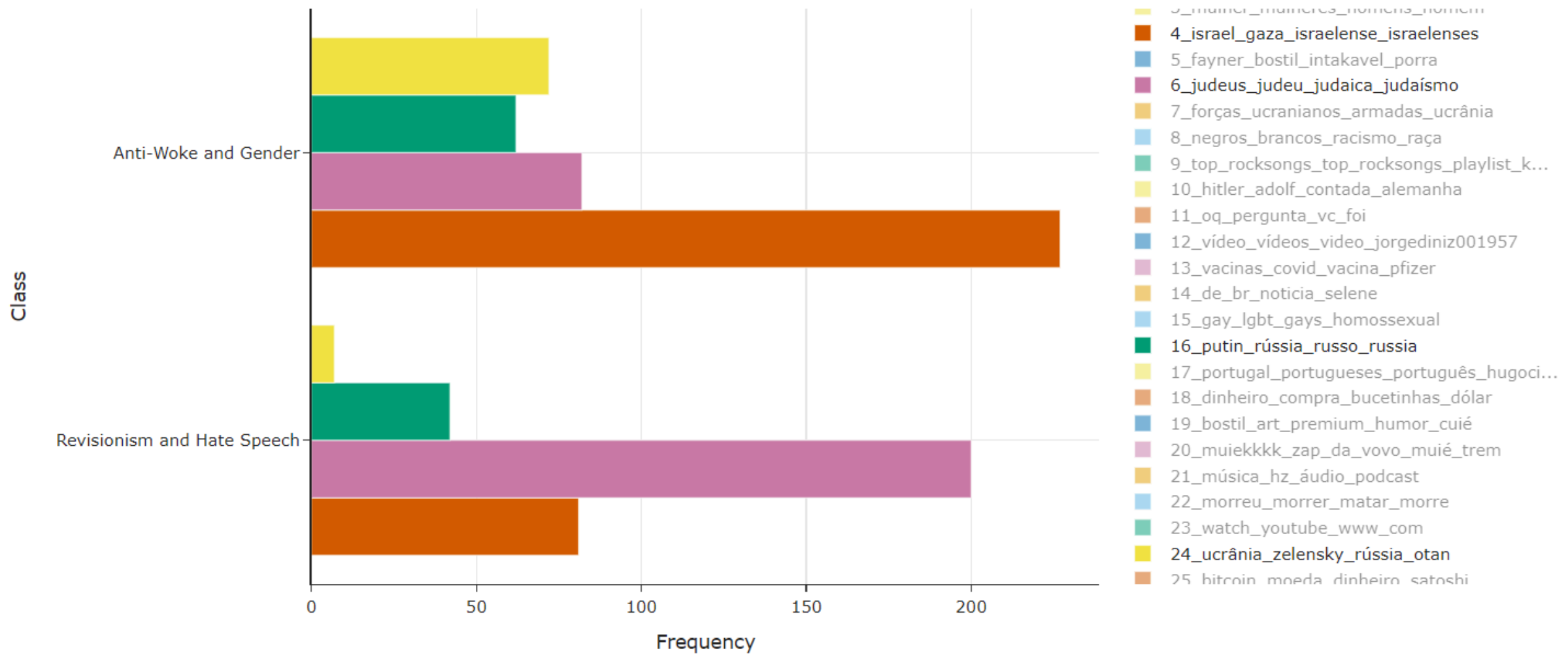

Fonte: Elaboração própria (2024).

A Figura 11 explora a interconexão entre discursos Anti-*Woke* e Revisionismo com temas de geopolítica e globalismo. Tópicos como “Putin”, “Rússia” e “Israel” aparecem com destaque, mostrando que essas comunidades frequentemente vinculam suas narrativas a eventos globais, utilizando essas questões para fortalecer suas agendas conspiratórias. A sobreposição com o globalismo sugere que essas comunidades veem conflitos geopolíticos como parte de uma conspiração maior para implementar uma agenda globalista, que supostamente ameaça seus valores e ideologias.

**Figura 12.** Temáticas de misoginia, racismo e anti-LGBT

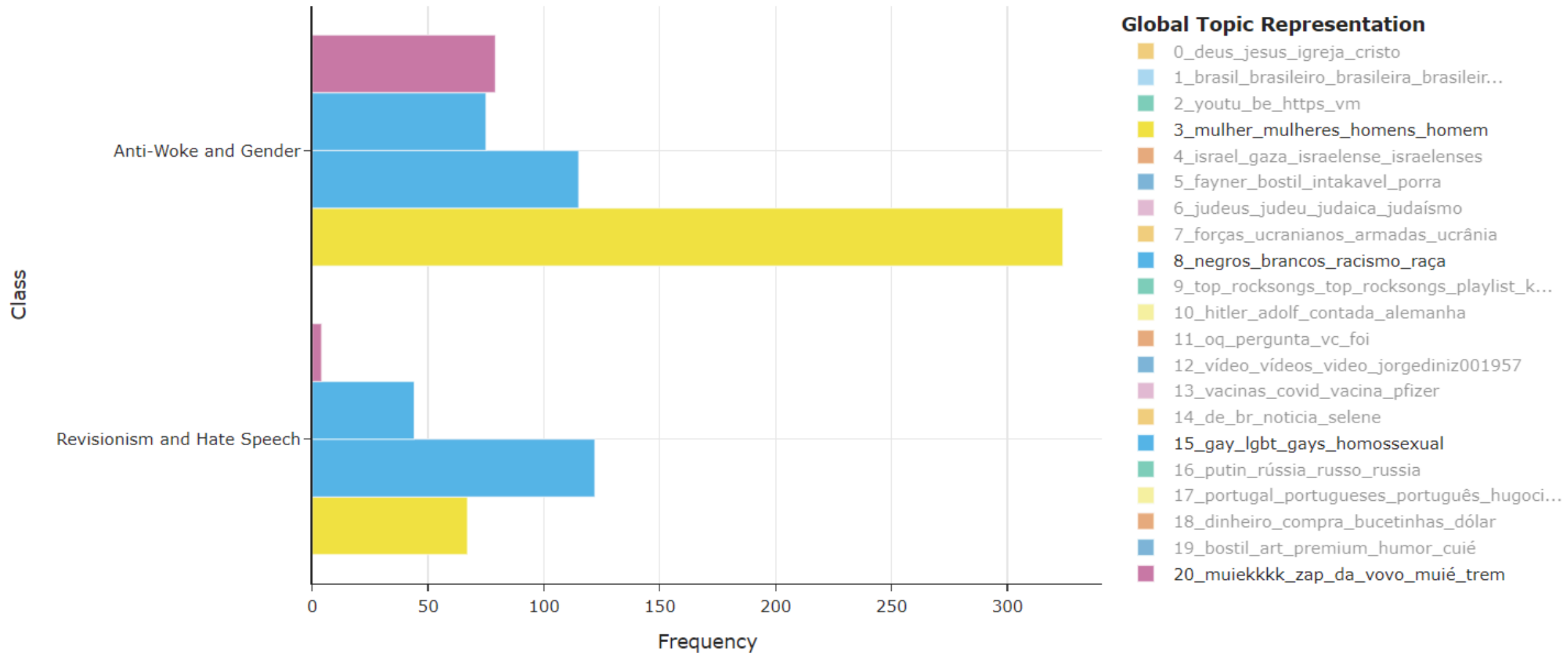

Fonte: Elaboração própria (2024).

Na Figura 12, são analisadas temáticas de misoginia, racismo e discursos anti-LGBT dentro das comunidades Anti-*Woke* e de Revisionismo. Tópicos como “mulher”, “homens”, e “raça” indicam uma forte presença de discursos que atacam minorias e promovem a desigualdade de gênero e raça. Essa sobreposição demonstra como o ódio e a discriminação são centrais para as narrativas dessas comunidades, utilizando questões de identidade e orientação sexual como alvos principais para reforçar seus discursos de ódio e exclusão.

**Figura 13.** Temáticas de negacionismo de vacinas

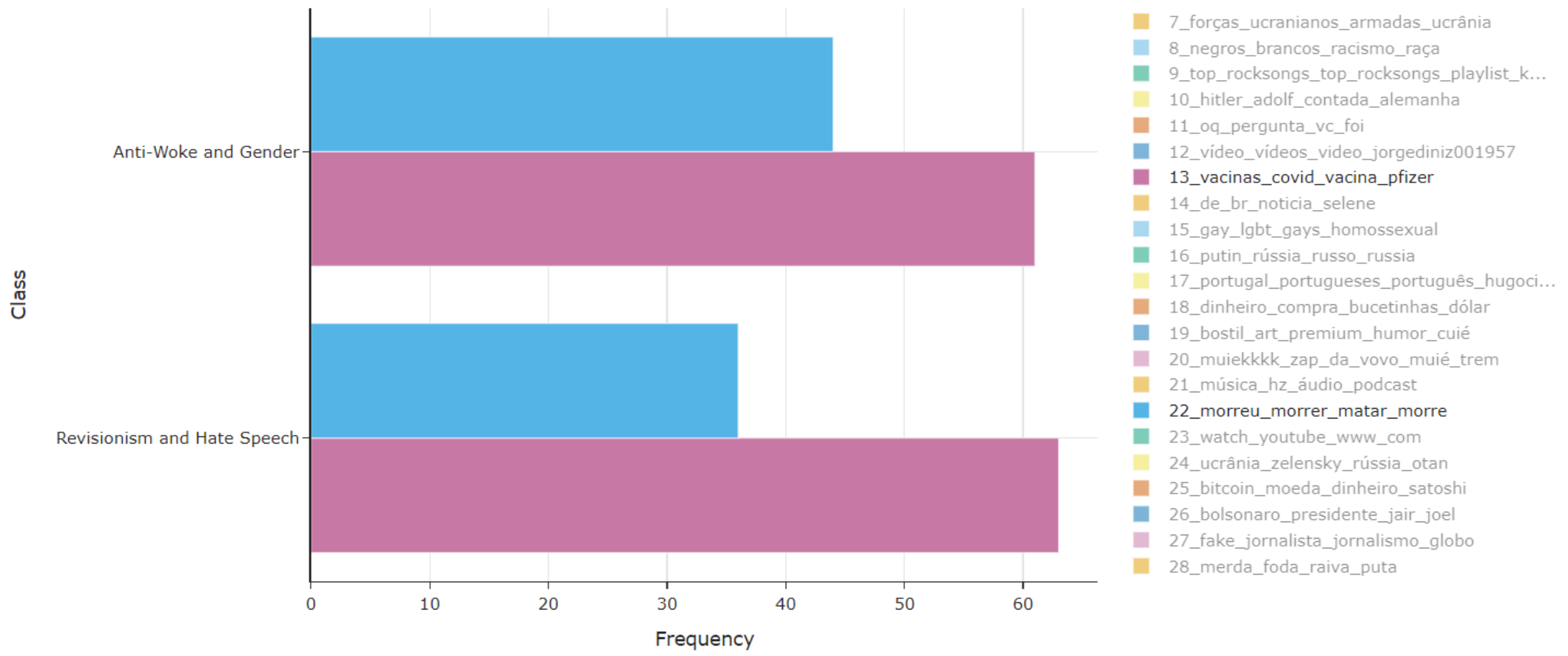

Fonte: Elaboração própria (2024).

A Figura 13 destaca como o negacionismo de vacinas está relacionado a discursos Anti-*Woke* e Revisionismo dentro dessas comunidades. Tópicos como “vacinas”, “Pfizer” e “covid” são prevalentes, sugerindo que essas comunidades utilizam o negacionismo de vacinas como uma extensão de suas agendas contra a ciência e as políticas públicas de saúde. A conexão com o Revisionismo evidencia uma tentativa de reescrever a narrativa científica, promovendo desinformação que alimenta a desconfiança em relação à vacinação e à ciência.

**Figura 14.** Temáticas de moralismo tradicionalista

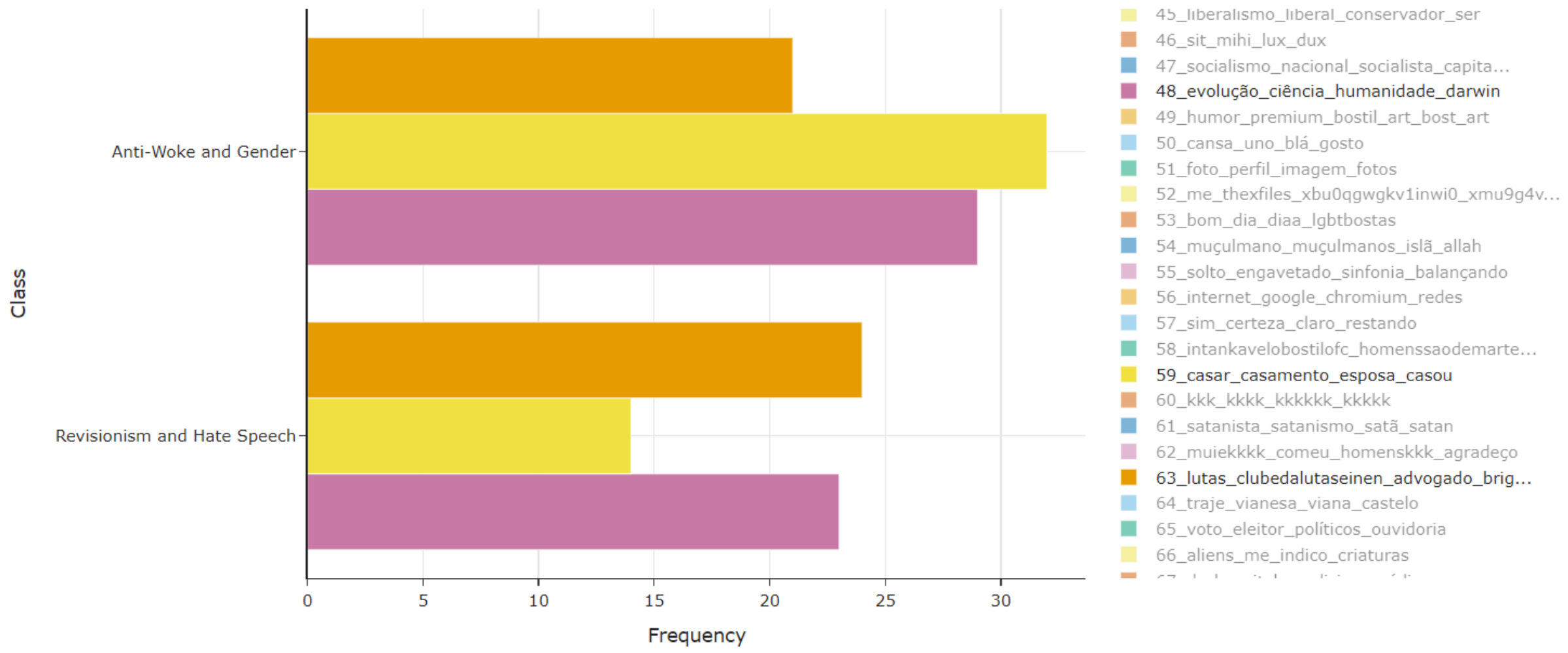

Fonte: Elaboração própria (2024).

Na Figura 14, observamos como as temáticas de moralismo tradicionalista se sobrepõem aos discursos Anti-*Woke* e Revisionismo. Tópicos como “evolução”, “darwin” e “liberalismo” indicam que essas comunidades utilizam uma narrativa moralista para justificar sua oposição a ideias progressistas e científicas. A presença de temas relacionados à moralidade e à ordem social sugere que esses discursos são utilizados para promover uma visão conservadora da sociedade, onde qualquer desvio das normas tradicionais é visto como uma ameaça à ordem estabelecida.

**Figura 15.** Temáticas de satanismo e fé

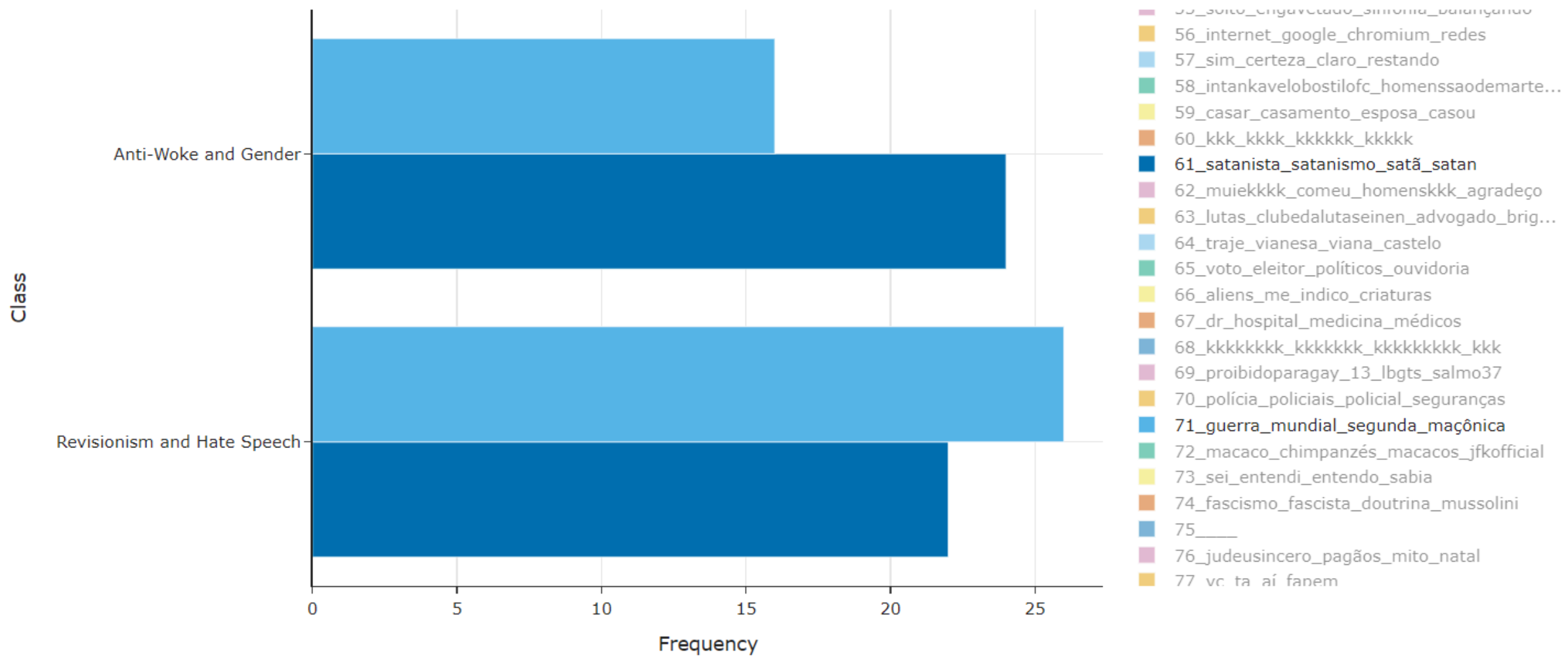

Fonte: Elaboração própria (2024).

A Figura 15 explora a associação entre temáticas de satanismo e fé com discursos Anti-*Woke* e Revisionismo. Tópicos como “satanismo”, “fé”, e “igreja” são frequentemente abordados, mostrando que essas comunidades utilizam o medo do satanismo e a manipulação religiosa para reforçar suas narrativas. A sobreposição com Revisionismo e Discurso de Ódio sugere que essas comunidades não só rejeitam movimentos sociais modernos, mas também buscam demonizar seus opositores, associando-os a práticas religiosas malignas.

## 4. Reflexões e trabalhos futuros

Para responder a pergunta de pesquisa “**como são caracterizadas e articuladas as comunidades de teorias da conspiração brasileiras sobre temáticas de anti-*woke*, anti-gênero, revisionismo e discurso de ódio no Telegram?**”, este estudo adotou técnicas espelhadas em uma série de sete publicações que buscam caracterizar e descrever o fenômeno das teorias da conspiração no Telegram, adotando o Brasil como estudo de caso. Após meses de investigação, foi possível extrair um total de 109 comunidades de teorias da conspiração brasileiras no Telegram sobre temáticas de anti-*woke*, anti-gênero, revisionismo e discurso de ódio, estas somando 1.640.149 de conteúdos publicados entre julho de 2016 (primeiras publicações) até agosto de 2024 (realização deste estudo), com 188.771 usuários somados dentre as comunidades.

Foram adotadas quatro abordagens principais: **(i)** Rede, que envolveu a criação de um algoritmo para mapear as conexões entre as comunidades por meio de convites circulados entre grupos e canais; **(ii)** Séries temporais, que utilizou bibliotecas como “Pandas” (McKinney, 2010) e “Plotly” (Plotly Technologies Inc., 2015) para analisar a evolução das publicações e engajamentos ao longo do tempo; **(iii)** Análise de conteúdo, sendo aplicadas técnicas de análise textual para identificar padrões e frequências de palavras nas comunidades ao longo dos semestres; e **(iv)** Sobreposição de agenda temática, que utilizou o modelo BERTopic (Grootendorst, 2020) para agrupar e interpretar grandes volumes de textos, gerando tópicos coerentes a partir das publicações analisadas. A seguir, as principais reflexões são detalhadas, sendo seguidas por sugestões para trabalhos futuros.

### 4.1. Principais reflexões

**A agenda anti-*woke* se consolida como uma das principais forças de resistência no ecossistema conspiratório brasileiro:** O estudo identificou que as comunidades anti-*woke* são centrais dentro do universo conspiratório no Telegram brasileiro. Com 1.293.430 publicações e 154.391 usuários ativos, essas comunidades não só rejeitam políticas progressistas de inclusão social, mas também promovem uma visão de mundo onde a diversidade é vista como uma ameaça à ordem estabelecida. A consolidação dessa agenda reflete uma resistência organizada contra mudanças sociais, tornando-se uma peça-chave nas narrativas de desinformação;

**Crescimento exponencial de menções a discursos de ódio e revisionismo durante crises globais:** Durante eventos críticos como as eleições de 2018 no Brasil e o auge da Pandemia da COVID-19, houve um crescimento acentuado nas menções a revisionismo e discursos de ódio, com um aumento de 1.100% entre 2019 e 2021. Esse aumento é um reflexo da polarização exacerbada durante esses períodos, onde a desinformação encontrou terreno fértil para se expandir e solidificar nas comunidades digitais;

**Comunidades nazistas brasileiras no Telegram propagam ideologias extremistas e glorificam Hitler:** Uma das descobertas mais alarmantes do estudo foi a identificação de dezenas de comunidades brasileiras no Telegram que explicitamente defendem ideologias nazistas, incluindo a glorificação de Hitler e a disseminação de discursos de ódio contra grupos minoritários. Essas comunidades representam uma grave ameaça, pois promovem a radicalização e a normalização de ideologias extremistas, comprometendo a segurança social e fomentando um ambiente de ódio;

**Interconectividade entre anti-*woke*, anti-gênero e revisionismo fortalece um ecossistema de ódio e exclusão:** As comunidades anti-*woke* e anti-gênero frequentemente se sobrepõem a discursos de ódio e revisionismo histórico, criando uma rede coesa que amplifica as ideologias extremistas. A análise de rede mostra que 1.729 links entre essas comunidades indicam um fluxo constante de ideias que reforçam uma narrativa de resistência contra as mudanças sociais, onde o ódio e a exclusão são normalizados;

**Portas de entrada: o anti-*woke* como caminho para outras narrativas conspiratórias:** As comunidades anti-*woke* funcionam como portas de entrada para outras teorias da conspiração, como Nova Ordem Mundial e revisionismo. Com 965 links direcionados a NOM e 599 para antivacinas, essas comunidades não apenas disseminam suas próprias ideologias, mas também introduzem seus membros a um espectro mais amplo de desinformação, criando uma rede de radicalização interconectada;

**Volume total e influência: o impacto das comunidades anti-*woke* e anti-gênero no Telegram:** O volume total de publicações relacionadas ao anti-*woke* e anti-gênero no Telegram alcançou 1.293.430, demonstrando a significativa influência dessas comunidades. Com 43 grupos e 154.391 usuários, essas comunidades atuam como *hubs* de desinformação, onde as narrativas contrárias à diversidade e inclusão são não só promovidas, mas também amplificadas para além de suas fronteiras digitais;

**A forte correlação entre revisionismo histórico e discursos anti-gênero:** O estudo revela que o revisionismo histórico está intimamente ligado às discussões anti-gênero, com 416 links de convites mostrando essa interconectividade. Esse alinhamento sugere que as mesmas forças que impulsionam a negação de fatos históricos, como o Holocausto, são as que rejeitam as políticas de igualdade de gênero, formando uma agenda coerente de oposição ao progresso social;

**Radicalização acelerada: o papel dos discursos anti-*woke* na polarização social:** A análise temporal das discussões anti-*woke* e revisionismo mostra uma aceleração na radicalização dos discursos, especialmente durante o período de 2020 a 2021, onde houve um aumento de 1.000% nas menções a essas temáticas. Esse crescimento reflete como as crises globais são utilizadas para intensificar a polarização e promover narrativas de ódio dentro dessas comunidades;

**Anti-*woke* como uma narrativa de resistência contra a globalização:** As comunidades anti-*woke* são fortemente interligadas com narrativas de resistência à globalização, mostrando uma sobreposição significativa com as temáticas de Globalismo e NOM. Com 978 links relacionados ao globalismo, essas comunidades posicionam a agenda anti-*woke* como uma defesa contra uma suposta imposição de valores globalistas que ameaçam a soberania e a identidade nacional;

**Discursos anti-gênero como facilitadores da disseminação de desinformação antivacinas:** O estudo destaca que as comunidades anti-gênero não só rejeitam as políticas de igualdade, mas também facilitam a disseminação de desinformação antivacinas. Com 599 links identificados, essas comunidades utilizam a narrativa antivacinas como uma extensão de sua oposição ao que percebem como uma ameaça global à liberdade individual, criando uma interseção entre saúde pública e ideologias extremistas.

### 4.2. Trabalhos futuros

Com base nos achados deste estudo, futuras pesquisas devem focar na exploração das dinâmicas internas das comunidades anti-*woke* e revisionistas, especialmente no que diz respeito à instrumentalização da fé e moralidade para promover narrativas de ódio e exclusão. Entender como essas comunidades utilizam a retórica religiosa e moralista para justificar suas posições pode revelar os mecanismos que fortalecem a resistência contra as mudanças sociais e a diversidade. Além disso, seria interessante investigar como essas narrativas são adaptadas e reformuladas em resposta a diferentes contextos culturais e crises globais, como a Pandemia da COVID-19 ou as eleições, para compreender a plasticidade e resiliência dessas ideologias.

Outro ponto crucial para futuras pesquisas é o mapeamento das interconectividades entre anti-*woke*, revisionismo e outras teorias conspiratórias, como o negacionismo de vacinas e o globalismo. A análise detalhada dessas conexões pode revelar padrões de radicalização que atravessam diferentes narrativas e identificar quais grupos ou indivíduos funcionam como "super spreaders" de desinformação. Essa investigação pode também explorar como essas redes de desinformação se estruturam, operam e se adaptam às intervenções externas, oferecendo insights sobre como intervenções específicas podem ser mais eficazes para desmantelar essas redes de desinformação.

Adicionalmente, a identificação e monitoramento das comunidades explicitamente nazistas no Telegram brasileiro devem ser uma prioridade em futuras pesquisas. Dada a gravidade dessa descoberta, estudos focados nessas comunidades poderiam explorar as estratégias utilizadas para recrutar e radicalizar novos membros, além de examinar o impacto dessas ideologias na sociedade em geral. Pesquisas que investiguem as interseções entre essas comunidades e outras redes de ódio, como as anti-*woke* e anti-gênero, podem fornecer uma visão abrangente das ameaças que essas ideologias extremistas representam.

Finalmente, estudos futuros devem considerar o impacto de eventos globais na intensificação dessas narrativas, como observado durante a Pandemia da COVID-19 e crises políticas significativas. Pesquisas que explorem a relação entre crises globais e o crescimento de discursos extremistas podem ajudar a desenvolver estratégias mais eficazes para mitigar a disseminação de desinformação em momentos de vulnerabilidade social. Além disso, seria relevante investigar a eficácia das intervenções digitais, como a moderação de conteúdo e campanhas de correção factual, para interromper a propagação de desinformação e discurso de ódio em plataformas como o Telegram.

Outras direções para pesquisas incluem a análise longitudinal dessas comunidades para identificar tendências e mudanças nas narrativas ao longo do tempo. Estudar como as comunidades reagem a intervenções, sejam elas tecnológicas ou políticas, pode oferecer uma compreensão mais profunda das dinâmicas de resistência e adaptação. Além disso, a pesquisa pode explorar o papel das comunidades internacionais e como influências externas moldam as discussões e práticas dentro das comunidades brasileiras. Entender essas interações globais pode ser crucial para o desenvolvimento de estratégias de combate à desinformação que sejam culturalmente sensíveis e mais eficazes.

Por fim, seria essencial investigar as consequências *offline* dessas comunidades, incluindo como as narrativas propagadas online podem influenciar comportamentos e atitudes no mundo real. Isso pode abranger desde a organização de eventos extremistas até atos de violência motivados por ideologias promovidas nesses espaços. Pesquisas que conectem as atividades online com impactos sociais tangíveis podem ajudar a moldar políticas públicas e intervenções que busquem mitigar os efeitos negativos dessas comunidades na sociedade.

## 5. Referências

## 6. Biografia do autor

**Ergon Cugler de Moraes Silva** possui mestrado em Administração Pública e Governo (FGV), MBA pós-graduação em Ciência de Dados e Análise (USP) e bacharelado em Gestão de Políticas Públicas (USP). Ele está associado ao Núcleo de Estudos da Burocracia (NEB FGV), colabora com o Observatório Interdisciplinar de Políticas Públicas (OIPP USP), com o Grupo de Estudos em Tecnologia e Inovações na Gestão Pública (GETIP USP), com o Monitor de Debate Político no Meio Digital (Monitor USP) e com o Grupo de Trabalho sobre Estratégia, Dados e Soberania do Grupo de Estudo e Pesquisa sobre Segurança Internacional do Instituto de Relações Internacionais da Universidade de Brasília (GEPSI UnB). É também pesquisador no Instituto Brasileiro de Informação em Ciência e Tecnologia (IBICT), onde trabalha para o Governo Federal em estratégias contra a desinformação. Brasília, Distrito Federal, Brasil. Site: https://ergoncugler.com/.